\documentclass{svjour3}  

\usepackage{fix-cm}
\smartqed  
\usepackage{graphicx}
\newif\ifSPACEHACK
\SPACEHACKfalse

\newif\ifDEBUG
\DEBUGfalse

\newif\ifANONYMOUS
\ANONYMOUSfalse

\newif\ifANNOTATED
\ANNOTATEDfalse

\usepackage{amsmath,amssymb,amsfonts}
\usepackage{graphicx}
\usepackage{textcomp}
\usepackage{xcolor}
\usepackage{booktabs} 
\usepackage{xspace} 
\usepackage[normalem]{ulem}
\usepackage{makecell}
\usepackage{tcolorbox}
\usepackage{enumitem}
\usepackage{siunitx}
\usepackage[vskip=1em,font=itshape,leftmargin=2em,rightmargin=2em]{quoting}
\usepackage{fontawesome}
\usepackage{hyperref}
\usepackage{lscape}
\usepackage{afterpage}

\usepackage{makecell}
\usepackage{multirow}
\usepackage{booktabs}
\usepackage[T1]{fontenc}

\usepackage{natbib}
\setcitestyle{aysep={}} 

\usepackage{subfigure}
\usepackage{url}
\usepackage{appendix}
\usepackage{svg}
\usepackage{orcidlink}

\usepackage[noadjust]{cite}

\usepackage{soul}

\ifDEBUG
    \newcommand{\JD}[1]{\textcolor{purple}{[JD:#1]}}
    \newcommand{\AG}[1]{\textcolor{olive}{[AG:#1]}}
    \newcommand{\WJ}[1]{\textcolor{teal}{[WJ:#1]}}
    \newcommand{\GKT}[1]{\textcolor{red}{[GKT:#1]}}
    
    \newcommand{\NV}[1]{\textcolor{red}{[NV: #1]}}
    \newcommand{\MK}[1]{\textcolor{blue}{[MK:#1]}}
    \newcommand{\DL}[1]{\textcolor{blue}{[DL:#1]}}

    \newcommand{\TODO}[1]{\hl{#1}}
\else
    \newcommand{\JD}[1]{}
    \newcommand{\AG}[1]{}
    \newcommand{\WJ}[1]{}

    \newcommand{\GKT}[1]{}
    \newcommand{\NS}[1]{}
    \newcommand{\NV}[1]{}
    \newcommand{\MK}[1]{}
    \newcommand{\DL}[1]{}
    \newcommand{\TODO}[1]{#1}
\fi

\newcommand{\myparagraph}[1]{\vspace{0.20cm}\noindent\textbf{#1} \noindent{}}

\ifSPACEHACK
    \titlespacing*\section{0pt}{5pt plus 1pt minus 1pt}{3pt plus 1pt minus 1pt}
    \titlespacing*\subsection{0pt}{4pt plus 1.5pt minus 1.5pt}{4pt plus 1.5pt minus 1.5pt}
    \titlespacing*\subsubsection{0pt}{3pt plus 1pt minus 1pt}{3pt plus 1.5pt minus 1.5pt}
    \titlespacing*\paragraph{0pt}{1pt plus 1.5pt minus 1.5pt}{2pt plus 1.5pt minus 1.5pt}
\fi

\usepackage{balance} 
\makeatletter
\def\cl@chapter{}
\makeatother
\usepackage{cleveref}

\crefformat{section}{\S#2#1#3}
\crefname{figure}{Figure}{Figures}
\crefname{appendix}{Appendix}{Appendices}
\crefname{table}{Table}{Tables}
\crefname{algorithm}{Algorithm}{Algorithms}
\crefname{listing}{Listing}{Listings}
\crefname{theorem}{Theorem}{Theorems}
\crefname{thm}{Theorem}{Theorems}
\crefname{lemma}{Lemma}{Lemmata}
\crefname{equation}{Eqt.}{Eqts.}
\crefformat{Grammar}{Grammar #1}

\usepackage{enumitem}
\usepackage{booktabs}
\usepackage{listings}
\usepackage{academicons} 
\usepackage{xcolor}      

\newcommand{\ie}{\textit{i.e.,} }
\newcommand{\eg}{\textit{e.g.,} }
\newcommand{\etal}{\textit{et al.}\xspace}

\newcommand{\myinlinequote}[1]{\emph{``#1''}}
\newcommand{\inlinequote}[1]{\myinlinequote{#1}}

\ifANNOTATED

    \newcommand{\revision}[1]{\textcolor{blue}{#1}}
    \newcommand{\AUTHOR}[1]{\textcolor{teal}{[AUTHORS:#1]}}
    \newcommand{\review}[2]{\textcolor{red}{\textbf{[#1]:}{#2}}}

    \newcommand{\secondReview}[1]{}

    \newcommand{\thirdReview}[1]{\textcolor{brown}{#1}}

\else

    \newcommand{\revision}[1]{#1}
    \newcommand{\review}[2]{}{}
    \newcommand{\AUTHOR}[1]{}
    \newcommand{\secondReview}[1]{}
    
    \newcommand{\thirdReview}[1]{}
    
\fi

\newcommand{\code}[1]{\texttt{#1}\xspace}

\newcounter{direction}

\newcommand{\ResponseRate}{5.4\%\xspace}
\newcommand{\HighQualityResponseRate}{4.9\%\xspace}
\newcommand{\SurveyTotalResNum}{119\xspace}
\newcommand{\SurveyNum}{108\xspace}
\newcommand{\SurveyTopicTwoNum}{52\xspace}
\newcommand{\mailNum}{1,985\xspace}
\newcommand{\NumofORGUsers}{1,985\xspace}
\newcommand{\NumofPROUsers}{228\xspace}
\newcommand{\SurveyCompensation}{\$10\xspace}

\newcommand{\PTMNum}{14,296\xspace}

\newcommand{\PeatmossPopularNum}{14,296\xspace}

\newcommand{\ElementMeasureAcc}{91\%\xspace}
\newcommand{\ConventionMeasureAcc}{89\%\xspace}
\newcommand{\KimsKappaInit}{0.72\xspace}
\newcommand{\KimsKappaFinal}{0.86\xspace}

\newcommand{\DARAFeatureDim}{4,304\xspace}
\newcommand{\DARAPTMNum}{4108\xspace}
\newcommand{\DARAModelTypeNum}{81\xspace}
\newcommand{\DARATaskNum}{22\xspace}
\newcommand{\DARAArchNum}{184\xspace}

\newcommand{\DARAModeltypeAcc}{99.0\%\xspace}
\newcommand{\DARATaskAcc}{76.3\%\xspace}
\newcommand{\DARAArchAcc}{72.0\%\xspace}

\newcommand{\CLModeltypeAcc}{85.61\%\xspace}
\newcommand{\CLTaskAcc}{46.40\%\xspace}
\newcommand{\CLArchAcc}{49.34\%\xspace}

\newcommand{\LongformerModeltypeAcc}{87.67\%\xspace}
\newcommand{\LongformerTaskAcc}{50.37\%\xspace}
\newcommand{\LongformerArchAcc}{50.66\%\xspace}

\newcommand{\DNNDiff}{DARA\xspace}

\newcommand{\InternalArchNum}{185\xspace}
\newcommand{\InternalModelNum}{8744\xspace}
\newcommand{\InternalMaxPairNum}{1212\xspace}

\newcommand{\ExternalArchNum}{56\xspace}
\newcommand{\ExternalModelNum}{368\xspace}
\newcommand{\ExternalMaxPairNum}{49\xspace}
\newcommand{\ExternalModelhubNum}{4\xspace}

\newcommand{\HF}{Hugging Face\xspace}

\usepackage[ruled, linesnumbered, vlined]{algorithm2e}
\usepackage{caption}

\newcommand{\AlgGets}{\ensuremath{\leftarrow}}

\newcommand{\Visited}{\ensuremath{\mathrm{Visited}}}
\newcommand{\EmptySet}{\ensuremath{\emptyset}}
\newcommand{\SortedLayers}{\ensuremath{\mathrm{SortedLayers}}}

\lstdefinestyle{json}{
    basicstyle=\ttfamily\small,
    numbers=left,
    stepnumber=1,
    numbersep=8pt,
    showstringspaces=false,
    breaklines=true,
    frame=single,
    backgroundcolor=\color{gray!10},
    tabsize=2,
    captionpos=b
}

\begin{document}




\newcommand{\TitleQuote}{``I see models being a whole other thing''}

\title{\textit{\TitleQuote}: Naming Practices of Pre-Trained Models on Hugging Face}

\title{\textit{\TitleQuote}: An Empirical Study of Pre-Trained Model Naming Conventions and A Tool for Enhancing Naming Consistency}

\authorrunning{Wenxin Jiang et al.}
\titlerunning{An Empirical Study of Pre-Trained Model Naming Conventions}

\newcommand{\orcid}[1]{\orcidlink{#1}}
\newcommand{\corresp}{\textsuperscript{*}}

\ifANONYMOUS
  \author{Anonymous Author(s)}
\else
  \author{
    Wenxin Jiang\orcid{0000-0003-2608-8576}\corresp \and
    Mingyu Kim\orcid{0009-0003-1308-6501} \and
    Chingwo Cheung\orcid{0009-0001-0599-7909} \and
    Heesoo Kim\orcid{0009-0008-5714-2866} \and
    George K. Thiruvathukal\orcid{0000-0002-0452-5571} \and
    James C. Davis\orcid{0000-0003-2495-686X}\corresp
  }

\fi

\begingroup
\renewcommand\thefootnote{}\footnotetext{* Corresponding authors: \texttt{jiang784@purdue.edu}, \texttt{davisjam@purdue.edu}}%
\addtocounter{footnote}{-1}
\endgroup

\institute{Wenxin Jiang \at
              Purdue University, West Lafayette, IN, USA \\
              \email{jiang784@purdue.edu}           
        \and
           Mingyu Kim \at
              Purdue University, West Lafayette, IN, USA \\
              \email{kim3118@purdue.edu}
        \and
        Chingwo Cheung \at
             Purdue University, West Lafayette, IN, USA \\
              \email{cheung59@purdue.edu}
        \and
           Heesoo Kim \at
              Purdue University, West Lafayette, IN, USA \\
              \email{kim2903@purdue.edu}
        \and
           George K. Thiruvathukal \at
              Loyola University Chicago, Chicago, IL, USA \\
              \email{gkt@cs.luc.edu}
        \and
            James C. Davis \at
                Purdue University, West Lafayette, IN, USA \\
              \email{davisjam@purdue.edu}
}

\date{Received: date / Accepted: date}

\maketitle

\begin{abstract}

As innovation in deep learning continues, many engineers are incorporating Pre-Trained Models (PTMs) as components in computer systems.
Some PTMs are foundation models, and others are fine-tuned variations adapted to different needs.
When these PTMs are named well, it facilitates model discovery and reuse.
However, prior research has shown that model names are not always well chosen and can sometimes be inaccurate and misleading. 
The naming practices for PTM packages have not been systematically studied, which hampers engineers' ability to efficiently search for and reliably reuse these models.

In this paper, we conduct the first empirical investigation of PTM naming practices in the Hugging Face PTM registry. 
We begin by reporting on a survey of \SurveyNum Hugging Face users, 
highlighting differences from traditional software package naming and presenting findings on PTM naming practices.
The survey results indicate a mismatch between engineers' preferences and current practices in PTM naming.
We then introduce DARA, the first automated \textit{D}NN \textit{AR}chitecture \textit{A}ssessment technique designed to detect PTM naming inconsistencies. Our results demonstrate that architectural information alone is sufficient to detect these inconsistencies, achieving an accuracy of 94\% in identifying model types and promising performance (over 70\%) in other architectural metadata as well. 
We also highlight potential use cases for automated naming tools, such as model validation, PTM metadata generation and verification, and plagiarism detection.
Our study provides a foundation for automating naming inconsistency detection.
%
Finally, we envision future work focusing on automated tools for standardizing package naming, improving model selection and reuse, and strengthening the security of the PTM supply chain.

 \vspace{2pt}
 \textit{``The main idea is to treat a program as a piece of literature,}\\
 \hspace*{0.4cm} \textit{addressed to human beings rather than to a computer''} \hspace{5mm} ---D. Knuth

\keywords{
    Software Reuse and Adaptation 
    \and Empirical Software Engineering
    \and Mixed-Methods Study 
    \and Machine Learning (ML)
    \and Deep Neural Networks (DNNs)
    \and Model Zoos 
    \and Package Registries
    \and Pre-trained Models
    \and Software Supply Chain
    }


\end{abstract}

\section{Introduction} \label{sec:Intro}

Software reuse helps engineers develop software more efficiently and at lower costs~\citep{griss1993softwareReusefromLibrarytoFactory, frakes2005SoftwareReuseResearch}.
A major form of software reuse is via software packages, which encapsulate functionality through an Application Programming Interface (API)~\citep{decan2019empiricalComparisonofDependencyNetworkEvolutioninSevenSWPkgEcosystems, frakes2005SoftwareReuseResearch}
This practice is common for traditional software~\citep{soto2019emergenceSWDiversityinMaven, Wittern2016JSPackageEcosystem, Abdalkareem2017TrivialPackages},
  \eg Java packages shared via Maven\footnote{See \url{https://mvnrepository.com/}.},
  JavaScript packages shared via NPM\footnote{See \url{https://www.NPMjs.com/}.},
  and
  Python packages shared via PyPI\footnote{See \url{https://pypi.org/}.}.
\revision{The same reuse practice is arising for ML-based software~\citep{Jiang2022PTMReuse}.}
In recent years, models based on deep neural networks (DNNs) have become more capable~\citep{abiodun2018SOTADNNAppSurvey, taye2023theoretical} and the necessary hardware resources have become more available~\citep{capra2020hardwareandSWOptimization}.
As a result, Pre-Trained Models (PTMs), which encapsulate the capabilities of DNNs trained for specific tasks, have become increasingly popular~\citep{Jiang2022PTMSupplyChain, Han2021PTM}.
PTMs are now commonly shared through PTM-specific package registries, the most prominent of which is \HF~\citep{Jiang2022PTMReuse}. The rapid growth of PTM packages on \HF, in both number and popularity, underscores their growing significance in modern software reuse practices~\citep{Jiang2024peatmoss, jones2024HFQualitativeValidation}.

\begin{figure}[t]
    \centering
    \includegraphics[width=0.9\columnwidth]{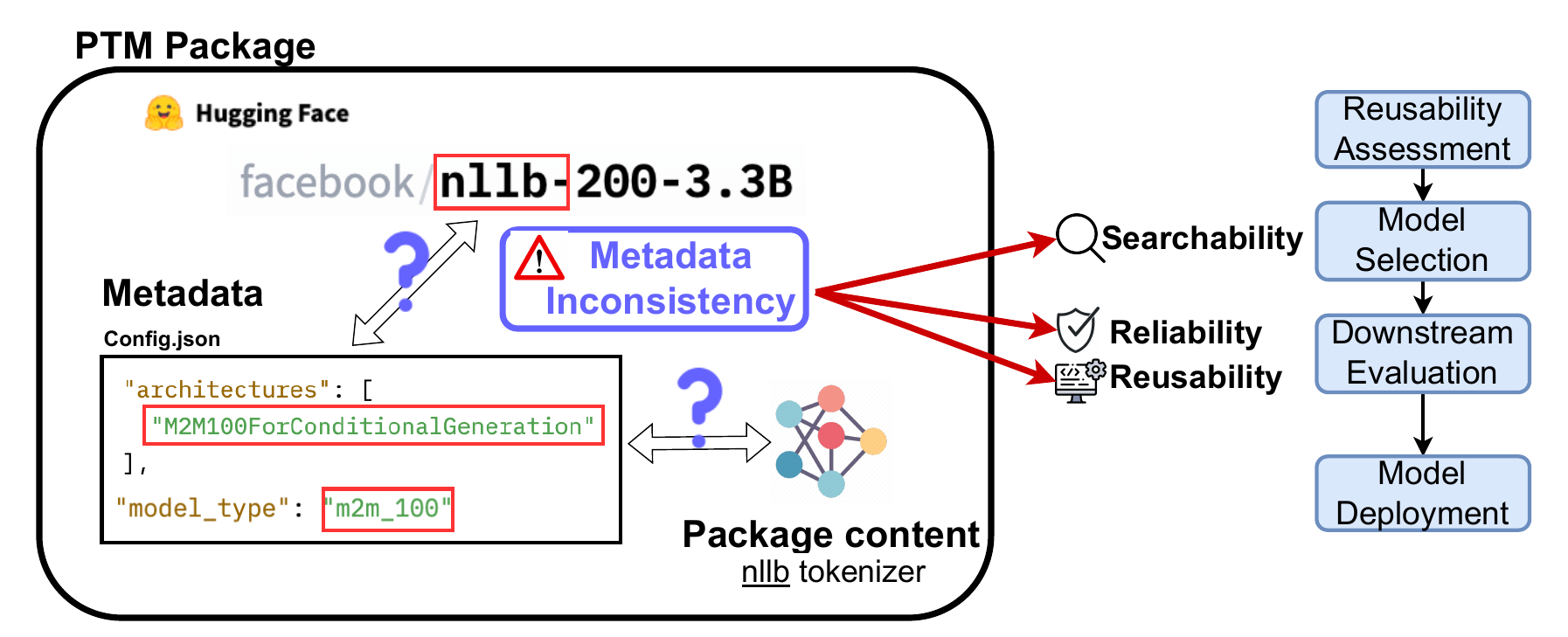}
    \caption{
    Example of a PTM \textbf{metadata inconsistency}
    for a PTM on Hugging Face.
    The top part shows the PTM identifier, and the bottom part displays the PTM metadata (\ie architecture and model type) and the package content. 
    On the right (blue boxes) is a four-step reuse process adapted from~\citep{Jiang2022PTMReuse}.
    In this case, Facebook (Meta) named the model \code{nllb} due to the introduction of a new tokenizer, while other models on Hugging Face typically name models based on architecture changes rather than changes in tokenizers.
    This mismatch between the identifier and the architecture metadata impacts the new model's searchability, perceived reliability, and ultimately reusability in the PTM reuse process.
    }
    \label{Def:ExamplePTMDefect}
\end{figure}

Consistent and standardized naming is a crucial element of software development and reuse~\citep{deissenboeck2006conciseandConsistentNaming, hofmeister2017shorterIdentifierNamesTakeLongertoComprehend, alpern2024NamingExperiments}. Over the years, research has consistently highlighted the challenges posed by naming components and software package reuse~\citep{griss1993softwareReusefromLibrarytoFactory, he2021learningtoFindNamingIssueswithBigCodeandSmallSupervision, Alsuhaibani2021NamingMethodsSurvey}. In the context of PTM reuse, the selection process becomes even more complex due to the high cost of evaluating packages and the abundance of models with overlapping functionality~\citep{Jiang2022PTMReuse, taraghi2024DLModelReuseinHF}. 
The naming practices and challenges for PTM packages have not been systematically studied. 
As shown in \cref{tab:top_packages}, the names for PTMs appear to be different in kind from the names for traditional software packages.
PTM names often encode critical information about a model's architecture, size, and dataset, enabling software engineers to infer significant details directly from the name.
As shown in \cref{Def:ExamplePTMDefect}, metadata in PTM packages is often embedded directly into the package, making it a critical component for effectively identifying and selecting models~\citep{Jiang2024peatmoss}. Therefore, in the context of PTMs, we regard metadata as an integral part of the package name.
Inaccurate names and inconsistent metadata can significantly hinder searchability, reliability, and reusability.

\begin{table}[t]
    \centering
    \caption{
    Top 10 package names by weekly downloads from NPM, PyPI, and Hugging Face.
    Compared to the NPM and PyPI packages (traditional software), the names of Hugging Face packages (PTMs) appear to convey more information about the package content.
    This paper is the first to study this phenomenon.
    Popularity data is as of September 2024.
    }
    \label{tab:top_packages}
    \begin{tabular}{p{0.25\textwidth}p{0.2\textwidth}|p{0.45\textwidth}}
        \toprule
         \textbf{NPM\footnotemark (JavaScript)} & \textbf{PyPI\footnotemark (Python)} & \textbf{\HF\footnotemark (Pre-Trained DNN Models)} \\
        \midrule
        ansi-styles & boto3 & ast-finetuned-audioset-10-10-0.4593 \\
        chalk & requests & chronos-t5-tiny \\
        supports-color & urllib3 & bert-base-uncased \\
        semver & botocore & fasttext-language-identification \\
        debug & certifi & all-MiniLM-L6-v2 \\
        has-flag & idna & clip-vit-large-patch14 \\
        color-convert & charset-normalizer & fashion-clip \\
        color-name & setuptools & clip-vit-base-patch32 \\
        tslib & packaging & wav2vec2-large-xlsr-53-english \\
        ms & python-dateutil & all-mpnet-base-v2 \\
        \bottomrule
    \end{tabular}
\end{table}

\footnotetext[4]{See \url{https://socket.dev/npm/category/popular}.}
\footnotetext[5]{See \url{https://socket.dev/pypi/category/popular}.}
\footnotetext[6]{See \url{https://huggingface.co/models?sort=downloads}.}

The goal of this work is to delineate and improve the naming practices of PTMs.
We focus on two themes: (1) \ul{empirical measurements of naming practices and rationales}; and (2) \ul{automated identification of naming inconsistencies}.
In the \textit{first theme}, our objective is to characterize PTM naming and compare it to the naming of traditional software packages.
We also want to understand what elements should be included in a ``good'' PTM name.
We take a mixed-methods approach here, combining a survey of \SurveyNum \HF users with a mining study of \PTMNum PTMs (1.7\% of \HF PTMs) from the PeaTMOSS dataset~\citep{Jiang2024peatmoss}.
For the \textit{second theme}, we designed and evaluated an automated tool to detect architecture-related naming inconsistencies. 
\revision{We experimented with various feature extraction methods, including n-gram features and advanced representations processed by CNNs and transformers. For transformer-based models, we explored continued pretraining, fine-tuning, and contrastive learning. We evaluated each variant in our pipeline in terms of both effectiveness and efficiency.}

Our findings reveal significant differences between PTM naming conventions and those of traditional software packages, highlighting unique challenges faced by engineers in the PTM domain.
For the \textit{first theme}, we report that PTM naming differs significantly from traditional software package naming. 
In this context, engineers follow distinct PTM naming conventions and face challenges unique to PTMs. 
Based on our survey data, we outline the key differences between traditional and PTM naming practices, while also presenting engineers' perspectives on these practices. 
A key distinction is the amount of information embedded in PTM names, which sets them apart from traditional software components.
Our empirical measurements further demonstrate that the actual distribution of naming elements in practice often diverges from engineers' stated preferences. 
\revision{For the \textit{second theme}, our automated tool, DARA, proves effective in detecting inconsistencies within PTM names. The best-performing architecture in terms of accuracy is the Longformer with contrastive learning, achieving up to 98\% accuracy for detecting \code{model\_type} inconsistencies and strong performance across other metadata types (\code{task} and \code{architecture}). In contrast, the N-gram + MLP classifier offers the lowest latency (0.19 ms per PTM) and highest throughput (over 28,000 PTMs/s), making it well-suited for real-time integration. These findings reveal a clear trade-off between performance and efficiency.}
We identified potential use cases for automated naming tools in addressing challenges related to model validation, PTM metadata generation and verification, and plagiarism detection.

\ul{To summarize our contributions}:
\begin{itemize}[leftmargin=0.5cm]
    \item 
    \textit{Theme 1: PTM Naming Practices}
    We presented PTM naming practices through a survey of \SurveyNum Hugging Face users (\cref{sec:RQ1-Results}--\cref{sec:RQ3-Results}).
    We triangulated the survey study with a repository mining study on \PTMNum PTM names from \HF (\cref{sec:RQ2-Results}). 
    \review{Reviewer 3-Concern 9}{Do the authors consider prompt-based extraction as a technical contribution? If yes, the prompts needed to be tested against their variants. It looks like multiple annotators were not needed to extract the model elements. The LLM was also able to extract 91\% of the items from the provided models. Thus, I am skeptical about the technical contribution of this part of the work.
}
\AUTHOR{We removed the prompt-based pipeline from this contribution point.}
    \item \textit{Theme 2: Naming Inconsistencies} We developed a novel algorithm, DNN Architecture Assessment (DARA), to detect naming inconsistencies (\cref{sec:RQ4-Results}). 
    \revision{DARA combines structured architectural feature extraction with multiple classification strategies (\cref{sec:DARA}) and achieves high accuracy with measurable efficiency trade-offs across metadata types (\cref{sec:DARA-Results}).}
    We describe three use cases of DARA: model validation, metadata label generation, and plagiarism detection (\cref{sec:DARA-UseCases}).
    \item
    \textit{Landscape of Future Work:} Our work envisions future research focused on developing automated tools for standardizing PTM package naming, supporting model selection and reuse, and enhancing the security of the PTM supply chain (\cref{sec:discussion}).
\end{itemize}

\noindent\ul{\textbf{Significance:}}
Prior work has shown that naming practices and conventions are critical in software engineering.
However, this phenomenon has not been examined in the context of pre-trained deep learning models (PTMs).
To fill that gap, we conducted the first empirical measurements on PTM naming conventions.
Our mixed-method approach describes engineers' current practices and rationales, and uses their insights to design the first tool to identify PTM naming mismatches.
Our findings provide engineers and researchers with a better understanding of PTM naming practices and offer an initial solution in this problem domain, informing the design of related tools.

\section{Background and Related Work}
\label{sec:backgroundAndRelatedWork}

This section presents the background information (\cref{sec:background}) and reviews related work on PTM reuse and software naming conventions (\cref{sec:RelatedWork}).


\subsection{Background}
\label{sec:background}

\label{sec:background-PTMNaming}
\review{Reviewer 2-Concern 12}{There is some redundancy in the content that could benefit from improvement.}
\review{Reviewer 3-MajorConern1}{Related work. Referring to the same few studies repetitively looks tiring. The related work section also contains duplicate and near-duplicate contents, which should be rephrased. }
\AUTHOR{We reorganized \cref{sec:backgroundAndRelatedWork} to eliminate redundancy. \Cref{sec:background} now focuses on defining the PTM naming hierarchy and practices, while \cref{sec:RelatedWork} consolidates all related work, with redundant content removed for clarity and conciseness.
}

\revision{In this section, we provide related contextual information and define key terms. We cover topics including software and package reuse (\cref{sec:background-SWReuse}) and how PTMs are named (\cref{sec: background-NamingPTMs})}.

\subsubsection{Reuse of Pre-Trained Models}
\label{sec:background-SWReuse}
Component-based development provides a systematic approach to constructing software systems from reusable parts, significantly reducing development costs and increasing software reuse~\citep{lau2006softwareComponentModels, kotonya2014teachingReuseDrivenSE, heineman2001componentBasedSE}.
This approach applies across various levels of software granularity, from variables and functions to larger entities such as libraries and packages, and ``commercial off-the-shelf'' software (COTS)~\citep{sommerville2015software}.
While some components can be used directly as black-box entities with well-defined interfaces, others require significant adaptation or customization to integrate effectively within specific application contexts~\citep{krueger1992SoftwareReuse, szyperski2002componentSoftware}.
Among these levels of abstraction, one of the most common modes of reuse is via software packages shared through software package registries such as NPM (JavaScript), PyPI (Python), and Maven (Java).
These registries serve a combined billions of packages monthly~\citep{Wittern2016JSPackageEcosystem}.

\revision{Similar to traditional software engineering,} for deep learning approaches, engineers may use a pre-trained model with a general understanding of the task (\eg text generation) and fine-tune it for the specific application (\eg machine translation, text summarization)~\citep{li2024PTLM4TextGenerationSurvey, Han2021PTM}. 
PTM reuse is facilitated through open-source package registries like Hugging Face, PyTorch Hub, and ONNX Model Zoo, where these models are shared as reusable components~\citep{Jiang2022PTMSupplyChain}.
\revision{As of April 2025, Hugging Face --- the largest open model registry~\citep{jones2024HFQualitativeValidation} --- hosts over 1.6 million PTM packages spanning domains like computer vision, natural language processing, audio, and reinforcement learning. 
These model registries have high industry engagement. 
For instance, Meta~\citep{facebook2024hf} and Google~\citep{google2024hf} have both published thousands of models, while DeepSeek~\citep{deepseek2024hf} has recently released 14 model collections with millions of downloads.}
%
\revision{In this work, we define PTMs as packages that contain deep neural networks (DNNs), which reflects the overwhelming majority of the PTMs on the Hugging Face registry.\footnote{Examining the NAME dataset from REFERENCE, we observe fewer than 0.01\% of all models on Hugging Face use techniques other than deep neural networks. Exceptions include graph neural networks, time series models (fewer than 60 total in the dataset, and other non-DNN models such as decision trees or random forests (typically fewer than 20 per category in the dataset)}.
Therefore, our approach focuses specifically on DNN-based PTM packages to ensure both methodological coherence and maximal applicability of our findings.}

\revision{
From a naming perspective, the key actors in the PTM supply chain~\citep{Jiang2022PTMSupplyChain} can be grouped into two roles: \textit{name producers} and \textit{name consumers}.
Names are produced by the researchers and developers who contribute new models and assign names to the resulting PTM packages. 
Names are consumed by the adopters and reengineers who reuse these models in downstream applications.
Name producers may introduce novel models or innovate on existing ones through various reengineering techniques, such as transfer learning, fine-tuning, and knowledge distillation~\citep{Han2021PTM, davis2023JVA, Jiang2023CVReengineering}. For example, reengineers often modify the model head --- such as the fully connected output layer --- to suit the target task~\citep{qi2023reusing}. As PTMs grow more powerful, particularly in LLMs, reuse strategies have expanded to include zero- and few-shot learning as well as prompt tuning~\citep{Brown2020gpt, Radford2019LMareUnsupervisedMultitaskLearners, Lester2021PromptTuning}.
PTM registries like Hugging Face facilitate this reuse by allowing consumers to retrieve and deploy PTMs via model names and APIs~\citep{Jiang2023CVReengineering}. These consumers rely heavily on naming conventions to search, compare, and select appropriate models. Thus, clearer naming benefits all stakeholders: it improves discoverability for authors, enhances selection and evaluation for reusers, and supports better user experience for registry operators. }

\subsubsection{\revision{Naming Pre-Trained Models}}
\label{sec: background-NamingPTMs}

\myparagraph{\revision{PTM Naming Hierarchy}}
\label{sec: background-PTMNamingHierarchy}
%
%
\cref{fig:Naming} presents the hierarchical naming framework in PTM registries.
We define this based on our study of documentation from open-source model registries, including \HF~\citep{HFDoc}, ONNX Model Zoo~\citep{ONNXModelZoo}, PyTorch Hub~\citep{PytorchHub}, and TensorFlow Hub~\citep{TFHub}.
These registries generally follow similar naming conventions, including information such as the problem domain, task, model identifier, and associated files~\citep{jiang2023ptmtorrent}.
\review{Reviewer 2-Concern 1}{The study focuses only on Hugging Face as a PTM registry and does not include other platforms (e.g., ONNX, PyTorch Hub, TensorFlow Hub). The authors should clarify why they chose to focus exclusively on Hugging Face and did not include other platforms to support their study.
}
\revision{In this work, we focus exclusively on the Hugging Face registry because 
    (1) it is, by far, the largest PTM registry, and the only viable open-source one~\citep{Jiang2022PTMSupplyChain};
    (2) it  provides a clear naming schema, a robust queryable API for large‑scale mining, and leverages an existing structured reuse dataset~\citep{Jiang2024peatmoss}.}

PTM registries employ a multi-tiered categorization system~\citep{HFDoc}. 
The top level of the hierarchy represents the model's domain, which refers to the broad area of application where a group of models is utilized, such as Natural Language Processing or Computer Vision~\citep{islam2023comprehensiveSurveyonTransformerApplications}.
Within a domain, \textit{tasks} are specific problems that models aim to solve, like Sentiment Analysis or Named Entity Recognition in the Natural Language Understanding domain.
A \textit{PTM} is an instance of a trained model for a specific task, complete with weights, parameters, and configuration metadata.
A PTM package's \code{config.json} file details the model's \textit{architecture} and specifies the \textit{model type} and the task it is designed for, \eg \code{Albert} for the model type and \code{QuestionAnswering} for the task.
\textbf{\textit{Definition 1}} summarizes these terms:

{
\small
\begin{figure}[t]
    \centering
    \includegraphics[width=0.95\columnwidth]{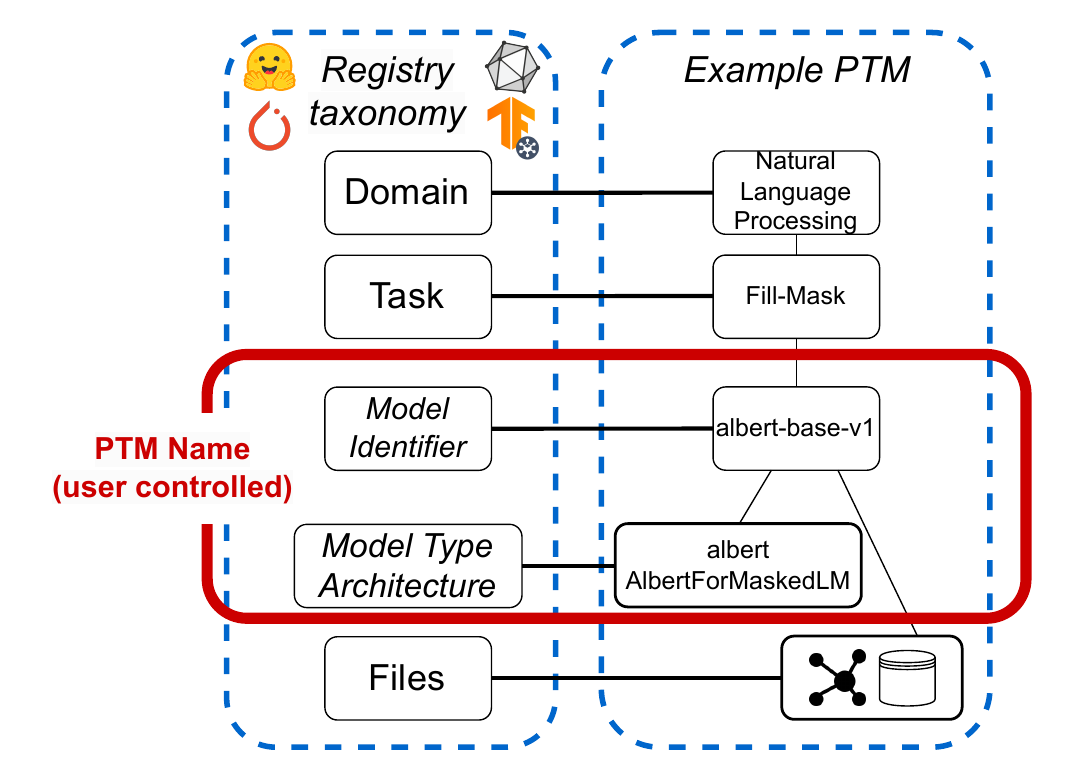}
    \caption{
    Hierarchy of PTM registry naming taxonomy across representative open-source PTM registries, including \HF, ONNX Model Zoo, PyTorch Hub, and TensorFlow Hub.
    Registries include multiple domains, such as natural language processing, computer vision, and multimodal; multiple tasks, such as depth estimation (vision), text generation (NLP); and many distinct models and model types/architectures for each task.
    The red box highlights the package name (model identifier and model type/architecture) --- this is the focus of our study.
    }
    \label{fig:Naming}
\end{figure}
}


\begin{tcolorbox} [title=Definition 1: Naming Hierarchy in PTM Registries, width=\linewidth, colback=blue!10!white, top=1pt, bottom=1pt, left=2pt, right=2pt]

    \begin{itemize}[left=3mm]
        \item \textbf{Domain and task}: High-level metadata that are part of the \textit{model type}.
        \item \textbf{\textit{PTM name}}: Defined as the combination of a package \emph{identifier} (\eg \code{albert-base-v2}) and the \emph{model type} or \emph{architecture} indicated in the metadata (\eg \code{AlbertForQuestionAnswering}). 
    \end{itemize}
\end{tcolorbox}


In the Hugging Face registry, we consider both the model metadata and its files to be objective factors because they are clearly defined by the model's architecture, dataset, and performance metrics, which are based on measurable attributes and standardized benchmarks. On the other hand, the package name is a subjective element, as it is manually crafted by PTM contributors based on their own interpretation of the model’s purpose, function, and approach, among other considerations.
This subjectivity arises from the flexibility contributors have in deciding how to represent key aspects of the model, allowing them to incorporate personal interpretation and preferences and leading to variations in naming conventions.

This subjectivity makes the naming process a critical point where inconsistencies or ambiguities can occur.
According to prior research~\citep{Jiang2022PTMReuse}, the engineers involved in the PTM ecosystem commonly rely on either package identifiers or architectural labels when searching for PTM packages appropriate for their specific tasks. 
The subjective components of a package involve judgment, both for the namer and for the potential re-user, which impacts how reengineers and adopters discover and assess a PTM.

\myparagraph{\revision{PTM Naming Elements and Conventions}}
\label{sec: background-PTMNamingElementsandConventions}
\revision{Two key concepts in PTM naming practices are naming \textit{elements} and how they are combined to make naming \textit{conventions}.}

In the context of PTM package naming, \textbf{naming elements} are essential components that make up the name of a model. For example, a model might have naming elements structured as “{Architecture}-{Size}-{Dataset Characteristic}” (\code{deberta-base-uncased}), where the name includes information about the model's architecture (\code{deberta}), size (\code{base}), and dataset characteristic (\code{uncased}).
\textbf{\textit{Definition 2}} formalizes this notion.

\begin{tcolorbox} [title=Definition 2: PTM Naming Elements, width=\linewidth, colback=blue!10!white, top=1pt, bottom=1pt, left=2pt, right=2pt]
\textbf{Naming elements:} The elements that comprise a model's name, representing different categories or attributes that are combined to form the full model name. 
\end{tcolorbox}

Another critical aspect of PTM naming practices is the \textbf{naming convention}, which refers to the norms and rules governing what elements are included in a package identifier. For example, in a model named \code{deberta-base-uncased}, the naming elements reflect the implementation details of the model. In contrast, a name like \code{fasttext-language-identification} includes both the \textit{implementation} unit (\ie the FastText model) and the \textit{application} goal (\ie language identification).
We define naming conventions for packages in \textbf{\textit{Definition 3}}.

\begin{tcolorbox} [title=Definition 3: PTM Naming Convention, width=\linewidth, colback=blue!10!white, top=1pt, bottom=1pt, left=2pt, right=2pt] \textbf{Naming convention:} Norms and rules that governs how elements are named and combined to form a package name, ensuring a clear link between the implementation details and the application's objectives~\citep{lawrie2006whatsinaName}.
\end{tcolorbox}

\subsection{Related Work}
\label{sec:RelatedWork}
We first present
  a broad review of known challenges in the PTM reuse process (\cref{sec:background-ptmReuseChallenges}),
  and then
  focus on prior work on naming in traditional software (\cref{sec:background-SWNamingConventions}).
  \revision{Prior work directly on PTM names is included in our problem statement (\cref{sec:ProblemStatement}).}

%


\subsubsection{Challenges in the PTM Reuse Process}
\label{sec:background-ptmReuseChallenges}

\AUTHOR{We have removed the redundant content from this section and merged the old content to a single paragraph.}
\revision{Reusing PTMs can speed up development, but it also brings unique challenges that differ from traditional software reuse.
 A common approach is fork-based reuse—cloning a model and modifying its architecture or retraining it to fit specific needs~\citep{Jiang2023CVReengineering, bhatia2023towardsMLpipelines}. However, this method complicates conceptual reuse, adaptation, and deployment~\citep{davis2023JVA}. 
Reusing PTMs presents several challenges, including missing attributes, inconsistent metadata, and security or privacy risks~\citep{Jiang2022PTMReuse}.
In the same study, Jiang \etal reported that many experienced Hugging Face users expressed concerns over model trustworthiness due to inconsistent or missing metadata, including names. 
 Other studies highlight challenges in analyzing PTM vulnerabilities~\citep{kathikar2023assessingHFVulnerabilities} and managing their lifecycle during reuse~\citep{castano2023analyzingEvolandMaintenanceofMLModelsonHF}. Taraghi \etal conducted a qualitative analysis of Hugging Face forum discussions and found that understanding and effectively reusing PTMs remains the most significant challenge~\citep{taraghi2024DLModelReuseinHF}.
    These issues underscore the need for standardized, descriptive naming conventions to support more effective PTM management, which can further enhance discoverability and reliability, and reduce both computational and financial costs in the reuse process~\citep{sens2024largeScaleStudyofModelIntegrationinMLSystem, Amazon2018MLModelManagementChallenge}.
}
Our work presents the first empirical study on PTM package naming practices and provides insights into standardized naming conventions to ensure consistency and reliability. 

\subsubsection{Software Naming Conventions}
\label{sec:background-SWNamingConventions}
The naming of software components is crucial for code readability, maintainability, and collaboration~\citep{seacord1998agoraASearchEngine4SWComponents, Lawrie2007IdentifierNamingStudy, Butler2012MethodNameConsistency, hofmeister2017shorterIdentifierNamesTakeLongertoComprehend}.
Existing studies have primarily concentrated on the naming of 
  variables~\citep{hofmeister2017shorterIIDNamesTakeLongertoComprehend},
  functions~\citep{Alsuhaibani2021NamingMethodsSurvey},
  and
  classes~\citep{butler_mining_2011}.
Researchers have documented two common strategies for naming:

  by the implementation approach, \ie ``\textit{how it works}'';
  and
  by the application goal, \ie ``\textit{what it does}''~\citep{henninger1994usingIterativeRefinementtoFindReusableSW, hofmeister2017shorterIIDNamesTakeLongertoComprehend}.
\revision{These strategies can be reflected in the package's naming elements (Definition 2) and the resulting naming convention (Definition 3), although the specific strategies used for PTMs have not previously been described.}

While naming strategies have been widely studied for fine-grained implementation elements such as variables and functions, the naming conventions for software packages have received less attention.
It is known that clear and descriptive names enhance the discoverability and reuse of packages~\citep{deissenboeck2006conciseandConsistentNaming,lawrie2006whatsinaName,abdellatif2020simplifyingtheSearchofNPMPackages, Alsuhaibani2021NamingMethodsSurvey}.
However, there has been no systematic work linking naming conventions directly to software reusability and reliability at the package level. Our study is the first to explore this connection, with a particular focus on the reuse of PTM packages.

Well-chosen names allow users to quickly grasp a package's purpose, aiding in the selection process.
As an illustration,~\cref{tab:top_packages} shows popular packages from NPM (JavaScript), PyPI (Python), and \HF (PTMs).
Anecdotally, looking at \cref{tab:top_packages}, 
we see an interesting phenomenon: the names of PTMs often convey more information than those of traditional software packages. 

\begin{itemize}
    \item 
    In \textbf{traditional software package registries} like NPM, package names often follow a convention of indicating ``what it does'',  typically focusing on the package's main feature or purpose, which helps developers easily search for and select packages based on their specific needs
    ~\citep{NPM_package_guidelines, pep8}.
    For example, Python’s PEP8 style guide recommends that package names be short, descriptive, and easy to understand, while avoiding unnecessary complexity~\citep{pep8}.
    This approach is also evident in names like \code{color-convert} and \code{color-name}, which clearly signal their color-related functionality.
    In contrast, a name like \code{chalk} is less clearly associated with color.
    This name has not kept \code{chalk} from being used --- it is one of the most popular packages on NPM --- but it does add a barrier for understandability.

\item In contrast, we observe that open-source \textbf{PTM registries} like Hugging Face, PyTorch Hub, and ONNX Model Zoo tend to follow a more detailed naming convention.
These names are usually longer compared to traditional package names frequently embed important metadata about the model, such as its architecture, training dataset, or performance metrics, offering users a more comprehensive understanding of the model's capabilities and purpose~\citep{Jiang2022PTMSupplyChain, Jiang2022PTMReuse, di2024automatedcategorizationofPTM}.
For example, in the Hugging Face column of~\cref{tab:top_packages},
  the PTM named \code{ast-finetuned-audioset-10-10-0.4593} is an \code{ast} (\ie Audio Spectrogram Transformer) model fine-tuned on the \code{AudioSet} dataset, achieving performance (mean Average Precision) of 0.4593.
  The two \code{10}s represent the frequency and time strides, respectively.
  Similarly, the PTM named \code{bert-base-uncased} is a base version of the \code{BERT} model without case sensitivity. 
Model names like \code{fasttext-language-identification} convey information about the task and application, helping users make informed decisions when selecting models~\citep{Han2021PTM}.


\end{itemize}

\section{Problem Statement}
\label{sec:ProblemStatement}
\label{sec:background-roleofPTMNamingPractices}

\JD{Consider making this section 3: Problem Statement? It is so important and it feels like it's in the wrong place as part of 2.1 --- because it cites lots of related work and has some novel examples in it. Then 2.1 could end with a figure that puts NPM/PyPI/etc. next to HF/ONNXHub/etc. side by side, to emphasize that there is a symmetry that leads to lots of related work filling in the similarities and differences. Then we can transition to 2.2 that shows what we know so far, and NOW the reader is ready to see the problem statement about understanding names (which follows nicely from 2.2.1 (where documentation and names are a problem) and 2.2.2 (where we know about names for conventional software but not PTMs).}
\WJ{TODO: update fig 2}
\JD{Maybe for final version but I think it's too much for the revision given the remaining time.}

\review{Reviewer 2-Concern 8}{ The paper also lacks case studies demonstrating how naming inconsistencies directly impact model reuse efficiency, searchability, or security in real-world scenarios. Providing such case studies would make the work more compelling and practically relevant.}
\review{Reviewer 3-Concern 2}{P11: Examples of how naming issues of PTM led to reusability or reliability issues will help motivate the idea. Right now, there are a lot of discussions without meaningful evidence.
}
\AUTHOR{We have revised this section to include clearer descriptions and concrete real-world examples to better motivate our study.}
\revision{Recently, there has been increased attention paid to the broader challenges associated with PTM naming. Prior research has highlighted difficulties in PTM reuse, specifically related to model selection and management~\citep{tan2024challengesofUsingPTMs, taraghi2024DLModelReuseinHF}. Addressing these challenges requires more consistent naming practices and enhanced transparency of PTMs~\citep{di2024automatedcategorizationofPTM}. Furthermore, industry stakeholders, including Google and HiddenLayer, have emphasized the importance of maintaining accurate inventories of PTMs, tracking their provenance to enable rapid responses to emerging vulnerabilities~\citep{Hepworth2024, HiddenLayer2024}.
}

An illustrative example of a PTM \textbf{naming inconsistency} is presented in \cref{Def:ExamplePTMDefect}.
While the appropriateness of a PTM name can be subjective, the PTM ecosystem currently lacks standardized guidelines to inform and evaluate naming practices, \revision{making model selection and downstream reuse more difficult}. 
\revision{For example:
\begin{itemize}
    \item NVIDIA engineers made a typo in the model card of their speech recognition model \code{Parakeet} (with 15K monthly downloads). The model\_config incorrectly specified \code{EncDecCTCModel} instead of \code{EncDecRNNTModel}. This inconsistent architecture metadata caused users significant debugging effort and confusion when attempting to load the model properly~\citep{nemo_issue10006}.
    \item Similarly, Longformer initially listed \code{BartModel} as its base architecture, which confused users who expected to load it using \code{LongformerModel}. The issue was later resolved when Hugging Face updated the configuration to correctly reference the \code{LongformerModel} class~\citep{longformer_issue185}.
    \item The LLaVA model also suffers from naming inconsistencies. For instance, users have expressed confusion over similarly named models such as \code{LLaVA-Plus}, \code{LLaVA-1.6}, and \code{LLaVA-Next}, with no clear guidance on which is the latest or performs best overall~\citep{llava_issue1136}.
\end{itemize}
}

These naming inconsistencies can negatively impact both the \textit{searchability} of models, making them harder to discover, and their \textit{reliability}, as misleading names may confuse the model's true functionality.
We discuss these two topics in more detail next.

\begin{tcolorbox} [title=Definition 4: PTM Naming Inconsistency, width=\linewidth, colback=blue!10!white, top=1pt, bottom=1pt, left=2pt, right=2pt]

\textbf{PTM Naming Inconsistency}: A deviation where the name (\ie identifier and architectural metadata, \textit{cf.} Definition 1) of a PTM package does not correctly reflect its underlying architecture or intended functionality, leading to potential confusion or misinterpretation by users. 
\end{tcolorbox}

\paragraph{Impact on Package Searchability}
\label{sec:background-impactonSearchability}

%
%
Model names play a crucial role in determining search capabilities within platforms like Hugging Face, as they serve as a primary source for generating metadata. 
However, user-generated metadata often contains errors, forcing engineers to manually inspect models to verify their suitability~\citep{2022JiangEmpirical, Jiang2022PTMReuse}. For instance, mislabeling ``\code{ResNet}'' as ``\code{VGGNet}'' is a clear naming mistake beyond mere subjectivity.
Missing details and inconsistencies in names can create significant challenges in PTM reuse, leading to difficulties in understanding model capabilities, maintaining code, and collaborating effectively~\citep{Butler2012MethodNameConsistency}.

The lack of standardized metadata and clear documentation significantly complicates the discovery, evaluation, and reuse of PTMs~\citep{jones2024HFQualitativeValidation, Jiang2024peatmoss}. 
According to prior research~\citep{taraghi2024DLModelReuseinHF, Jiang2022PTMReuse}, engineers often depend on PTM package names as primary indicators of a model's architecture and functionality, particularly when model cards (\ie \code{READMEs}) and metadata are frequently insufficient.
Inconsistent or misleading names can lead to the selection of unsuitable or incompatible models, resulting in wasted time and effort during the development process.

One recurring challenge in PTM reuse is the inefficiency caused by unclear or inconsistent naming conventions~\citep{tan2024challengesofUsingPTMs, Jiang2022PTMReuse, taraghi2024DLModelReuseinHF}. 
Well-structured naming practices enhance transparency, streamline searchability, and reduce the need for manual verification by ensuring that names accurately convey essential information such as model architecture, size, and dataset characteristics~\citep{seacord1998agoraASearchEngine4SWComponents,deissenboeck2006conciseandConsistentNaming}.
For example,~\cref{Def:ExamplePTMDefect} illustrates a PTM metadata inconsistency where there is a mismatch between the model identifier (\eg ``\code{nllb}''), package content (\eg ``\code{nllb}'' tokenizer) and its architecture metadata (\eg ``\code{m2m}''). 
While the appropriateness of PTM names can be subjective, the absence of standardized guidelines within the PTM ecosystem complicates the evaluation of naming practices. Our study examines the methods practitioners use to identify such inconsistencies and introduces an automated tool designed to detect incorrect model architecture and types.

To address these challenges, researchers have proposed solutions like model nutrition labels and queryable model zoos, which aim to provide comprehensive, standardized metadata about models, including their intended use, limitations, and performance metrics~\citep{chmielinski2022datasetNutritionLabel, li_metadata_2022, IBM2020AIMMX}. However, recent studies indicate that these approaches have not been fully adopted in practice, and many PTM models still lack key attributes and metadata~\citep{Jiang2024peatmoss, Jiang2022PTMReuse}. Consequently, engineers often rely heavily on PTM names to make initial decisions and communicate about models. 
Our study optimizes PTM naming practices to improve searchability by identifying the information that engineers want to be conveyed in names, and we provide an automated tool to detect some kinds of naming defects (inconsistencies between model architecture and metadata), further enhancing model selection.


\paragraph{Impact on Package Reliability and Security}
\label{sec:background-impactonReliability}


In addition to searchability, accurate naming in PTM packages is also critical for ensuring reliability and security. Prior research on traditional software packages highlights risks such as typosquatting and package confusion, where misleading or poorly chosen names can result in the unintentional download of malicious content~\citep{taylor_defending_2020, neupane2023beyondTyposquat}. Similarly, PTM packages with unclear or inconsistent names can obscure vulnerabilities like backdoors, introducing significant risks to downstream applications~\citep{kathikar2023assessingHFVulnerabilities, Wang2022EvilModel2, Gu2019BadNets}. The rapid proliferation of new models and the complexity of PTM architectures make clear and consistent naming essential to prevent these security threats. Accurate names that align with the package's content, architecture, and metadata can mitigate these risks and improve overall package reliability.

Manually verifying naming consistency in PTM reuse is not only time-consuming but also prone to errors. In the context of PTM integration, models are often incorporated with minimal inspection, making reliance on manual checks inefficient and risky. This can lead to the oversight of critical issues such as mislabeled models or hidden backdoors, which pose security threats. Although automated PTM reuse, facilitated by task-based assessments, is becoming more feasible, these tools still require engineers to pre-select models based on names and metadata~\citep{You2021RankingandTuningPTMs, zhang2024modelSpider}. This dependency underscores the need for clear and consistent naming conventions, as they directly influence the accuracy and security of automated reuse processes. Prior research has shown that secure software reuse and effective searchability are significantly enhanced by standardized and validated metadata practices~\citep{khalid2024repairingRawMetadata, gonzalez2004standard}. 
Our work provides an automated way to verify metadata consistency, which can reduce manual effort and improve the security and reliability of PTM reuse.



\begin{tcolorbox} [title=Summary of contribution in light of prior work, width=\linewidth, colback=yellow!30!white, top=1pt, bottom=1pt, left=2pt, right=2pt] 

Prior work shows naming conventions are crucial for the software reuse process, including PTM packages. PTM packages are known to have distinct naming challenges, but the details of those challenges have not been elaborated, nor have tools been developed to mitigate them. Our work is the first empirical study on PTM naming practices and includes the development of an automated tool to address metadata inconsistency, with a vision for future improvements. By promoting more \textbf{accurate and standardized naming conventions}, our study aims to enhance (1) the management and validation of PTM packages in registries like Hugging Face, improving organization and discoverability, and (2) the user experience by streamlining model search, reducing manual verification, and minimizing engineering overhead.

\end{tcolorbox}

\section{Research Questions and Justification} \label{sec:RQ}


\subsection{Research Questions in light of the Previous Literature}

To address this gap and promote more accurate and standardized PTM naming conventions, we present the first study in this area. Our research focuses on two themes: (1) \ul{empirical measurements of naming practices} to enhance standardization, and (2) \ul{automated identification of naming inconsistencies} to improve naming accuracy.
We ask:

\vspace{0.5mm}
\noindent \textbf{\textit{Theme 1: Measuring Naming Practices}}

\begin{itemize}[leftmargin=26pt, rightmargin=5pt]
     \item[\textbf{RQ1}]{How is PTM naming different from traditional software package naming?}
    \item[\textbf{RQ2}]{What elements should be included in a PTM identifier?}
\end{itemize}

\noindent \textbf{\textit{Theme 2: Detecting Naming Inconsistencies}}
\begin{itemize}[leftmargin=26pt, rightmargin=5pt]
   \item[\textbf{RQ3}]{How do engineers identify naming inconsistencies?}
   \item[\textbf{RQ4}]{How can naming inconsistencies be detected automatically? Specifically, how well can architectural metadata inconsistencies be detected?}
\end{itemize}

\revision{We note that RQ1-3 primarily focus on the PTM package identifier, while RQ4 shifts attention to the architectural metadata which is defined as part of PTM name (\textit{cf.} Definition 1). Although this may appear as a shift in terminology -- from naming practices to architectural metadata -- the scope remains consistent with our definition of PTM name as a combination of both elements. RQ4 specifically investigates whether the architectural metadata aligns with the model’s actual behavior, which we consider a critical aspect of naming reliability and transparency in the ecosystem.
}

\subsection{How are the Answers Expected to Advance the State of Knowledge?}

Our study has both \ul{\textit{theoretical}} and \ul{\textit{practical}} motivations that contribute to advancing the state of knowledge in software engineering and machine learning.

Theoretically, understanding developer behavior, particularly in the context of software naming practices, is a fundamental area of software engineering research. Such software engineering-theoretic knowledge holds intrinsic value, even if it does not lead to immediate applications. Our work extends prior research on naming practices, such as Feitelson’s studies on code-level naming~\citep{feitelson2020HowDeveloperChooseNames, alpern2024NamingExperiments}, by focusing on package-level naming within the pre-trained model (PTM) ecosystem.
The research literature is silent on the naming of higher-order entities such as software packages, and no comparison has been performed between the naming of traditional packages and of PTM packages.
Addressing this gap, our research questions (RQ1) and subsequent analyses (RQ2-3) aim to provide novel insights into how naming conventions evolve in these distinct contexts.

Practically, our work was originally motivated by the findings of \citet{Jiang2022PTMReuse}, one of the first qualitative studies on PTM reuse from the software engineering perspective.
Their results highlighted that engineers often depend on model names for key information when searching for and discovering models, especially when metadata and documentation are insufficient. 
Our research specifically addresses three major issues within the PTM ecosystem:

\begin{itemize}
    \item \textit{Transparency, discoverability, and reusability}: Previous work has shown that the documentation available for Hugging Face models is often inadequate, leading to reduced transparency, discoverability, and reusability of PTM packages~\citep{taraghi2024DLModelReuseinHF, Jiang2022PTMReuse}. Since model names and metadata are primary sources of information for users~\citep{di2024automatedcategorizationofPTM, Hepworth2024}, inconsistencies or omissions in these identifiers can hinder effective reuse. The alternative—manually inspecting model architectures—is more resource-intensive and time-consuming.
    Our research offers an automated approach to detecting mismatches and discrepancies in model metadata, a critical aspect of PTM naming, to enhance transparency, discoverability, and reusability of PTM packages. This represents an initial solution to the problem, we do not claim that it fully resolves all challenges related to metadata inconsistencies.

    \item \textit{Adversarial attacks on PTMs}: Architectural adversarial attacks, such as model backdoors, are closely tied to inaccuracies in model metadata. For example, an attacker could embed a backdoor within a model that triggers incorrect outputs for specific inputs~\citep{bober-irizar_architectural_2023, langford2024architecturalNeuralBackdoors}. Our work has the potential to improve the accuracy and standardization of PTM package naming conventions, thereby reducing risks associated with potential backdoors and model plagiarism by identifying discrepancies. This, in turn, facilitates a more reliable and efficient PTM package reuse process.

    \item \textit{Supply chain attacks}: Supply chain attacks, such as typosquatting, are another threat to PTM reliability. In such cases, a model might claim to be a well-known architecture like Llama but actually be a different model, potentially embedding malicious code. 
    Our approach detects inconsistencies by analyzing the architecture and generating more accurate metadata labels, thereby improving efficiency and reducing risks in model selection and validation.

\end{itemize}

Theme 1 describes PTM package naming practices in \HF, and Theme 2 develops practical methods for detecting naming inconsistencies.
Practitioner interest, as revealed through our survey, underscores the relevance of this work, particularly in establishing naming conventions and standards within the PTM ecosystem.
Our findings contribute to improving naming practices and addressing reuse challenges, laying the foundation for more efficient and secure PTM reuse.

\subsection{Structure of the rest of this paper}

We answer RQ1-3 through a survey and mining study.
We follow conventional empirical software engineering research methods for these questions.
\cref{sec:RQ1-3-Methods} describes our methodology and \cref{sec:RQ1-3-Results} shares our results for these questions.

Our goal in RQ4 is to develop an automated system to detect architectural metadata inconsistencies.
In \cref{sec:RQ4-Results}, we describe the requirements, our system design, and the results of our evaluation.

The remainder of the paper describes the threats to the validity of our work (\cref{sec: threats}),
discusses the implications of our findings for engineers and future research (\cref{sec:discussion}),
and concludes (\cref{sec:conclusion}).

\section{Methodology for Research Questions 1-3}
\label{sec:RQ1-3-Methods}

This section presents our methodology to answer Research Questions 1, 2, and 3.
Because these questions involve both understanding developer perceptions (RQ1-3) and analyzing naming practices at scale (RQ2), we apply a mixed-method approach~\citep{storey2025guidingPrinciples4MixedMethodsResearchinSE, MixedMethodsResearch} using typical empirical software engineering methods. 
We conducted a survey study (\cref{sec:theme1-method-survey}), and complemented it with a repository mining study on \HF (\cref{sec:theme1-method-measure}).

\begin{figure}[h]
    \centering
    \includegraphics[width=0.98\linewidth]{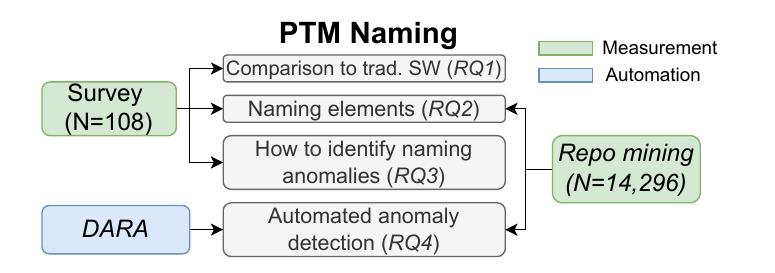}
    \caption{
    Relationship between research questions (RQs) and methodology. We conducted a survey study to address RQ1-3 and developed an automated tool, DARA, to address RQ4. Additionally, a repository mining study was conducted to support answers for RQ2 and RQ4.
    }
    \label{fig:RQ-method}
\end{figure}

\cref{fig:RQ-method} shows the relationship between RQs and methods.
Because RQ4 applies a system design methodology, we defer its answer to~\cref{sec:RQ4-Results}.

\subsection{Method for Survey Study (used for RQs 1-3)}
\label{sec:theme1-method-survey}

This section presents our methods for the survey portion of our investigation of RQs 1-3:
  preliminary analysis (\cref{sec:Method-Survey-PrelimAnalysis}),
  instrument design (\cref{sec:Method-Survey-Design}),
  recruitment (\cref{sec:Method-Survey-Recruitment}),
  and
  analysis (\cref{sec:Method-Survey-Analysis}).


\subsubsection{Preliminary Analysis}
\label{sec:Method-Survey-PrelimAnalysis}

We conducted a preliminary analysis on 200 \HF models stratified by domains to build an initial understanding of the naming conventions in PTMs packages.
Based on those results, we identified 12 naming elements (\cref{table:NamingComponents}) and 4 naming conventions (\cref{table:NamingConventions}). 
We expanded our preliminary analysis to an additional 50 models and observed no emergence of new categories. 
These categories informed the design of our instrument, ensuring that we captured the key naming elements and conventions prevalent in the domain while providing clear definitions that align with engineers' familiar terminology.

As described in \cref{sec:backgroundAndRelatedWork}, in Hugging Face, models are categorized by (\cref{fig:Naming}) the overarching domain and the tasks within them. To control for potential variability along these dimensions, we collected a stratified random sample along task types (\eg depth estimation, question answering) within each major Hugging Face domain (\eg Computer Vision, Natural Language Processing, and Multimodal) to ensure coverage of diverse modeling goals and application areas.
This method allows for representation across different model categories and domains, reducing the potential for selection bias. Given the size and diversity of the total model population on Hugging Face (approximately 974K PTMs as of September 2024), our sample size provides a 99\% confidence level with an 8\% margin of error, indicating strong statistical reliability. These results lead us to conclude that the identified naming conventions and elements are comprehensive, as no new patterns emerged in the expanded sample. However, the practical distribution of these elements and engineers' preferences remain unclear.


\subsubsection{Survey Instrument Design}
\label{sec:Method-Survey-Design}

{
\small
\renewcommand{\arraystretch}{0.6}
\begin{table*}[t]
\caption{
    Definition and examples of PTM naming components.
    All components were identified from our preliminary study of 200 PTM packages, with no additional occurrences observed in the subsequent analysis of 50 packages. The boldfaced components were specifically identified in our preliminary study's second half (\ie the last 100).
    An example real name is \code{bert-base-uncased-finetuned-spam}. which is of the form \code{A-S-C-F-D}. Another example is \code{Meta-Llama-3.1-8B-Instruct}, which is of the form \code{O-A-V-P-F}.
    }
\label{table:NamingComponents}
\begin{tabular}{p{0.32\linewidth}
 p{0.4\linewidth} p{0.18\linewidth}
}
\toprule
\textbf{Naming component} & \textbf{Definition}    & \textbf{Example}                 \\
\midrule
Architecture [A] 
& The foundational structure or design of the deep learning model.
& bert, albert, resnet
  \\
    \\
Model size [S]
& The model complexity \eg \# of parameters or architectural units.
& 50, 101, base, large, xxlarge  \\
    \\
Dataset [D] 
& The specific dataset used for training or evaluation of the model.
& squad, imagenet
  \\
    \\
Dataset characteristics [C] 
& The traits of the dataset, \eg text casing, image dimensions.
& case, uncased, 1024-1024, 244
\\
    \\
Model versioning [V] 
& The version of the model.
& v1, v2
  \\
    \\
Language [L] 
& The human language for which the model is trained on.
& english, chinese, arabic
  \\
    \\
Task [T] 
& The problem or functionality the model is designed to address.
& qa, cls, face-recognition
  \\
    \\
\textbf{Training process} [R] 
& The algorithm or approach used for training the model.
& pretrain, sparse
  \\
    \\
\textbf{Reuse method} [F]
& The approach employed to adapt a PTM for downstream tasks.
& fine-tune, distill, few-shot, instruct
  \\ 
    \\
\textbf{Number of layers} [Y] 
& The depth of the neural network in terms of layers.
& L-12, L-24
  \\
    \\
\textbf{Number of parameters} [P] 
& The total number of trainable parameters in the model.
& 100M, 8B
  \\
    \\
Other [O] 
& A portion of the model name not classified above categories.
& demo, test
\\
    \\

\bottomrule
\end{tabular}
\end{table*}
}

{
\small
\renewcommand{\arraystretch}{0.8}
\begin{table*}[t]
\caption{
    Definition and examples of PTM naming conventions.
    These conventions were identified from our preliminary study of 200 PTM packages. 
    }
\label{table:NamingConventions}
\begin{tabular}{p{0.18\linewidth}
 p{0.5\linewidth} p{0.22\linewidth}
}
\toprule
\textbf{Naming \newline conventions} & \textbf{Definition}    & \textbf{Example}                 \\
\midrule
Implementation Unit
& The detailed configuration and design aspects of a model that shape its structural and operational framework.
& bert-base-uncased, whisper-large-v2-pt-v3
  \\
    \\
Application/Task
& The real-world applications or specific tasks that the models are designed for.
& fake-news-detector, question-answering\\
    \\
Implementation with App./Task
& The model name that includes both the implementation unit and the application/task it is designed for.
& distilroberta-base-finetuned-fake-news-detection\\
\\
Other
& The model name that does not fall into non of the above categories or have unclear meaning.
& potat1, csproject

\\
\bottomrule
\end{tabular}
\end{table*}
}
%
%
We developed our survey instrument following Kitchenham and Pfleeger's guidelines for survey design~\citep{kitchenham2008personalSurvey}. 
\revision{They recommend six steps. In this section we cover the first four -- defining objectives, designing the survey, developing the instrument, and survey evaluation.
The remaining two steps -- data collection and analysis -- are discussed in \cref{sec:Method-Survey-Recruitment} and \cref{sec:Method-Survey-Analysis}, respectively.}


\myparagraph{Instrument Development}
\ul{Step 1: Search the relevant literature:} 
We began our study with a comprehensive literature review focused on naming practices within software engineering. We conducted the literature review using a keyword search method on the DBLP database.\footnote{See \url{https://dblp.org/}.} 
\review{Reviewer 3-Concern 4}{I also wonder about the purpose of the review.}
\revision{The goal of this review was to inform the survey design by identifying existing naming practices and reuse challenges relevant to PTMs.}
\review{Reviewer 2-Cocern 14}{How did the authors ensure that the survey questions were sufficient to support the study, similar to the use of keyword searches in DBLP for the literature review?}
\revision{
To identify papers on software naming practices, we searched using the following Boolean keyword query:
\begin{quote} \texttt{("software name" OR "naming convention" OR "naming practices")} AND
\texttt{("software" OR "programming" OR "software package")}
\end{quote}
We also included related terms like \texttt{"empirical study"} and \texttt{"software naming"} to broaden the scope. Separately, to identify relevant work on PTM reuse, we used the following query:
\begin{quote} 
\texttt{("pre-trained model" OR "machine learning model" OR "AI model" OR "deep learning model") AND "reuse"}
\end{quote}
}
\noindent
\review{Reviewer 3-Concern 3}{How many studies were returned from the systematic review? How were they filtered and used for subsequent analysis? The paper does not provide many details after the study collection.}
These search queries resulted in \revision{22} papers on software package naming (\eg \citet{seacord1998agoraASearchEngine4SWComponents, Lawrie2007IdentifierNamingStudy, Butler2012MethodNameConsistency}) and \revision{2 papers on }PTM reuse \citep{Jiang2022PTMReuse, Jiang2022PTMSupplyChain} at the time of designing our instrument.
\revision{We excluded 16 papers that primarily focused on naming at the method or class level, which were less relevant to our focus on package-level and model-level naming conventions (\eg \citet{allamanis_suggesting_2015, Wang2023PreImplementationMethodNamePrediction4OOP, Butler2012MethodNameConsistency}).
}

\review{Reviewer 3-Concern 5}{Were different sets of documents analyzed for multiple rounds of coding during thematic analysis? More details should be provided.}
\revision{The remaining 8 papers were analyzed in depth by the lead author, who was responsible for survey and instrument design. Through multiple close readings and reflection, the author synthesized key themes on naming motivations and reuse challenges in traditional software and considered how they may transfer to the PTM ecosystem. This thematic synthesis set the objectives of our study and informed the development of our research questions (\cref{sec:RQ}).}

\interfootnotelinepenalty=10000
\ul{Step 2: Construct an instrument: }The initial instrument includes five sections: demographic questions, comparison between PTM naming and traditional software engineering (RQ1), PTM naming elements and conventions (RQ2), and practices on naming inconsistency detection (RQ3).\footnote{For completeness, we note that the survey instrument also collected data on software engineers' use of PTM interoperability tools, such as the Open Neural Network Exchange (ONNX), with results reported in \citet{jajal2023ONNXFailureStudy}. Those questions were presented on a separate page \ul{after} all questions related to PTM naming, so we do not anticipate any priming effects or interactions between the two parts of the survey.}
\revision{The survey was designed to take approximately 15 minutes and primarily consisted of multiple-choice questions, with a few Likert-scale and open-ended questions included to gather more detailed insights from participants.}

\ul{Step 3: Evaluate the instrument:}
\review{Reviewer 2-Concern 14}{How did the authors ensure that the survey questions were sufficient to support the study, similar to the use of keyword searches in DBLP for the literature review?}
\revision{To ensure the survey instrument was appropriate and aligned with our research goals, we conducted a pilot study in two rounds.}
\review{Reviewer 2-Concern 3}{The survey was conducted using only 1.7\% of the Hugging Face dataset, with a low 5.4\% r\textit{}esponse rate predominantly from experienced users. This approach potentially introduces significant methodological biases, which could impact the validity of the findings. The authors should explain how they addressed these biases.}
\revision{Our target population consists of practitioners and researchers familiar with PTMs, particularly those who have used or published models on Hugging Face.}

\revision{In \textit{Round 1 (internal pilot)}, we invited three PhD students conducting ML research with prior experience using PTMs from Hugging Face. They completed the survey and then participated in interviews focused on (1) whether the survey comprehensively covered relevant naming and reuse issues, and (2) identifying confusing, redundant, or unnecessary questions.
After this round, the survey was revised to clarify wording, increase conciseness, and align it with the terminology and activities of users on Hugging Face. We reviewed the changes with these participants to confirm the changes addressed their concerns.}

%
\revision{In \textit{Round 2 (external pilot)}, we distributed the survey to 30 \HF PRO users and 100 ORG users. However, the response rate was very low (1\%).  In response, we substantially simplified the survey. We consolidated numerous questions and restructured the survey. For example, we removed several open-ended questions, eliminated all Likert-type questions (which required significant time and effort from pilot participants), and converted as many questions as possible into structured multiple-choice questions (based in part on the kinds of responses received in the pilots). Retesting with the 3 subjects from the internal pilot, the revised survey took 5--8 minutes to complete and still produced data relevant to our research questions.}
We allowed respondents to leave answers blank to improve the response rate, which reduced the volume of responses per question but increased the total number of responses.


{
\renewcommand{\arraystretch}{1.2}
\begin{table}[h]
\centering
\caption{
  Example survey questions.
  The survey had 13 questions in total.
  The survey used a mix of
    multiple-choice questions (Topics 1 and 2),
    checkbox questions (Topics 1 and 3),
    and
    open-ended responses (Topic 2 and 4).
  The full instrument is available in the supplemental material (\cref{sec:Artifact}).
  } 
\label{tab:ExampleSurveyQuestions}
\small
\begin{tabular}{p{0.22\columnwidth} p{0.7\columnwidth}}
\toprule
    \textbf{Topic} & \textbf{Example questions} \\
\midrule
(1) Demographics & 
(1) What is the size of your organization? \\
& (2) What deployment contexts do you work on? \\
& (3) How many PTM packages have you used/created from model registries? \\
\midrule
(2) Comparison to Traditional Software & 
(1) How is PTM naming similar/different from naming traditional software packages such as those on NPM or PyPi? Why do you think that is? \\
\midrule
(3) Naming practices & 
(1) Which naming convention do you prefer when reusing PTMs from model registries like Hugging Face? \\
& (2) Here is a list of PTM naming elements. Check each box if you think it would be important to include that element in a name. \\
& (3) Machine learning models are often adapted for improved performance or specific needs. These modifications might involve changing the architecture, training regime, and dataset. What kinds of modifications might necessitate a new model type/architecture for the model (\eg ``distilBERT''), as opposed to continuing to hyphenate the name (\eg ``albert-v2-50M'')? \\
\midrule
(4) Naming challenges & 
(1) In your experience, do the PTMs available in model registries accurately describe their behavior/content? What discrepancies have you experienced? Please explain. \\
& (2) In your regular practice, do you think you would notice if a PTM had an incorrect name? Check the box if you think YOU WOULD NOTICE if the naming element were incorrect. \\
& (3) If you think you would typically notice one or more of these naming elements being incorrect, please tell us what process you follow to do so (e.g., reading model card, reading source, etc.). \\
\bottomrule
\end{tabular}
\end{table}
}

\ul{Step 4: Document the instrument:} \cref{tab:ExampleSurveyQuestions} presents example questions from the final survey used in this study, which includes demographic questions and queries on PTM naming practices. To encourage higher response rates, all questions were optional. Additionally, the survey provided a definition of PTM naming (\cref{sec:background-PTMNaming}) for clarity. 
The final instrument is available in~\cref{sec:Artifact}.
Our study was approved by institutional IRB \#2022-606.

\subsubsection{Recruitment}
\label{sec:Method-Survey-Recruitment}

%
%
We compiled email addresses from Hugging Face's PRO and ORG (\ie organization) user accounts. \footnote{In our judgment, this does not violate \HF's terms of service~\citep{huggingface2024termsofService}. These terms of service permit usage under strict compliance with applicable laws and regulations. Our study, being controlled and approved by our institutional IRB, is in compliance with these terms. We used sampling to ensure our activities are neither excessive nor disruptive and to respect Hugging Face's data privacy and user rights.}
PRO accounts mean that the users actively pay for the features provided by Hugging Face.
To filter for practitioners, we collected users from organizations which were labeled as \code{company}, \code{community}, and \code{non-profit}, excluding \code{university}, \code{classroom}, and no labeled organizations. 

Our survey aimed to achieve a 95\% confidence level with a margin of error of 10\%. Previous statistics indicate that Hugging Face had over 1.2 million users by 2022~\citep{garg2023powerofHFAI}, and it kept increasing in recent years. Therefore, we needed at least 96 participants to ensure statistical significance~\citep{SurveySystem2021}. 


\review{Reviewer 2-Concern 3}{The survey was conducted using only 1.7\% of the Hugging Face dataset, with a low 5.4\% response rate predominantly from experienced users. This approach potentially introduces significant methodological biases, which could impact the validity of the findings. The authors should explain how they addressed these biases.}
\revision{To ensure we targeted the appropriate population (\cref{sec:Method-Survey-Design}), we randomly shuffled the collected PRO and ORG users. We then distributed the survey in batches of 100 emails until reaching our desired sample size.}
In total, we distributed our survey to \NumofPROUsers PRO users and \NumofORGUsers organization members.
As an incentive, each participant was compensated with a \SurveyCompensation gift card.
\review{Reviewer 2-Concern 2}{The authors apply statistical methods to ensure a representative sample for the study, but it is unclear how they filtered responses for high quality.}
\revision{To ensure the quality of our survey responses, we began the survey with a screening question: \textit{``Do you regularly interact with a model exchange platform such as Hugging Face, PyTorch Hub, or an internal registry?''}. We also included a CAPTCHA to verify that responses were submitted by humans.}
\SurveyTotalResNum users responded to our survey, resulting in a response rate of \ResponseRate.
Of these, \SurveyNum actively interacted with \HF, \revision{which we consider as high-quality responses.
We used this subset in our analysis, yielding an effective response rate of \HighQualityResponseRate.}
\review{Reviewer 2-Concern 3}{The survey was conducted using only 1.7\% of the Hugging Face dataset, with a low 5.4\% response rate predominantly from experienced users. This approach potentially introduces significant methodological biases, which could impact the validity of the findings. The authors should explain how they addressed these biases.
}
\review{Reviewer 2-Concern 16}{How did the authors ensure that their findings are generalizable?}
\revision{Our recruitment strategy achieved the desired statistical significance at a 95\% confidence level by specifically targeting experienced users with in-depth insights into PTM naming conventions. However, we acknowledge that this focus on experienced users may introduce biased responses.
We discuss this further in \cref{sec: threats}.}
\cref{tab:demographic} gives subject demographic information. 

A key consideration in our analysis was whether to differentiate between respondents who only use PTMs and those who also create them. In our study, all respondents have experience using PTMs, but only 34 out of 108 (31.5\%) have not created a PTM. As a result, approximately 70\% of respondents have experience both using and creating PTMs. Given the small size of the non-creator group, making statistical distinctions between these two groups would be infeasible.

{

\renewcommand{\arraystretch}{0.4}
\begin{table}[h]
\centering
\caption{
  Participant demographics (N=\SurveyNum but some left blank). 
  The majority of participants have intermediate or expert experience in both ML and SE. All participants are PTM users.
  } 
\label{tab:demographic}
\small
\begin{tabular}{
    p{0.19\columnwidth}p{0.71\columnwidth}
}
\toprule
    \textbf{Type} &
    \textbf{Break-down}
    \\
\midrule
ML Exp.  & $<$1 yr. (11), 1-2 yr. (24), 3-5 yr. (32), $>$5 yr. (36)\\
\\
SE Exp.  & $<$1 yr. (6), 1-2 yr. (12), 3-5 yr. (22), $>$5 yr. (63)\\
\midrule
Org. Size  &  Small (1-50 employees, 58), Medium (51-250 employees, 19), Large (over 250 employees, 26) \\
\\
Deployment Env.  & Web application (68), Desktop application (21), Cloud and data center (61), IoT/embedded systems (12), Mobile (15), Other (6)\\
\midrule
\# of PTMs used  & 0 (0), 1-5 (38), 6-10 (18), 11-20 (19), $>$ 20 (28)\\
\\
\# of PTMs created  & 0 (34), 1-5 (37), 6-10 (15), 11-20 (5), $>$ 20 (12)\\ 
\bottomrule
\end{tabular}
\end{table}
}

\subsubsection{Survey Analysis} 
\label{sec:Method-Survey-Analysis}

The first step in survey analysis is to filter low-quality responses. We determined that it was unnecessary to implement techniques for filtering low-quality or erroneous responses, such as attention checks, knowledge checks, or manual reviews. Since we specifically targeted \HF users with direct experience in reusing PTMs, the responses inherently met our quality standards. 
However, to ensure accuracy, we manually inspected all data to confirm the consistency of the results and validate our assumptions about response quality.

The majority of our survey questions were formulated as multiple-choice/checkbox, which are simple to analyze.
Two open-ended textual responses required subjective analysis. We analyzed these as follows: 
\begin{enumerate}[leftmargin=10mm]
    \item The \ul{first} open-ended question, for RQ1, asked: \textit{``How is PTM naming similar/different from naming traditional software packages such as those on NPM or PyPI? Why do you think that is?''}.
 Responses were lengthy and variable, so two analysts worked together to analyze them following thematic coding methods~\citep{guest2011applied}.
 \begin{itemize}
     \item First, the analysts independently performed memoing~\citep{saldana2021coding} to develop initial themes. They initially agreed on $\sim$65\% of codes. 
     \item 
     Next, they met to iteratively refine the themes (initial codebook), assessing their reliability, merging entangled codes, and discarding rare ones. 
     \item After a second coding round, they met to resolve disputes and adjust codes until consensus was reached. For instance, ``CONFIG'' and ``REUSE'' were merged due to overlap and unclear definitions. After refining the themes, they reviewed all previous codes and achieved unanimous agreement.
 \end{itemize}
 \item 
The \ul{second} open-ended question, for RQ3, asked respondents to describe their practices for identifying naming inconsistencies.
Two analysts reviewed the data and determined that the analysis was more straightforward and showed less variability than the first question. As a result, thematic coding methods were deemed unnecessary, and the analysis was conducted by a single researcher for efficiency.

\end{enumerate}

\subsection{Method for Repository Mining in \HF (used for RQ2)}
\label{sec:Repo-Mining}
\revision{The survey data permitted us to answer RQ1 and RQ3. 
For RQ2, we complemented the survey with a repository mining study.}
This section details the methods of our mining study (\cref{sec:theme1-method-survey}) which offers a more comprehensive understanding of practical PTM naming conventions and evaluates whether these practices align with engineers' stated preferences.
\label{sec:theme1-method-measure}

\subsubsection{Data Collection}
We use the latest PTM dataset named PeaTMOSS~\citep{Jiang2024peatmoss}, which was collected in August 2023.
In our study, we focused on the identifiers of PTM packages from PeaTMOSS dataset which has over 50 downloads. This included \PeatmossPopularNum PTMs (5\% of the full dataset), representing the PTM packages actively used by engineers.
\review{Reviewer 2-Concern 3}{The survey was conducted using only 1.7\% of the Hugging Face dataset, with a low 5.4\% response rate predominantly from experienced users. This approach potentially introduces significant methodological biases, which could impact the validity of the findings. The authors should explain how they addressed these biases.}
    \revision{ We excluded models with low download counts because (by definition) they do not reflect common usage. Similar to the long tail in traditional software package registries, such models are often experimental, student work, or throwaways. Lower usage may also be correlated with lower model quality, which may be associated with poorer naming practices. However, we acknowledge that this exclusion may introduce potential bias, as discussed in \cref{sec: threats}.}

\begin{figure}[h!]
    \centering
    \includegraphics[width=0.85\columnwidth]{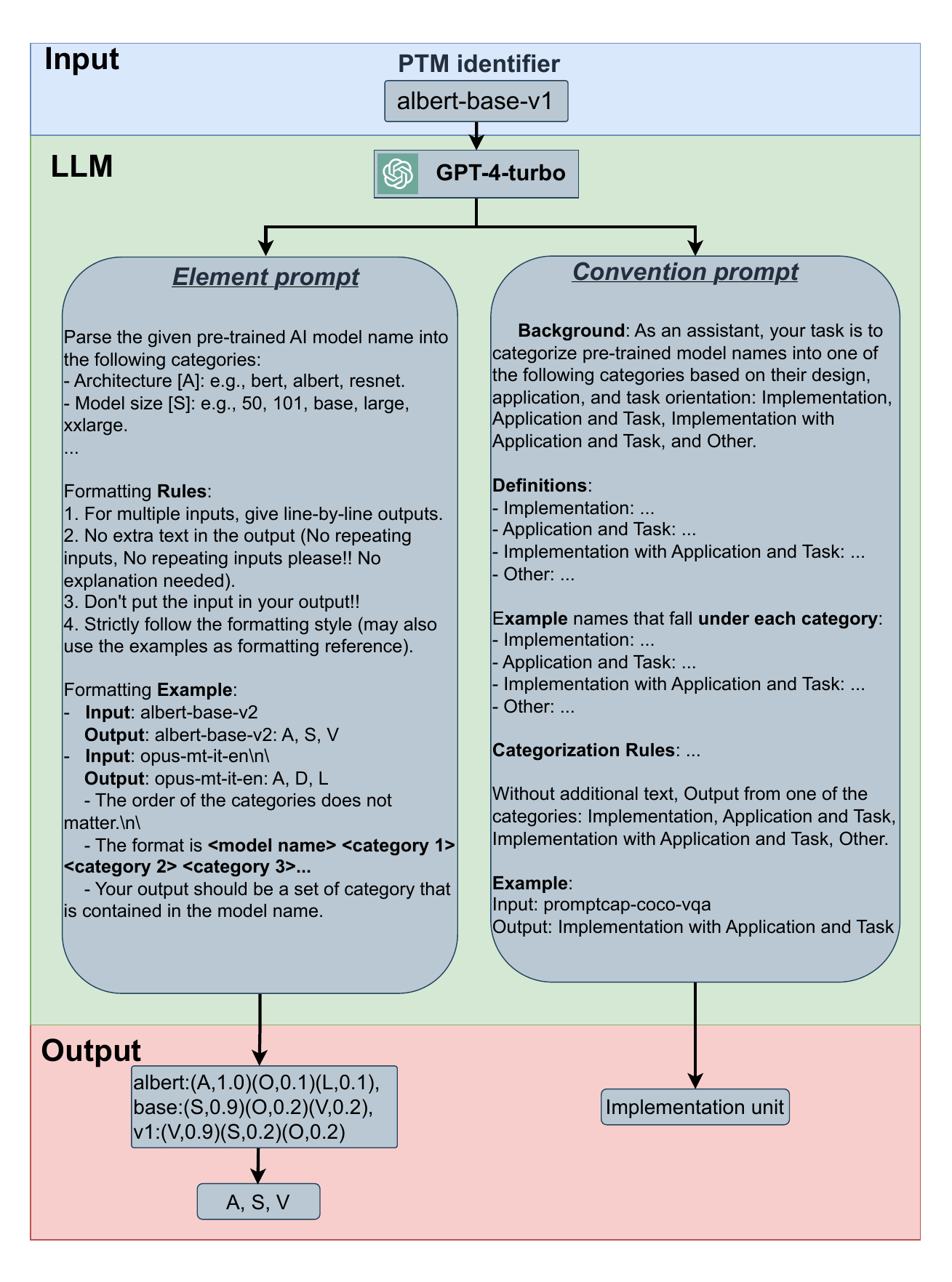}
    \caption{
    Example of input/output of GPT-4. Input is a batch of PTM package names. The prompts include both the definition of naming elements (\cref{table:NamingComponents}) and conventions (\cref{tab:NamingConventionMeasure}). 
    In this example, A represents the \textbf{\ul{A}}rchitecture (\eg Albert), S represents the model \textbf{\ul{S}}ize (\eg base), and V represents the \textbf{\ul{V}}ersion (\eg v1). Full versions of both prompts are available in \cref{sec:Artifact}. \review{Reviewer 3-Concern 7}{The prompt design is not clear. For example, how the few-shot learning was applied to LLM is not clear.  How were the two tables included in the prompt? The detailed prompt needs to be shown somewhere. The relevant reference is also missing there.
} \AUTHOR{This figure now includes the prompt examples to illustrate the prompt design, as requested by the reviewers.} 
    }
    
    \label{fig:PromptDesign}
\end{figure}

\subsubsection{Repository Mining Pipeline Design Using LLM}

The aim of our repository mining study is to extract naming elements and conventions from open-source PTM package names. Since these names often reflect nuanced, context-specific choices by engineers, the pipeline requires strong natural language understanding.
Large language models have demonstrated strong capabilities in this area~\citep{Bubeck2023AGISystem}. 
\review{Reviewer 3-Concern 9}{Do the authors consider prompt-based extraction as a technical contribution? If yes, the prompts needed to be tested against their variants. It looks like multiple annotators were not needed to extract the model elements. The LLM was also able to extract 91\% of the items from the provided models. Thus, I am skeptical about the technical contribution of this part of the work.
}
\AUTHOR{We clarified that the use of LLMs in our work serves as a practical tool for mining rather than a core research contribution, and we now provide more detail on the rationale behind this choice.
}
\revision{To analyze naming practices at scale, we designed a rule-guided prompt engineering approach using the ChatGPT API (\code{gpt-4-turbo}). 
We chose to apply LLM technology to this language processing task because of its ability to parse diverse naming patterns, infer implicit conventions, and handle edge cases that are difficult to capture with regular expressions or static heuristics. Our pipeline was then built around this capability, with a focus on interpretability and reproducibility through clear task instructions and output formatting constraints.}

\cref{fig:PromptDesign} illustrates the LLM pipeline we developed. In this process, we input the name of a PTM package and employ two distinct prompts
(\cref{sec:method-MeasurementEval}) 
to deduce the naming conventions and elements associated with that name. 


Following the methodology suggested by Zamfirescu \etal~\citep{zamfirescu2023johnny}, we utilized a structured process to design the prompt for our measurements. 
For each measurement, we first crafted a textual input (\ie prompt) and conducted a preliminary analysis on 20 randomly selected models. 
\textit{First}, we incorporated the definitions from our survey into the prompt to clarify naming elements and conventions. 
\textit{Then}, we enriched the prompt with definitions and examples for each specified component to unify semantic naming patterns. 
\textit{Finally}, we added several rules to regulate the output format, and apply few-shot prompting, presenting example inputs alongside desired outputs\review{Reviewer 3-Concern 8}{The relevant reference is also missing there.
}~\revision{\citep{brown2020languageModelsAreFewShotLearners}}. 

This process involved iterative refinement focused on two primary objectives by adjusting the \textit{output format instructions} and refining the \textit{definitions} used in each prompt:
    (1) Ensuring the generated output adheres to a well-defined, easily parsable format; 
    (2) Optimizing the performance of the output, evaluated through manual inspection.

\review{Reviewer 3-Concern 7}{The prompt design is not clear. For example, how the few-shot learning was applied to LLM is not clear.  How were the two tables included in the prompt? The detailed prompt needs to be shown somewhere.}
\revision{\cref{fig:PromptDesign} presents the structure and some content of each prompt. We also briefly describe each prompt as follows:
\begin{itemize}
    \item \textbf{Naming Elements Prompt:} The prompt leverages definitions from \cref{table:NamingComponents} along with illustrative examples to guide few-shot prompting within the LLM pipeline. The prompt details the output formatting rules and provides examples to promote compliance.
    \item \textbf{Naming Convention Prompt:} The prompt begins by defining all four naming convention categories (\cref{tab:NamingConventionMeasure}) and specify the relevant naming elements under each convention (\eg architecture, model size, dataset, versioning) that contribute to each category. A curated list of examples further supports few-shot reasoning, and the prompt concludes with explicit output rules and an end-to-end example to ensure consistent formatting.
\end{itemize}
}
The final prompts are available in our artifact (\cref{sec:Artifact}).
We report our final evaluation results in \cref{sec:method-MeasurementEval}.


\subsubsection{Method of Evaluation}
\label{sec:method-MeasurementEval}

\review{Reviewer 3-Concern 10}{I also wonder why the experimental findings (Tables 5 and 6) are a part of the study methodology. It makes the paper hard to read/understand. If they are study findings, they should be sent to the next section.
}
\AUTHOR{We have moved the original Table 5 and 6 into the result section (\cref{sec:RQ2-EvalofMiningMethod}) and rewritten the relevant prose. We also removed the prompting from the claimed contributions in \cref{sec:Intro} to more appropriately focus those on our results.}
\revision{We evaluated the reliability of our mining approach by comparing the LLM-generated outputs to manually labeled data for both naming elements and naming conventions. 
Two manually labeled datasets were used to assess the accuracy of the LLM pipeline.
}

\revision{For \ul{\textit{naming elements}}, we randomly sampled 300 models from the full list, providing a 95\% confidence level. 
One researcher labeled the naming components by referring to model cards and consulted a second researcher for ambiguous cases. This labeling process was straightforward due to clear patterns in the model names and documentation.}


\review{Reviewer 3-Concern 9}{Do the authors consider prompt-based extraction as a technical contribution? If yes, the prompts needed to be tested against their variants. It looks like multiple annotators were not needed to extract the model elements. The LLM was also able to extract 91\% of the items from the provided models. Thus, I am skeptical about the technical contribution of this part of the work.
}
\AUTHOR{We address this concern by providing more rationale in the following paragraph.}
\revision{Given the context-sensitive nature of \ul{\textit{naming conventions}}, two annotators followed our predefined labeling criteria and independently annotated 200 PTM package names. For instance, the naming element ``language'' could be categorized as either an implementation unit or an application, depending on the model context.
The initial agreement between the two researchers, measured by Cohen's Kappa, was \KimsKappaInit (substantial agreement). Most disagreements were from confusion over abbreviations or borderline cases. 
The researchers also agreed to classify unclear cases as ``other'' category. After resolving these issues through discussion, they refined their labels and reached a final agreement of \KimsKappaFinal (almost perfect agreement).
}
\section{Results and Analysis for RQs 1-3}
\label{sec:RQ1-3-Results}

Here we present the results for Research Questions
  1 (\cref{sec:RQ1-Results}),
  2 (\cref{sec:RQ2-Results}),
  and
  3 (\cref{sec:RQ3-Results}).

\subsection{\textbf{RQ1}: How is PTM naming different from traditional software package naming?}
\label{sec:RQ1-Results}

\begin{tcolorbox} [title=Finding 1: PTM naming is different from traditional naming., width=\linewidth, colback=yellow!30!white, top=1pt, bottom=1pt, left=2pt, right=2pt]
88\% of respondents perceive PTM naming to be different from traditional package naming.
The main difference is the greater semantic knowledge embedded in PTM names.
The main explanations for this difference were:
  (1) differences in evolution practices;
  and
  (2) different information needed for reuse decisions.
  and
\end{tcolorbox}

A total of \SurveyTopicTwoNum respondents (out of \SurveyNum) replied to the relevant subset of questions (Topic 2 in \cref{tab:ExampleSurveyQuestions}).
We focused on the common case (88\%): responses describing differences.
We analyzed responses in two ways:
  \textit{how} the names differ,
  and
  \textit{why} the names differ.

\cref{table:Survey-HowPTMNamesDiffer} summarizes the codes and response frequency related to \textit{how} names differ between PTMs and traditional software packages.
The primary difference was in the greater amount of semantic knowledge in PTM names, as depicted in~\cref{table:NamingComponents}.
A representative answer was ``\textit{Pre-trained model names often include more specific information about the model's architecture, training data, and performance. This...detail is not typically included in traditional...package names.}''
Other distinctions were that
  subjects observed greater variation in PTM names
    (``\textit{The versioning system for models seems less well-developed}''),
  and that
  subjects saw different kinds of names between PTMs and traditional packages:
    ``\textit{PyPI can be sensible names, based on archetypal action, or fun names, based on authors' aesthetics}''.

%
%
{
\small
\renewcommand{\arraystretch}{0.5}
\begin{table}[h]
\centering
\caption{
  Induced themes and definitions for \textit{how} PTM names differ from traditional names.
  Based on code-able responses from 34 participants.
    The key difference in PTM names is their higher semantic content compared to traditional packages.
  }
\label{table:Survey-HowPTMNamesDiffer}
\small
\begin{tabular}{llc}
\toprule
\textbf{Theme}  & \textbf{Definition} & \textbf{\# Participants (\%)} \\
\midrule
SEM-KNOW & Contain more semantic knowledge & 30 / 34 (88\%) \\
MORE-VARIATION & More variation in PTM naming practices & 5 / 34 (15\%) \\
DIFF-KINDS  &
Kinds: Archetypal vs. aesthetic names & 4 / 34 (12\%) \\
\bottomrule
\end{tabular}
\end{table}
}

\cref{table:Survey-WhyPTMNamesDiffer} summarizes the codes and response frequency related to \textit{why} names differ between PTMs and traditional packages.
The two main explanations were about evolution practices and reuse practices.
One subject described the influence of evolutionary practices on names as follows:
  ``\textit{[Traditional] packages are not usually as directly built on previous ones, with small extensions added like for PTM. Finetuning models generally means that the outputs are similar to the base, unlike with traditional packages.}''
With respect to the information needed for reuse, a representative quote was:
  ``\textit{Traditional software packages are supposed to contain an entire library...[with] various functions...and configuration...PTMs are generally trained for a single objective...Therefore, [PTM] naming is critical for readability and avoiding mistakes.}''.


{
\small
\renewcommand{\arraystretch}{1}
\begin{table}[h]
\centering
\caption{
  Induced themes and definitions for \textit{why} PTM names differ from traditional package names, based on coded responses from 37 participants. 
  Half of the respondents identified evolution practices and essential reuse information as the key factors driving these differences.
  }
\label{table:Survey-WhyPTMNamesDiffer}
\small
\begin{tabular}{llc}
\toprule
\textbf{Theme}  & \textbf{Definition} & \textbf{\# Participants (\%)} \\
\midrule
FORK & Evolution practice (``forking'') & 18 / 37 (49\%) \\
REUSE & Information needed for reuse & 18 / 37 (49\%) \\
COMMS & Conventional location of semantic information & 6 / 37 (16\%) \\
VERSION & Definition of versioning & 4 / 37 (11\%) \\
NO-STD & No standardization, ``Wild West'' & 5 / 37 (14\%) \\
OTHER & Low frequent categories  & 7 / 37 (19\%) \\
\bottomrule
\end{tabular}
\end{table}
}

\subsection{\textbf{RQ2}: What elements to include in PTM identifiers?}
\label{sec:RQ2-Results}


\revision{This section first validates our mining method (\cref{sec:RQ2-EvalofMiningMethod}), and then gives an integrated answer to RQ2 combining our survey data and mining data (\cref{sec:ComparisontoMethodNaming}).}

\begin{tcolorbox} [title=Finding 2: Participants prefer naming by architecture/function., width=\linewidth, colback=yellow!30!white, top=1pt, bottom=1pt, left=2pt, right=2pt]
Survey participants preferred naming pre-trained models based on architectural characteristics (``how it works'') and intended functions (``what it does'').
There was less interest in incorporating training details (``how it was made'').
\end{tcolorbox}

\subsubsection{Validation of Mining Method}
\label{sec:RQ2-EvalofMiningMethod}

As described in \cref{sec:method-MeasurementEval}, we manually labeled naming elements and convention data, using these annotations as ground truth. We then compared the pipeline outputs against the manually labeled dataset to validate our pipeline.
\cref{tab:NamingElementMeasure} shows that the naming element extraction achieved an F1-score of \ElementMeasureAcc, while \cref{tab:NamingConventionMeasure} shows an F1-score of
\ConventionMeasureAcc for naming convention classification.
These results indicate that the pipeline produces reliable outputs. Therefore, we applied the pipeline at scale to systematically measure naming practices in real-world PTMs, and compare with our survey results, as detailed in \cref{sec:ComparisontoMethodNaming}.

{
\begin{table}[h]
\centering
\caption{
Evaluation of naming element measurements.
We evaluated the results from 300 randomly selected names, achieving a 95\% confidence level with a 5\% margin of error.
The definition of naming elements are presented in \cref{table:NamingComponents}.
\textit{NOTE: No existing tools or data are available, so our manual analysis serves as the baseline.}
}

\begin{tabular}{lcccc}
\toprule
\textbf{Element} & \textbf{Precision} & \textbf{Recall} & \textbf{F1-Score} \\ 
\midrule
Architecture [A] & 0.98 & 0.98 & 0.98 \\
Model size [S] & 0.96 & 1.00 & 0.98 \\
Dataset [D] & 0.94 & 0.89 & 0.92 \\
Dataset characteristics [C] & 1.00 & 0.74 & 0.85 \\
Model versioning [V] & 1.00 & 0.94 & 0.97 \\
Reuse method [F] & 1.00 & 0.87 & 0.93 \\
Language [L] & 0.95 & 0.92 & 0.94 \\
Task [T] & 0.90 & 0.91 & 0.91 \\
Training process [R] & 0.80 & 0.80 & 0.80 \\
Number of layers [Y] & 0.67 & 1.00 & 0.80 \\
Number of params. [P] & 1.00 & 0.96 & 0.98 \\
Other [O] & 0.87 & 0.81 & 0.84 \\
\midrule
\multicolumn{1}{l}{\textit{Average}} & 0.92 & 0.90 & 0.91\\ 

\bottomrule
\end{tabular}
\label{tab:NamingElementMeasure}
\end{table}
}

%
%
%

{
\small
\begin{table}[h]
\centering
\caption{Evaluation of naming convention measurements. We evaluated the results of 200 randomly selected names, achieving a 95\% confidence level with a 7\% margin of error. The data were manually labeled, and the results were discussed by two researchers to ensure quality. \textit{NOTE: No existing tools or data are available, so our manual analysis serves as the baseline. Data is available in our artifact (\cref{sec:Artifact}).} 
}
\small
\begin{tabular}{lcccc}
\toprule
\textbf{Naming Convention} & \textbf{Precision} & \textbf{Recall} & \textbf{F1-Score} \\ 
\midrule
Impl. unit & 0.79 & 0.94 & 0.86 \\
App. or Task & 1.00 & 1.00 & 1.00 \\
Impl. + App./Task & 0.92 & 0.73 & 0.81 \\
Other & 1.00 & 1.00 & 1.00 \\
\midrule
\multicolumn{1}{l}{\textit{Average}} & 0.90 & 0.89 & 0.89\\ 

\bottomrule
\end{tabular}
\label{tab:NamingConventionMeasure}
\end{table}
}

\subsubsection{Comparison of Mining and Survey Results
}
\label{sec:ComparisontoMethodNaming}

\begin{figure}[h]
    \centering
    \includegraphics[width=0.6\columnwidth]{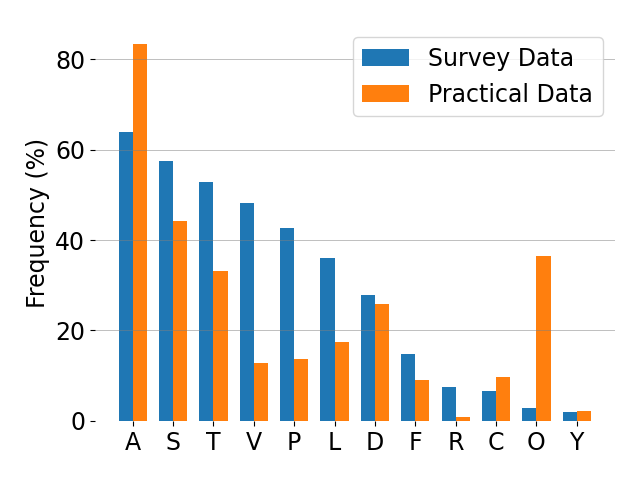}
    \caption{
    The surveyed (blue)
      and measured (orange)
      distribution of naming elements.
      The x-axis refers to the codes from~\cref{table:NamingComponents}.
      The figure shows that naming practices do not consistently align with engineers' naming preferences.
      }
    \label{fig:NamingElements}
\end{figure}

\cref{fig:NamingElements} illustrates survey participants' preferences (blue bars) for including specific elements in the naming of pre-trained models (PTMs). The elements most favored are related to the model's architectural lineage: ``Architecture'' (69/108, 63.9\%), ``Model size'' (62/108, 57.4\%), and ``Task'' (57/108, 52.8\%). 
This highlights a strong preference for names that convey the model's high-level implementation details, facilitating easy identification and distinction, This is usually a starting point of PTM reuse as indicated by Jiang \etal~\citep{Jiang2022PTMReuse}.
A smaller yet significant number of users want low-level details, including model
versioning (52/108, 48.1\%), number of parameters (46/108, 42.6\%), language (39/108, 36.1\%), and training dataset (30/108, 27.8\%). 
These elements are relevant to model selection and evaluation. 

\cref{fig:NamingElements} also presents the practical distribution (orange bars) of naming elements. Comparing survey participant preferences (blue) with real PTM names (orange), we see that naming practices do not always match naming preferences.
Almost all models ($\sim$85\%) include Architecture (A) information, though only $\sim$60\% of respondents prefer it.
As an example in the reverse direction, many respondents value information such as task (T), version (V), and parameter count (P), but these elements are comparatively rare in real model names, appearing in only 32\%, 11\%, and 12\% of names, respectively.

\begin{tcolorbox} [title=Finding 3: Design purpose is important in PTM naming., width=\linewidth, colback=yellow!30!white, top=1pt, bottom=1pt, left=2pt, right=2pt]
The survey suggests a preference for names that combine details about both the implementation and the application or task (60.2\%), while actual PTM packages tend to be named solely based on implementation units (59.0\%). \end{tcolorbox}

\begin{figure}[h]
    \centering
    \includegraphics[width=0.6\columnwidth]{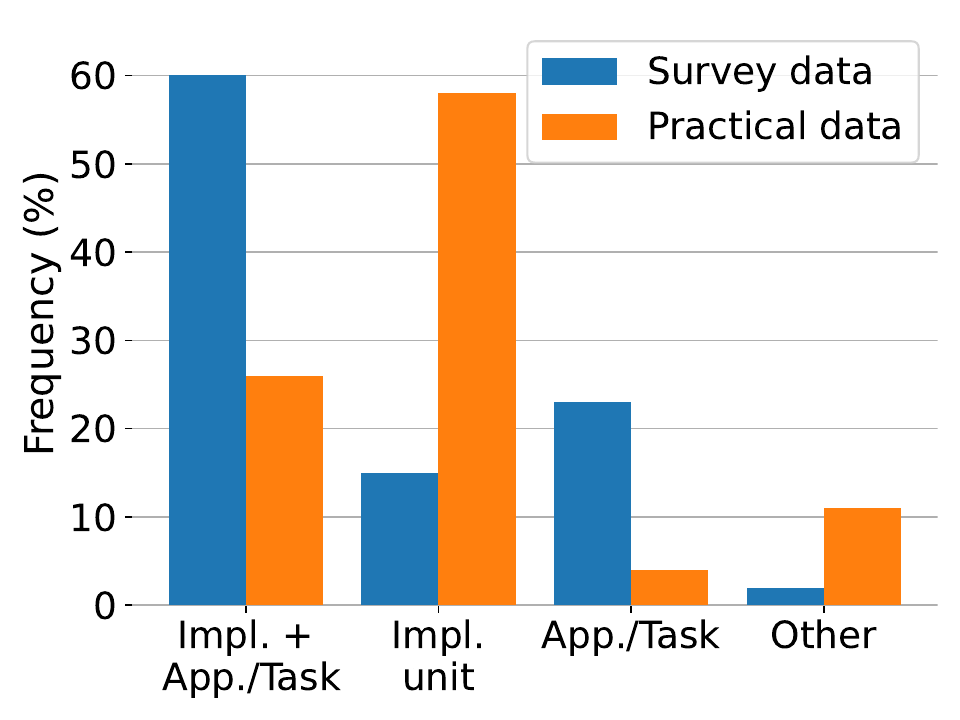}
    \caption{
      The surveyed (blue)
      and measured (orange)
       distribution of naming conventions.
       Survey results indicate that design purpose is considered important in PTM naming, yet it is not widely reflected in practice.
    }
    \label{fig:NamingConventions}
\end{figure}

\cref{fig:NamingConventions} displays engineers' preferences for naming conventions in the context of reusing PTMs, along with the measured practices using our LLM-based automated extraction pipeline.
The most preferred convention (58, 60.2\%) combines ``implementation + application/task'' (\eg \code{Llama-2-7b-chat-hf}), with ``Application/task'' following (41, 23.3\%). This suggests engineers favor names that reflect both the intended purpose and the implementation specifics of PTM packages.
These patterns are rare in practice (only $\sim$35\% and $\sim$5\% of models) --- PTM names tend to share only implementation details (59\%). 
PTM users' preferences are misaligned with naming practices. 

\subsection{\textbf{RQ3}: How do engineers identify inconsistent names?}
\label{sec:RQ3-Results}



\begin{tcolorbox} [title=Finding 4: Users manually verify naming consistency., width=\linewidth, colback=yellow!30!white, top=1pt, bottom=1pt, left=2pt, right=2pt]
PTM users identify naming inconsistencies by reading both internal (69\%) and external (11\%) information provided by the PTM package. Around half of users (46\%) inspect it, \eg via measurement or visualization. They inspect both model metadata and architecture.
\end{tcolorbox}

A total of 52 respondents replied to the survey question corresponding to this RQ (\cref{sec:Method-Survey-Analysis}).
\cref{table:AnomalyProcesses} 
summarizes the codes and response frequency
for how to identify PTM naming inconsistencies.
PTM users described two kinds of approaches:
  (1) reading internal/external information about the PTM;
  and
  (2) downloading and inspecting the PTM.

{
\small
\renewcommand{\arraystretch}{0.5}
\begin{table}[h]
\centering
\caption{
  Actions for identifying PTM naming inconsistencies, mentioned by multiple participants.
  The third column shows the number of participants who mentioned each process.
  52 (out of \SurveyNum) participants answered this question. 
  Users typically rely on manual metadata validation or model inspection to verify the correctness of a PTM name.
  }
\label{table:AnomalyProcesses}
\small
\begin{tabular}{lc}
\toprule
\textbf{Action}  & \textbf{\# Participants (\%)} \\
\midrule
\addlinespace[4pt] 
Read metadata included with PTM
& 36 / 52 (69\%) \\
\addlinespace[2pt] 
Read external information about PTM 
& 11 / 52 (21\%) \\
\addlinespace[2pt] 
Inspect the PTM 
& 24 /52 (46\%) \\
\bottomrule
\end{tabular}

\end{table}
}

Most participants mentioned that they check for inconsistencies by reading information provided by the PTM packages. 
A large proportion of the participants (36/52, 69\%) mentioned that they identify the naming inconsistencies by \inlinequote{reading the model cards} and other metadata included with PTM, and comparing the data with the elements in PTM names.
The metadata available in the \code{config.json} file is another primary source for verifying the accuracy of naming elements. That metadata is \inlinequote{the ultimate source of truth because models can lack model cards}. 
Some participants (11/52, 21\%) also mentioned that they refer to external information about PTM, including research papers, discussion forums, and issue reports. The external information can provide the users with more detailed information, \eg whether other people have already verified the naming elements for that PTM.

Approximately half of the participants (24/52, 46\%) mentioned that they inspect PTMs to identify naming inconsistencies. This inspection often involves \inlinequote{visualizing/summarizing the model architecture}, \inlinequote{running [the model] on actual data/examples}, and \inlinequote{examining the memory usage}. They noted that visualization or manual inspection makes it \inlinequote{obvious if the architecture is different to what the model name says}. When running inference, users can verify if \inlinequote{the memory usage matches their expectations} or \inlinequote{if there is any substantially decreased performance}. Detecting naming inconsistencies related to the task and application goals is also clear \inlinequote{if the model is unable to perform the given task/achieve the mentioned goal}. Although costly, these hands-on approaches ensure that the model's capabilities align with its description.

\section{\textbf{RQ4}: How can architectural metadata inconsistencies in PTM naming be automatically detected?}
\label{sec:RQ4-Results}

\AUTHOR{We substantially revised our presentation of this RQ's methods and results (and the methods themselves!) in response to comments from multiple reviewers. The relevant comments are listed below. To summarize the changes: (1) we moved this content into a standalone section to clearly distinguish between the survey and mining studies, as also reflected in \cref{sec:RQ}; (2) we added more explanation about the rationale behind each design decision in \cref{sec:DARA}; (3) we provided more detailed implementation information in \cref{sec:DARA}; (4) we conducted a more comprehensive evaluation, including five categories of experiments with detailed measurements of DARA’s effectiveness and efficiency; (5) we updated all figures and tables for clarity and to reflect the latest results; and (6) we expanded the discussion of DARA’s use cases in \cref{sec:DARA-UseCases}.}
\begin{tcolorbox} [title=Finding 5: Architectural information suffices to detect some kinds of naming inconsistency., width=\linewidth, colback=yellow!30!white, top=1pt, bottom=1pt, left=2pt, right=2pt]
We developed an automated tool (\DNNDiff) to identify naming inconsistencies in PTM metadata based solely on architectural information. Our results show that the tool achieves high accuracy for \code{model\_type} detection (\DARAModeltypeAcc), and performs moderately well for \code{architecture} (\DARAArchAcc) and \code{task} (\DARATaskAcc). These findings suggest that architectural features are reliable for broad model categorization but insufficient alone for more fine-grained classification tasks.

\end{tcolorbox}

\revision{From RQ1 and RQ2, we found that PTM naming differs from traditional package naming, with a preference for names that reflect architectural characteristics (\ie architecture and model type) and intended functionality (\ie task).}
From RQ3,
we learned that
  PTM users are concerned about naming inconsistencies,
  and that their current checking approaches are either cursory (reading metadata and external information) or costly (visual inspection or measurement of the PTM).

Here we investigate the automatic detection of naming inconsistencies.
Early approaches to finding anomalous names in software were predominantly rule-based~\citep{host2009debuggingMethodNames, pradel2011detectingAnomaliesEquallyTypedMethodArgs, liu2016ExploringandExploitingSimBetweenArgandParameterNames}.
More recently, machine learning-based approaches have achieved state-of-the-art performance~\citep{pradel2018deepbugs,winkler2021replication,he2021learningtoFindNamingIssueswithBigCodeandSmallSupervision}.
For instance, Pradel \etal approached the problem of naming-based bug detection using binary classification~\citep{pradel2018deepbugs}.
They proposed a method for identifying and correcting naming-related issues in source code by mining name patterns from extensive codebases.
This approach uses a classifier trained with a small dataset to achieve high precision in real-world repositories~\citep{he2021learningtoFindNamingIssueswithBigCodeandSmallSupervision}.

Inspired by these approaches, we propose a machine learning-based approach to detect anomalous names in PTM packages. 
We discuss
  feature selection (\cref{sec:DARA-FeatureSelection}),  
  feature extraction (\cref{sec:DARA}),
  evaluation approach (\cref{sec:DARA-Eval}),
  and
  results and analysis (\cref{sec:DARA-Results}).

\subsection{Feature Selection} \label{sec:DARA-FeatureSelection}

Our survey instrument included a set of questions on ``\textit{What kinds of modifications necessitate a new model type or architecture}'' (\cref{tab:ExampleSurveyQuestions}).
\cref{table:Survey-NewName} presents a compilation of factors influencing the creation of new model types or architectures by PTM developers.
More than one-third of the survey respondents believe that alterations such as modified training protocols, changes in input/output dimensions, layer additions or removals, modifications to the tokenizer, and shifts in the training dataset significantly contribute to the generation of new PTM names (\cref{table:Survey-NewName}).
If the typical engineer acts on these beliefs, then their naming function could be mathematically represented as:

\begin{equation}
\mathcal{F}\left(
\begin{array}{@{}c@{}}
\text{layer connections}\\
\text{layer parameters}\\
\text{tokenizer}\\
\text{pre-trained weights} \\
\text{training regime}\\
\text{dataset}
\end{array}
\right) = 
\left(
\begin{array}{@{}c@{}}
\text{model\_type} \\
\text{architecture} \\
\text{task}
\end{array}
\right)
\end{equation}

{
\small
\renewcommand{\arraystretch}{0.3}
\begin{table}[h]
\centering
\caption{
  Factors influencing the need for new PTM names from reusers' perspectives. Over 30\% of participants consider training regime, tensor shape, and layers in the main body as key factors for PTM name changes, while others hold no strong opinion.
  }
\label{table:Survey-NewName}
\small
\begin{tabular}{p{0.7\columnwidth}c}
\toprule
\textbf{Theme}  & \textbf{\# Part. (\%)} \\
\midrule
Modified training regime & 43/108 (40\%)\\
\\
Modified tensor shape & 34/108 (31\%) \\
\\
Modified layers in the main body & 33/108 (31\%) \\
\\
Addition/deletion of layers in the main body & 33/108 (31\%) \\
\\
Changed training dataset & 31/108 (29\%)\\
\\
Modified tokenizer & 31/108 (29\%) \\
\\
Addition/deletion of layers in IN/OUT layers & 26/108 (24\%) \\
\\
Other & 4/108 (4\%) \\

\bottomrule
\end{tabular}
\end{table}
}

The survey data suggests that the \code{model\_type} and \code{architecture} are determined by several factors \cref{sec:RQ3-Results}. However, prior studies on PTM reuse suggested that there is a lot of missing metadata from the model cards on \HF~\citep{jiang2023ptmtorrent, Jiang2024peatmoss}. 
\HF does not provide enough details about training regime and much dataset information is missing; the non-transparency reduced the trustworthiness and reusability of PTM packages~\citep{Jiang2022PTMReuse}.
The primary information provided by Hugging Face encompasses tokenizer details, pre-trained weights, and the labels for \code{model\_types} and \code{architecture}.
Given this ``data of convenience'', we therefore formulate the following engineering hypothesis:

\begin{tcolorbox} [width=\linewidth, colback=black!10!white, top=1pt, bottom=1pt, left=2pt, right=2pt]
\textbf{Engineering hypothesis:} Using only architectural information, we can effectively detect architectural naming inconsistencies.
In other words, we can learn the following function $\mathcal{F}$:

\begin{equation}
\mathcal{F}\left(
\begin{array}{@{}c@{}}
\text{layer connections}\\
\text{layer parameters}\\
\end{array}
\right) \approx 
\left(
\begin{array}{@{}c@{}}
\text{model\_type} \\
\text{architecture} \\
\text{task}
\end{array}
\right)
\end{equation}

\end{tcolorbox}



Following the guidance of~\citet{melegati2019hypotheses} and~\citet{olsson2014opinions}, we frame this as an engineering hypothesis.
Engineering hypotheses are formulated to guide the design and development of systems and can be evaluated through empirical testing~\citep{melegati2019hypotheses}.
They are design tools to be tested through system-building and evaluation, not traditional scientific hypotheses to be tested through statistical tests.
Thus, to test this hypothesis, we designed and evaluated an automated technique that extracts layer connection and parameter information, and then learns the relation between these features and the human labels from \HF.

\revision{We assess this hypothesis individually for each key component of a model's name -- model type, architecture, and task.}
We describe our basic design of the automated technique in~\cref{sec:DARA},
and further feature extraction methods in~\cref{sec:DARA-OtherFeatureExtraction}.
We consider a performance of over 75\% (heuristic) as supportive of the hypothesis, indicating that further refinement could make the approach feasible for a model registry to use to automatically detect naming inconsistencies.

\subsection{Design of DNN Architecture Assessment (DARA)} \label{sec:DARA}


\subsubsection{Overview}

\begin{figure*}[ht]
    \centering\includegraphics[width=0.7\textwidth]{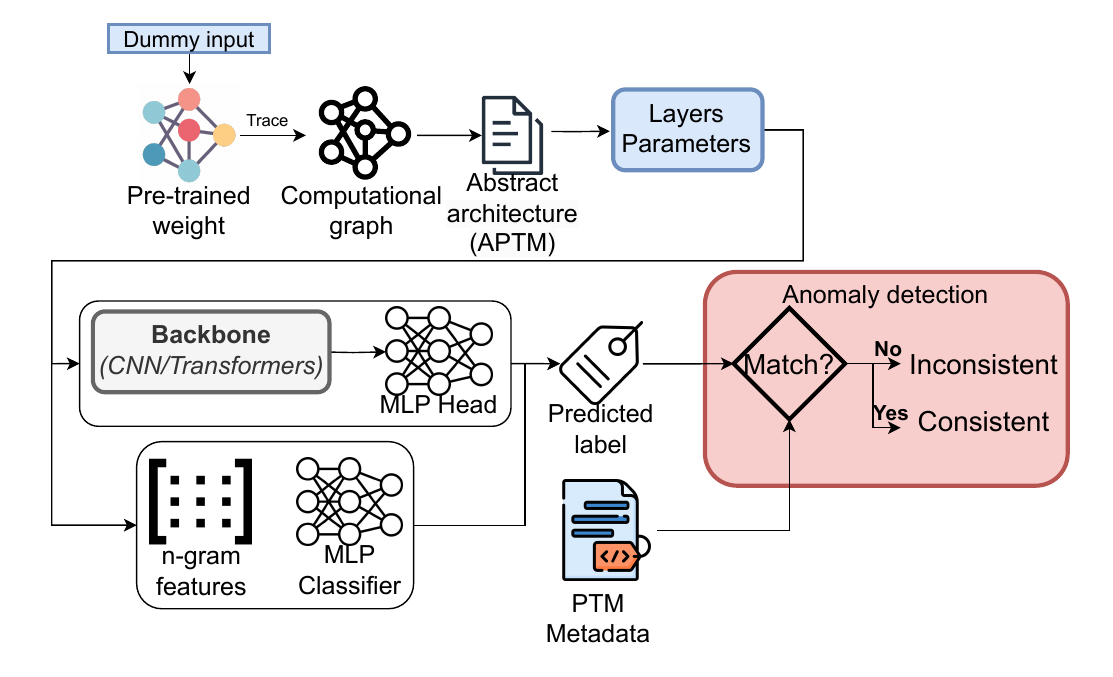}
    \caption{
      To improve PTM naming accuracies, we propose the \ul{D}NN \ul{AR}chitecture \ul{A}ssessment pipeline (\DNNDiff), an anomaly detection algorithm designed to automatically identify discrepancies in PTM names.
      The approach involves detecting architectural variations across multiple PTMs, training a classifier, and using it to spot naming inconsistencies.
    Metadata inconsistencies are defined as mismatches between the actual architecture of a PTM and its corresponding metadata.
      In the DARA pipeline, these inconsistencies manifest as differences between the classifier's prediction and the labels provided by the PTM creator.
      We evaluate DARA using three key labels: \code{model\_type}, \code{task}, and \code{architecture} (\cref{fig:Naming}).
      \review{Reviewer 3-Concern 11}{One way to tackle the inconsistency is to look at the metadata/model cards. However, it was not clear how the metadata information was used as a part of the classification algorithm. Its omission looks counterintuitive. There should be some ablation study to demonstrate that the model's architecture is better than metadata for the classification task.} 
\review{Reviewer 3-Concern 16}{Given the current workflow and classification/CNN architecture, it is not clear how DARA will generate labels for a PTM during its registration. Looks like an unsubstantiated claim.}
\review{Reviewer 3-Concern 13}{What makes a label in the proposed method's training data? Is it the whole name of the PTM or one of the three components? Model type, task, and architecture -- are they considered as training features or the labels for classification? According to my understanding, the labels should be binary -- consistent or inconsistent, where the metadata/model feature and PTM names should make the features.}
\AUTHOR{We updated this figure to explicitly indicate how we integrate the PTM metadata in the anomaly detection process to address the above three confusions from Reviewer 3.}
    }
    \label{Method:Tool}
\end{figure*}

\cref{Method:Tool} presents the \ul{D}NN \ul{AR}chitecture \ul{A}ssessment (\DNNDiff) workflow.
The process begins with open-source pre-trained weights. First, we load each weight and feed dummy inputs to outline the graph architecture of each Pre-trained Model (PTM). Next, we transform the computational graph into an abstract architecture and implement n-gram feature extraction. Utilizing these features, we train a CNN classifier to identify naming inconsistencies.


\review{Reviewer 1-Concern 1}{Specifically, I am curious whether the authors compared DARA to any baselines. My understanding is that developing this approach required substantial engineering, as the model undergoes various architectural transformations starting from open-source pre-trained weights. My question is: As you did with the method built around GPT-4 for extracting naming conventions, what were the obstacles that prevented you from using a solution that might require less engineering effort from the developer's perspective?}
\revision{We made the decision to develop DARA pipeline to detect naming anomalies for the following three reasons:
\begin{enumerate}
    \item Prior work has shown that model metadata and identifiers on Hugging Face are often manually created and may be incomplete or inconsistent~\citep{Jiang2022PTMReuse, Montes2022DiscrepanciesAmongPTNN}. Consequently, relying solely on such metadata is unreliable. \item \review{Reviewer 3-Concern 11}{One way to tackle the inconsistency is to look at the metadata/model cards. However, it was not clear how the metadata information was used as a part of the classification algorithm. Its omission looks counterintuitive. There should be some ablation study to demonstrate that the model's architecture is better than metadata for the classification task.
} 
\AUTHOR{This point was added for the concern from Reviewer 1, but it is also relevant to the preceding concern from Reviewer 3.}
We view the pre-trained weights as the core components of PTM packages, similar to code modules in traditional software, since models function as executable code~\citep{zhao2024modelsarecode, ayodele2010typesofMLAlgorithms}. Thus, the pre-trained weight file can be relied upon as the trustworthy element in a PTM package --- not that it can be trusted, necessarily, but that it is the incontrovertible fact beneath the PTM's name.
    \item There is no off-the-shelf approach that could be directly applied to pre-trained model packages for anomaly detection tasks.
\end{enumerate}
}

\noindent
\revision{Therefore, we developed a pipeline from scratch that can take pre-trained weight files as input and identify naming inconsistencies with an anomaly detection approach (\cref{sec:DARA}).}

In this section, we present our DNN architecture assessment pipeline, including computational graph conversion, delineation of abstract architecture, multiple feature selection methods, and CNN classifier head for anomaly detection.

\begin{figure}
    \centering
    \includegraphics[width=0.7\textwidth]{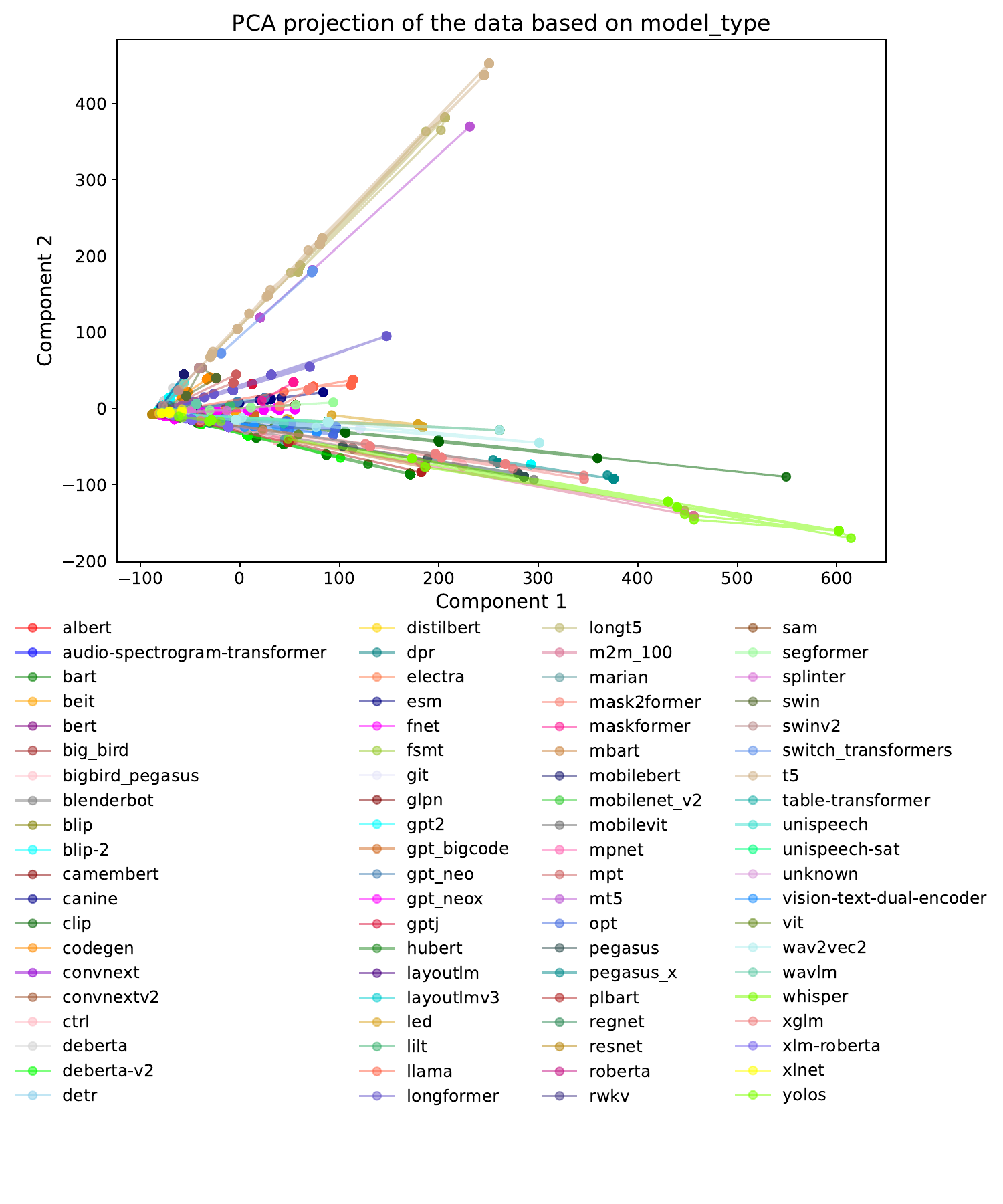}
    \caption{
    PCA projection of our high-dimensional feature vector (\DARAFeatureDim dimensions in our dataset of \DARAPTMNum PTM packages), visualizing \code{model\_type}. The plot demonstrates that at least some \code{model\_types} should be distinguishable using our feature extraction method.
    }
    \label{fig:DARAFeatureVisualization}
\end{figure}


\subsubsection{Computational Graph Conversion} \label{sec:GraphConversion}
We developed an automated pipeline to transform pre-trained weights into computational graphs.
Existing work shows that PyTorch is the dominant framework in \HF ~\citep{castano2023analyzingEvolandMaintenanceofMLModelsonHF}, and using \HF API we will load the PyTorch model by default~\citep{HFDoc}.
Recent studies confirm that PyTorch is the most commonly used framework on Hugging Face, followed by TensorFlow and JAX~\citep{castano2023analyzingEvolandMaintenanceofMLModelsonHF}. According to the PeaTMOSS dataset~\citep{Jiang2024peatmoss}, 96\% of models use frameworks such as PyTorch, TensorFlow, or SafeTensor --- all primarily designed for DNN development. This estimate is based on a representative sample with a 99\% confidence level and a 5\% margin of error.
\revision{Our pipeline supports both dynamic graphs (\eg PyTorch), and static graphs (\eg TensorFlow, ONNX). }

\myparagraph{\revision{General Approach}}
During graph conversion, our method is to visit each node in the model architecture and traverse the edges.
\revision{The detailed algorithm is shown in Algorithm \ref{alg:sorting}.}
When converting models with cycles like \textit{Gated Recurrent Neural Networks}~\citep{chung2014GRU}, we skip tracing upstream or backward edges after their initial visit. This approach turns every network into a \textit{Directed Acyclic Graph} (DAG) with complete reachability, forming a \textbf{rooted DAG} in which each node lies on a path from an input to an output.
This approach allows us to maintain the integrity and functionality of the graph while ensuring all nodes are reached efficiently.

\myparagraph{\revision{For Dynamic Graphs (PyTorch)}}
Due to PyTorch's dynamic structure, we introduce \emph{dummy inputs} to the models to construct and trace the computational graphs effectively~\citep{PyTorchGraphs2021}. 
We refined our automated pipeline by iteratively adjusting \emph{dummy inputs}: if a conversion failed due to the input, we incorporated a new input type into our input list. Consequently, our finalized pipeline employs fixed language inputs while utilizing random tensors for models in the vision and audio domains.
The associated threat is discussed in \cref{sec: threats}.

\myparagraph{\revision{For Static Graphs (\eg TensorFlow, ONNX)}} 
\revision{We load each framework’s serialized model file and parse its graph definition directly via the native API (\eg \code{tf.GraphDef} for TensorFlow, \code{onnx.ModelProto} for ONNX). Since these formats explicitly enumerate nodes and their inputs, we extract the node list and adjacency information without runtime tracing. Control‑flow constructs (loops, conditionals) are unfolded where supported, and unsupported ops are logged and omitted. This direct graph parsing yields a complete DAG representation ready for downstream analysis.
}




\begin{algorithm}[htbp]
\captionsetup{width=\textwidth}
\caption{\small{Deterministic layer sorting for model serialization in DARA. This algorithm computes hash values for each layer in a DNN based on its structure and dependencies, enabling a consistent serialization order. Determinism ensures reproducibility and meaningful comparison across models.}}
\label{alg:sorting}
\small
\KwIn{Pre-trained weights of a DNN model}
\KwOut{List of layers sorted deterministically}

\SetKwFunction{FTraverse}{traverse}
\SetKwProg{Fn}{Function}{:}{}

\Visited \AlgGets \EmptySet

model \AlgGets \texttt{load\_model}(weights)

\Fn{\FTraverse{layer}}{ \tcp{Recursively traverse the model graph to assign structural hash values to each layer}
    \If{layer $\in$ \Visited}{
        \Return
    }
    \Visited $\leftarrow \Visited \cup \{layer\}$

    \eIf{layer.connections is empty}{
        layer.hash $\leftarrow$ hash(layer)
    }{
        \ForEach{next\_layer $\in$ sort(layer.connections)}{
            \FTraverse{next\_layer}
        }
        layer.hash $\leftarrow$ hash(layer, sum(next\_layer.hashes))
    }
}

\ForEach{input\_layer $\in$ model}{
    \FTraverse{input\_layer}
}

\SortedLayers $\leftarrow$ sort\_by\_hash(model.layers)

\Return{\SortedLayers}
\end{algorithm}

\subsubsection{Abstract Architecture}
\label{sec:AbstractArch}
\revision{To enable consistent model representation and facilitate downstream analysis such as feature extraction and reuse, we introduce the concept of an ``\textit{abstract architecture}''.}
Prior work proposed \emph{abstract neural network} (ANN)~\citep{Nguyen2022Manas}.
We propose a similar concept, \emph{abstract PTM architecture} (APTM).
This new concept not only includes information about layers, parameters, and input/output dimensions (as ANN did), but also incorporates detailed information about the connections between nodes, offering a fuller depiction of the model's architecture. 
This enhancement is pivotal for our n-gram feature selection method.
As part of our evaluation, we ablate to show the value of the APTM representation on this problem.

\subsubsection{Feature Extraction Methods}
\label{DARA:FeatureExtraction}
\review{Reviewer 2-Concern 4}{Regarding the choice of models, the authors selected RoBERTa, despite the availability of other transformers with better accuracy. Similarly, they chose Convolutional Neural Network (CNN) classifiers for anomaly detection. The reasoning behind these choices needs further elaboration.}
\revision{To extract features from the abstract architecture (\ie the edge connectivity in model graphs),} we explore two types of features extraction in this work: (1) basic \textit{n-gram} feature extraction, and (2) advanced feature extraction utilizing pre-trained \textit{transformers and contrastive learning}. \revision{Both approaches are typical representation learning approaches~\citep{bengio2013representationLearning}.}

\revision{We first experimented with using the raw architecture graph directly. This motivated our use of n-gram features, which efficiently capture frequently occurring local patterns in PTM architectures.}

\revision{Next, we linearized the architecture graphs by applying traversal and topological sorting, producing consistent and semantically meaningful sequences of layer names. 
To evaluate how well these sequences preserve architectural semantics, we considered multiple classifier architectures. Specifically, we experimented with CNNs and transformer-based models. These linearized sequences can be processed analogously to natural language, where CNNs are effective at capturing \ul{local} architectural patterns~\citep{lecun2015deepLearning}, while transformers model \ul{global} dependencies across the sequence~\citep{vaswani2017attention}.}

\revision{While many powerful transformer variants exist, our goal in this study was to assess the effectiveness of different feature extraction paradigms -- rather than exhaustively benchmark all possible model backbones. }


\paragraph{\ul{(1) Basic Feature Extraction: N-gram}} \label{sec:Vectorization}
We propose a method for vectorizing PTM architectural information using two fundamental components that define their architecture: layer connections,
and layer parameters. 
These aspects include the structural configuration of each PTM, such as pairs of connected layers (\eg \code{(Conv2D, BatchNormalization)}), and specific layer attributes (\eg \code{<kernel\_size: [1, 1]>} in a Conv2D layer).
Following prior works~\citep{ali2020malgra, terdchanakul2017bugReportClsusingNgram, sabor2017durfex}, we apply an N-gram feature extraction method to convert the architectural characteristics of each DNN into vector form.
Specifically, we employ 2-grams for analyzing layer connections and 1-grams for detailing layer parameter information. 
Our goal is to find a good unsupervised feature for PTM architectures accurately and efficiently~\citep{zhu2015unsupervised}.
We also experimented with 3-gram methods for feature extraction; however, we found that the process was notably slow, and the resulting features were overly sparse and high-dimensional, suggesting this method is unsuitable for extracting features from PTM architectures.
Subsequently, we apply padding to standardize the length of features across all PTMs.
This innovative approach provides a structured and detailed basis for analyzing and comparing PTM architectures.
\revision{An example of the extracted n-gram features is shown in Listing \ref{lst:ngram-feature}. These features are then converted into a one-hot encoded vector over all observed connection types. 
A visualization of the resulting feature representations across our sampled data is provided in \cref{fig:DARAFeatureVisualization}.}

We chose n-gram graph representation over Graph Neural Networks (GNNs)~\citep{wu2020comprehensiveSurveyonGNN} for three primary reasons:
\begin{itemize}
    \item \textit{Simplicity}: Our n-gram graph method is equivalent to a simple graph neural network~\citep{ma2022GNN}. Initially, we pursued this simpler approach to test our hypothesis, and it performed well enough that we decided to report it to the broader community. The n-gram approach yielded informative results for different task types, guiding future research directions. N-grams effectively represent sequential text data, which translates well to capturing the architectural information of PTM packages in our context. Our method mirrors approaches used in other domains, such as molecular data analysis~\citep{liu2019ngram} and text classification over high-frequency data streams~\citep{violos2018textclassificationUsingNgram}. In these instances, n-grams offer a compact representation of graphs equivalent to a basic GNN but without the necessity for extensive training. Additionally, as reported by~\citet{liu2019ngram}, the representation construction time using the n-gram method is faster than that of GNNs, significantly reducing training times for our DARA implementation.

\item \textit{Computational complexity}: We considered using GNNs for our analysis. While GNNs can manage moderate graphs with reasonable node and edge counts, their computational efficiency decreases as the graph size increases, especially with models like LLMs that have thousands of nodes and edges~\citep{ma2022GNN}. 
For instance, the \code{Mistral-7B-v0.3} model\footnote{See \url{https://huggingface.co/mistralai/Mistral-7B-v0.3}.} has 1577 nodes and 1864 edges.
This model, the smallest among the Mistral family at time of writing, illustrates the complexity that would burden GNNs at scale. 
Given the size of our dataset, we believe that the GNN method would introduce more computational overhead. Since one use case for DARA is as a check at PTM upload time, we think that registries such as Hugging Face would prefer using an automated tool for verification of the uploaded artifact. This is similar to the malicious package detection on NPM~\citep{sejfia2022practical} or automatically measuring performance goals in other registries such as the Android app store~\citep{wu2017appcheck}.

\item \textit{Feature dimension}: GNNs primarily extract features from the graph structure itself, making it difficult to incorporate specific layer parameters like kernel size in convolutional layers and dropout rate in dropout layers. In contrast, n-gram features can capture both layer connection information and detailed layer parameter information. These parameters are essential architectural details that we aim to consider in our analysis.
\end{itemize}

\begin{lstlisting}[style=json, 
caption={\revision{Example of extracted n-gram features from the Hugging Face model \texttt{Salesforce/blip2-opt-6.7b-coco}. The ``l'' field encodes the frequency of consecutive layer-type pairs (\eg (Linear, ReLU)), while the ``p'' field captures parameter-level patterns within individual layers. This representation is then converted into a one-hot vector and used as input for our downstream anomaly detection analysis.}}, 
label={lst:ngram-feature}]

{
  "l": {
    "([INPUT], to)": 2,
    "(LayerNorm, Linear)": 259,
    "(Linear, ReLU)": 32,
    "(to, Conv2d)": 1,
    "(Conv2d, flatten)": 1,
    "...": "..."
  },
  "p": {
    "...": ...,
    "Linear ['<in_features, 4096>', '<out_features, 4096>']": 128,
    "LayerNorm ['<normalized_shape, (4096,)>', '<eps, 1e-05>', '<elementwise_affine, True>']": 65,
    "Conv2d ['<in_channels, 3>', '<out_channels, 1408>', '<kernel_size, (14, 14)>', '<stride, (14, 14)>', '<padding, (0, 0)>', '<dilation, (1, 1)>', '<transposed, False>', '<output_padding, (0, 0)>', '<groups, 1>', '<padding_mode, zeros>']": 1,
    "...": ...
  }
}
\end{lstlisting}

\paragraph{\ul{(2) Advanced Feature Extraction: CNN, Transformers, and Contrastive Learning}}
\label{sec:DARA-OtherFeatureExtraction}
%
%
To achieve better representation of model architectures, we explored various feature extraction methods, including transformer-based backbones and contrastive learning approaches, to replace the N-gram approach. Our process of computational graph conversion and abstract architecture remain the same in this advanced approach. 

Transformers excel at capturing complex dependencies in sequences using attention mechanisms, enabling them to model intricate relationships between model layers~\citep{vaswani2017attention}. Contrastive learning, on the other hand, is effective for learning discriminative representations by maximizing the similarity between positive pairs and minimizing it between negative pairs~\citep{jain2021contrastive, saieva2023contrastive, huang2023cigar}.
In our approach, we represent model architectures as text sequences composed of their layer names. These sequences are then processed by transformers, while contrastive learning is applied to capture meaningful representations of the architectures.

\paragraph{Transformer-based Feature Extraction}
Since APTM is already sorted by Algorithm \ref{alg:sorting}, we can directly extract a sequence of ordered layer names from each model, as illustrated in Listing \ref{lst:advanced-feature}.
Transformer-based models are highly effective at modeling the complex, contextual relationships within sequences, making them well-suited for processing tokenized PTM layer names in such format. Following prior work~\citep{huang2023cigar}, we chose \code{RoBERTa} due to its efficiency and adaptability~\citep{liu2019roberta}. RoBERTa’s deep contextual embeddings enable our model to capture subtle differences in PTM layer naming conventions, which is essential for downstream classification tasks.

However, RoBERTa has a maximum input token limit of 512, restricting the extracted features to only certain parts of the PTM package architecture. To overcome this, we experimented with the \code{Longformer} model, which supports a maximum token length of 4096~\citep{beltagy2020longformer}. By increasing the token limit, Longformer eliminates the need for input trimming, preserving more crucial layer information. This approach proved more effective than RoBERTa, as it enhanced performance by retaining essential details that would otherwise be lost due to token truncation.


\begin{lstlisting}[style=json, 
caption={\revision{Example of our advanced sequential features and associated architectural metadata extracted from the Hugging Face model \texttt{Salesforce/blip2-opt-6.7b-coco}. 
The listing shows a linearized sequence of layer names, which we use to represent architectural structure in a format suitable for downstream classification.}}, 
label={lst:advanced-feature}]
{
  "Salesforce/blip2-opt-6.7b-coco": {
    "layers": "Input to Conv2d flatten transpose cat add LayerNorm Linear reshape permute __getitem__ matmul mul softmax Dropout matmul permute reshape Linear add ... Output Input Embedding to",
    "model_type": "blip-2",
    "arch": "Blip2ForConditionalGeneration",
    "task": [
      "text2text-generation", 
      "image-to-text", 
      "visual-question-answering"
    ]
  }
}
\end{lstlisting}

\paragraph{Contrastive Learning}
Our approach leverages Contrastive Learning (CL) to group together PTMs with the same labels and to separate those with different labels in the embedding space.
To construct features for the CL pipeline, we extract and tokenized the names of individual layers from each PTM. Preprocessing begins by separating layer names with whitespace, followed by tokenization.

The core of our CL feature selection strategy is built on transformer-based encoders, augmented with task-specific linear classification head to predict model type, task, and architecture. We extract the final hidden representation of the special \code{[CLS]} token from each sequence and apply $l_2$ normalization to obtain the embedding used in the loss computation.


\revision{To leverage the architectural metadata in PTM packages, we adopt Supervised Contrastive Learning (SupCon) loss~\citep{khosla2021supervisedcontrastivelearning}, which extends traditional contrastive learning by leveraging label supervision to pull together examples from the same class while pushing apart those from different classes. This approach allows each anchor to be contrasted against multiple positive samples, in addition to a diverse set of negatives, leading to more discriminative and robust representations.}

\revision{
\begin{align}
    \mathcal{L}_{\text{SupCon}} &= \sum_{i=1}^{N} \frac{-1}{|P(i)|} \sum_{p \in P(i)} 
    \log \left( \frac{\exp\left(z_i \cdot z_p / \tau \right)}{
    \sum_{a \in A(i)} \exp\left(z_i \cdot z_a / \tau \right)} \right)
    \label{eq:CLloss}
\end{align}
}


\noindent
\revision{Here, $N$ is the batch size, with \( \left\{ x_i, y_i \right\}_{i=1}^N \) denoting input-label pairs. The embedding $z_i \in \mathbb{R}^d$ is the $l_2$ normalized final hidden representation of the special \code{[CLS]} token from the encoder. For each anchor $z_i$, $P_i \subseteq A(i)$ denotes the set of positive indices, where $P(i)= \left\{ p\in A(i): y_p = y_i \right\}$, representing the indices of all positive samples in the batch that share the same sample as $y_i$ (excluding $y_i$ itself). $A(i) = \{1, \dots, N\} \setminus \{i\} $ is the set of all other samples in the batch. $\tau$ is the temperature, a hyperparameter used to scale the logits, and can be tuned to scale the penalties on negative samples.}

\revision{For \code{task} metadata, some models can have multiple tasks. Therefore, we consider it as a multi-label classification problem for detecting \code{task} inconsistencies. 
We employ Multi-Label Supervised Contrastive (MultiSupCon) loss~\citep{zaigrajew2022contrastive}, which relaxes the constraint of positive instances by treating samples as positives if they share any overlapping label. The set of positive indices are defined as $P'(i)= \left\{ p\in A(i): s_{i,p} \geq c \right\}$, where $c \in [0,1]$ is a hyperparameter representing the threshold value that indicates how similar sets of labels should be considered positive. MultiSupCon is defined as}

\revision{ 
\begin{align}
    \mathcal{L}_{\text{MultiSupCon}} &= \sum_{i=1}^{N} \frac{-1}{|P'(i)|} \sum_{p \in P'(i)} 
    s_{i,p} \cdot \log \left( \frac{\exp\left(z_i \cdot z_p / \tau \right)}{
    \sum_{a \in A(i)} \exp\left(z_i \cdot z_a / \tau \right)} \right)
    \label{eq:CLloss}
\end{align}
}
\noindent
\revision{where $s_{i,p}$ represents the weight in the assimilation value. This value indicates how similar two sets of ground truth labels are, with the value of 0 meaning zero overlap and 1 meaning perfect match. }

\revision{To train the model, we adopt a joint training strategy that combines CL with cross-entropy (CE) loss. 
The CL component encourages semantically similar samples to cluster in the embedding space, while the CE component optimizes for label accuracy~\citep{gunel2020supervised}:
For \code{model type} and \code{architecture} metadata inconsistency detection, we use SupCon loss ($\mathcal{L}_{SupCon}$) and standard cross-entropy loss ($\mathcal{L}_{CE}$):
\begin{equation}
    \mathcal{L} = \lambda \cdot \mathcal{L}_{SupCon} + (1-\lambda) \cdot \mathcal{L}_{CE}
\end{equation}
}

\noindent
\revision{For \code{task} metadata inconsistency detection, we apply MultiSupCon loss ($\mathcal{L}_{MultiSupCon}$) and binary cross-entropy loss ($\mathcal{L}_{BCE}$):
\begin{equation}
    \mathcal{L} = \lambda \cdot \mathcal{L}_{MultiSupCon} + (1-\lambda) \cdot \mathcal{L}_{BCE}
\end{equation}}

\subsection{Evaluation}
\label{sec:DARA-Eval}


In this section, we demonstrate the \emph{effectiveness} of our proposed anomaly detection method (DARA).
We note that this is the first known approach to detect PTM naming inconsistencies so there is no baseline for comparison.
Our evaluation focuses on three key naming labels: \code{model\_type}, \code{task} and \code{architecture} (\cref{sec:background-PTMNaming}). 

\subsubsection{Evaluation Dataset}
We evaluate DARA on sampled data from the PeaTMOSS dataset by calculating the accuracy of the prediction of the trained classifier~\citep{Jiang2024peatmoss}.
There are totally 219 architectures which have over 20 PTM instances. For those with more than 20 downloads, we randomly selected models to match 30 PTM instances. If fewer than 30 instances were available, we randomly sampled models with fewer than 20 downloads.
During feature extraction, we were unable to load the PTMs from four architectures, \eg \code{Bloom}~\citep{le2022bloom}, because their model sizes are too large to be loaded.
We were also unable to load another 2212 PTMs across these categories because either our dummy inputs could not be used on those models, missing configuration files or their architecture included customized code which we cannot fully trust.

\revision{For \code{model\_types} and \code{architectures} labels, we utilized metadata available in the \code{config.json} file. 
For \code{task} label, 
we identify supported tasks for each PTM by parsing the task tags listed in the model card metadata.
Any data entries without task tags were excluded from the analysis. 
}

\review{Reviewer 3-Concern 14}{Was the dataset split across architecture, task, and model type? Table 11 suggests something like this. If yes, how was the 5-fold cross-validation performed? How was the balance between different subsets established during training? The details are vague.
}
\revision{
Overall, we collected \DARAPTMNum PTM packages from \DARAArchNum \code{architectures}, \DARAModelTypeNum \code{model\_types}, and \DARATaskNum \code{tasks}.
We implemented an 80-20 split for training and evaluation purposes~\citep{geron2019hands}. Each model in the dataset has its corresponding APTM and architectural metadata as labels (\ie model\_type, architecture, task). To ensure robust validation, we employed 5-fold cross-validation on the shuffled dataset.
}

\subsubsection{Evaluation Design}

\review{Reviewer 2-Concern 4}{Regarding the choice of models, the authors selected RoBERTa, despite the availability of other transformers with better accuracy. Similarly, they chose Convolutional Neural Network (CNN) classifiers for anomaly detection. The reasoning behind these choices needs further elaboration.}
\revision{
\cref{tab:DARA-Eval-Design} summarizes the model variants evaluated in our study.
When designing DARA's pipeline, we systematically evaluated the contributions of each key component, including feature extraction methods, model architectures, and training strategies.
We started with a simple classifier using n-gram features, and then incrementally modified individual components to assess their impact.
We summarized all ablation experiments below:
\begin{enumerate}
    \item \textbf{N-gram + MLP Classifier:} Training a simple FCN classifier with N-gram features.
    The classifier consists of four fully connected layers. We systematically implemented a grid search to optimize the hyperparameters for DARA. 
    The final training parameters include 50 epochs, a learning rate of $1e-3$, Cross Entropy loss, and a step-wise LR scheduler. Batch sizes are 256 for training and 32 for evaluation. ``\code{l}'' denotes layer connections; ``\code{p}'' denotes layer parameters. For our ablation study, we evaluated two configurations: \code{l+p} and \code{l} only.
    \item \textbf{CNN:} Applying a Convolutional Neural Network to capture local patterns in the sequential model architecture. This setup treats the architecture as a 1D sequence and extracts spatially localized features using convolutional filters. 
    Training hyper-parameters are identical to the N-gram approach.
    \item \textbf{Transformers (Vanilla Finetuning):} Fine-tuning using cross-entropy loss alone, serving as a strong baseline for classification. The model is trained for a maximum of 50 epochs, a learning rate of 5e-5, Cross Entropy loss, step-wise LR scheduler, and early stopping with a patience of 5 epochs. For RoBERTa, batch sizes are 256 for training and 32 for evaluation. For Longformer, batch sizes are 8 for training.
    \item \textbf{Transformers (Continued Pretraining):} Continued pre-training of transformer models on our data to adapt general representations to solve the model architecture problem. We continue pre-training using a masked language modeling objective. A pre-trained masked language model is further trained on our sequential PTM layer names using 15\% token masking (\citep{devlin2019bert, liu2019roberta}). The model is trained with vanilla fine-tuning with the same training hyper-parameters.
    \item \textbf{Transformers (Finetuning w/ CL):} Fine-tuning using contrastive learning combined with cross-entropy loss to enhance representation quality. Training parameters are identical to vanilla finetuning setup. We performed a grid search over 8 combinations of the temperature parameter $\tau \in \{0.1, 0.3,0.5,0.7\}$ and $\lambda \in \left\{ 0.1, 0.3\right\}$. The best performance was observed with $\tau = 0.1$ and $\lambda = 0.1$, which we used for all \code{model\_type}, \code{task}, \code{architecture} labels.
\end{enumerate}  
}


\begin{table}[h]
\centering
\caption{\revision{Summary of DARA model variants and the rationale for considering each.}}
\label{tab:DARA-Eval-Design}
\begin{tabular}{p{4cm} p{7.5cm}}
\toprule
\textbf{Model Variant} & \textbf{Design Rationale} \\
\midrule
\textbf{N-gram + MLP Classifier~\citep{wang2012baselines, zhang2015character}} & Serves as a simple baseline using hand-crafted n-gram features (\code{l}, \code{p}) with a multi-layer perceptron. Fast, interpretable, and useful for benchmarking. \\
\\
\textbf{CNN~\citep{kim2014cnn}} & Applies 1D convolution over the architecture sequence to capture localized patterns in model structure. Effective for modeling short-range dependencies. \\
\\
\textbf{Transformers (Vanilla Finetuning)~\citep{devlin2019bert, liu2019roberta}} & Fine-tunes pre-trained transformer models (\eg RoBERTa, Longformer) using cross-entropy loss. Provides strong general-purpose representations for classification tasks. \\
\\
\textbf{Transformers (Continued Pretraining)~\citep{gururangan2020dontstoppretraining}} & Continues masked language model pretraining on PTM sequences to better align transformer representations with domain-specific patterns. \\
\\
\textbf{Transformers (Finetuning w/ CL)~\citep{khosla2021supervisedcontrastivelearning, zaigrajew2022contrastive, gunel2020supervised}} & Enhances feature discrimination by combining contrastive learning with cross-entropy loss. Improves representation quality through structured training objectives. \\
\bottomrule
\end{tabular}
\end{table}

\subsubsection{Anomaly Detection}
\review{Reviewer 3-Concern 16}{Given the current workflow and classification/CNN architecture, it is not clear how DARA will generate labels for a PTM during its registration. Looks like an unsubstantiated claim.}
\review{Reviewer 3-Concern 13}{What makes a label in the proposed method's training data? Is it the whole name of the PTM or one of the three components? Model type, task, and architecture -- are they considered as training features or the labels for classification? According to my understanding, the labels should be binary -- consistent or inconsistent, where the metadata/model feature and PTM names should make the features.}
\revision{As illustrated in \cref{Method:Tool}, \DNNDiff applies the MLP classification head to prediction on the \code{model\_type}, \code{architecture}, and \code{task} based on the features from the computational graph to make. detect anomalies in the PTM dataset by comparing predicted labels with the original PTM metadata. This strategy is motivated by its effectiveness in related tasks, such as time-series and log-based anomaly detection~\citep{ren2019timeSeriesAnomalyDetection, le2022logBasedAnomalyDetection}.}




\subsection{Results and Analysis} 
\label{sec:DARA-Results}

\subsubsection{Results}
We conducted all our training and evaluation using a single \code{A100-80GB} GPU.
\review{Reviewer 2-Concern 6}{The authors should report the false positive rate of anomalies reported by DARA}
\revision{\cref{tab:DARAEval-TypeArch} and \cref{tab:DARAEval-Task} present 5-fold cross-validation results for DARA on \code{model\_type}, \code{architecture}, and \code{task} metadata prediction.}


\revision{DARA achieves high accuracy in detecting inconsistencies in \code{model\_type}, with top models (\ie Longformer with CL) reaching up to 99.0\% accuracy. For \code{architecture}, DARA shows moderate performance, with the best transformer-based models achieving up to 72.0\% accuracy, notably outperforming n-gram and CNN approaches.}

\revision{In the \code{task} prediction task, DARA shows more modest accuracy. N-gram-based models outperform transformer-based ones on Top-1 predictions, achieving up to 56.5\% F1 and 72.3\% accuracy. This trend persists in Top-2 and Top-3 predictions, where N-gram models reach 61.0\% F1 at 82.1\% accuracy (Top-2) and 53.0\% F1 at 90.0\% accuracy (Top-3), outperforming transformer variants. These results suggest that shallow lexical features may better capture task-relevant patterns than deeper contextualized representations in this setting.}

\begin{landscape}
\begin{table}[t]
\centering
\caption{\revision{5-fold cross-validation results for predicting \code{model\_type} and \code{architecture} using DARA. Transformer-based approaches consistently outperform N-gram and CNN methods for both architectural metadata. Best performance is indicated in bold.}}
\small
\begin{tabular}{l|rrrr|rrrr}
\toprule
DARA type & \multicolumn{4}{c|}{model\_type} & \multicolumn{4}{c}{architecture}  \\
\cmidrule(lr){2-5} \cmidrule(lr){6-9} 
 & Recall/(\%) & Prec./(\%) & F1/(\%) & Acc./(\%) & Recall/(\%) & Prec./(\%) & F1/(\%) & Acc./(\%) \\
\midrule
N-gram + MLP Classifier (l only) & 96.2$\pm1.1$ & 96.4$\pm1.1$ & 96.1$\pm1.0$ & 95.7$\pm0.4$ & 60.9$\pm0.8$ & 56.1$\pm2.2$ & 55.4$\pm1.4$ & 62.2$\pm0.7$ \\
N-gram + MLP Classifier (l+p) & 97.6$\pm1.2$ & 97.6$\pm1.3$ & 97.5$\pm1.2$ & 98.8$\pm0.3$ & 62.8$\pm1.2$ & 59.8$\pm1.6$ & 58.2$\pm1.3$ & 63.7$\pm1.3$ \\
\midrule
CNN & 88.5$\pm1.5$ & 89.1$\pm2.0$ & 88.0$\pm1.8$ & 92.1$\pm1.1$ & 65.1$\pm2.6$ & 64.9$\pm2.4$ & 62.1$\pm2.4$ & 66.2$\pm1.4$ \\
\midrule
RoBERTa (Vanilla Finetuning) & 97.3$\pm2.0$ & 97.4$\pm1.8$ & 97.3$\pm1.9$ & \textbf{99.0$\pm$0.3} & \textbf{70.4$\pm$1.4} & 68.9$\pm2.0$ & 67.3$\pm1.4$ & 71.5$\pm0.6$ \\
Longformer (Vanilla Finetuning) & 97.8$\pm2.1$ & 97.8$\pm2.0$ & 97.7$\pm2.1$ & \textbf{99.0$\pm$0.3} & 70.3$\pm2.3$ & 69.2$\pm3.3$ & 67.4$\pm2.6$ & 71.6$\pm1.1$ \\
\midrule
RoBERTa (Continued Pretraining) & 97.4$\pm2.1$ & 97.4$\pm1.9$ & 97.3$\pm2.0$ & 98.8$\pm0.3$ & 69.7$\pm2.8$ & 67.3$\pm1.6$ & 66.3$\pm1.1$ & 71.1$\pm1.6$ \\
Longformer (Continued Pretraining) & 97.8$\pm$2.1 & 97.9$\pm$2.1 & 97.8$\pm$2.2 & 99.0$\pm$0.5 & 70.2$\pm1.6$ & 68.8$\pm1.7$ & 67.4$\pm$1.5 & \textbf{72.0$\pm$0.3} \\
\midrule
RoBERTa (Finetuning w/ CL) & 97.6$\pm1.9$ & 97.5$\pm2.2$ & 97.4$\pm2.0$ & 98.7$\pm0.3$ & 70.3$\pm1.3$ & 68.1$\pm1.9$ & 67.1$\pm1.5$ & 71.3$\pm0.6$ \\
Longformer (Finetuning w/ CL) & \textbf{98.0$\pm$1.7} & \textbf{98.1$\pm$1.6} & \textbf{97.9$\pm$1.7} & \textbf{99.0$\pm$0.3} & 70.2$\pm1.2$ & \textbf{69.2$\pm$2.2} & \textbf{67.5$\pm$1.5} & 71.9$\pm$1.3 \\
\bottomrule
\end{tabular}
\label{tab:DARAEval-TypeArch}
\end{table}

\vspace{1em}

\begin{table}[t]
\centering
\caption{\revision{5-fold cross-validation results for predicting \code{Task} metadata using DARA. N-gram-based approaches outperform transformer-based methods for task classification. Best performance is indicated in bold.}}
\small
\resizebox{0.99\linewidth}{!}{
\begin{tabular}{l|rrrr|rrrr|rrrr}
\toprule
DARA type & \multicolumn{12}{c}{task} \\
\cmidrule(lr){2-13}
 & \multicolumn{4}{c}{Top@1} & \multicolumn{4}{c}{Top@2} & \multicolumn{4}{c}{Top@3} \\
\cmidrule(lr){2-5} \cmidrule(lr){6-9} \cmidrule(lr){10-13}
 & Recall/(\%) & Prec./(\%) & F1/(\%) & Acc./(\%) & Recall/(\%) & Prec./(\%) & F1/(\%) & Acc./(\%) & Recall/(\%) & Prec. & F1 & Acc./(\%) \\
\midrule
N-gram + MLP Classifier (l only) & \textbf{56.2$\pm$0.5} & \textbf{59.2$\pm$1.2} & \textbf{56.5$\pm0.7$} & 72.3$\pm1.2$ & \textbf{77.1$\pm$1.3} & \textbf{53.6$\pm$0.22} & \textbf{61.0$\pm$2.1} & 82.1$\pm0.8$ & \textbf{85.3$\pm$1.3} & 42.9$\pm1.2$ & 53.0$\pm1.5$ & 90.0$\pm0.9$  \\
N-gram + MLP Classifier (l+p) & 55.6$\pm0.5$ & 57.7$\pm1.3$ & 55.9$\pm0.8$ & 71.8$\pm0.5$ & 76.0$\pm0.8$ & 51.0$\pm1.6$ & 58.8$\pm1.0$ & 82.9$\pm0.4$ & 85.7$\pm1.9$ & 40.5$\pm2.5$ & 51.3$\pm1.9$ & 90.5$\pm1.0$ \\
\midrule
CNN & 52.6$\pm2.6$ & 58.1$\pm3.5$ & 53.8$\pm2.6$ & 73.5$\pm1.3$ & 75.1$\pm4.1$ & 45.5$\pm3.1$ & 54.0$\pm3.1$ & 82.1$\pm1.7$ & 81.4$\pm3.3$ & 34.3$\pm2.7$ & 44.5$\pm2.3$ & 88.3$\pm1.0$ \\
\midrule
RoBERTa (Vanilla Finetuning) & 51.8$\pm0.8$ & 52.6$\pm1.7$ & 51.9$\pm0.9$ & 74.5$\pm1.3$ & 74.4$\pm2.5$ & 51.9$\pm2.9$ & 59.2$\pm2.6$ & 83.7$\pm1.0$ & 82.1$\pm1.9$ & 43.2$\pm3.1$ & 53.0$\pm2.6$ & 90.5$\pm1.0$ \\
Longformer (Vanilla Finetuning) & 54.4$\pm2.7$ & 58.0$\pm4.2$ & 55.1$\pm2.8$ & \textbf{76.3$\pm1.4$} & 75.1$\pm1.8$ & 52.2$\pm1.4$ & 59.4$\pm0.9$ & \textbf{84.5$\pm0.6$} & 83.5$\pm2.4$ & 43.3$\pm2.9$ & 53.6$\pm1.9$ & 90.6$\pm0.8$ \\
\midrule
RoBERTa (Continued Pretraining) & 52.0$\pm3.3$ & 53.2$\pm3.5$ & 52.2$\pm3.5$ & 75.2$\pm1.2$ & 74.7$\pm2.5$ & 51.9$\pm3.0$ & 58.9$\pm2.7$ & 84.2$\pm0.5$ & 82.9$\pm1.5$ & \textbf{44.0$\pm$3.1} & \textbf{53.7$\pm2.4$} & \textbf{90.8$\pm0.7$} \\
Longformer (Continued Pretraining) & 53.4$\pm2.5$ & 57.0$\pm2.7$ & 53.8$\pm2.3$ & 75.0$\pm4.5$ & 75.5$\pm3.5$ & 51.1$\pm1.9$ & 58.4$\pm2.3$ & 83.0$\pm4.2$ & 82.3$\pm4.5$ & 42.4$\pm2.4$ & 52.4$\pm2.5$ & 89.6$\pm3.0$ \\
\midrule
RoBERTa (Finetuning w/ CL) & 50.4$\pm2.9$ & 51.0$\pm3.3$ & 50.3$\pm3.3$ & 75.7$\pm2.6$ & 69.9$\pm4.7$ & 47.0$\pm1.0$ & 52.7$\pm2.5$ & 84.2$\pm1.6$ & 79.9$\pm3.6$ & 39.0$\pm1.4$ & 48.4$\pm1.9$ & 90.6$\pm1.4$ \\
Longformer (Finetuning w/ CL) & 45.8$\pm$5.7 & 46.8$\pm5.9$ & 45.4$\pm6.2$ & 68.9$\pm7.3$ & 65.3$\pm$8.8 & 39.9$\pm$5.9 & 46.8$\pm$6.7 & 80.2$\pm$5.4 & 75.9$\pm$7.9 & 33.7$\pm$3.3 & 42.7$\pm$5.0 & 88.0$\pm$5.1 \\
\bottomrule
\end{tabular}
}
\label{tab:DARAEval-Task}
\end{table}
\end{landscape}



\begin{figure}[h]
    \centering
    \includegraphics[width=1.0\textwidth]{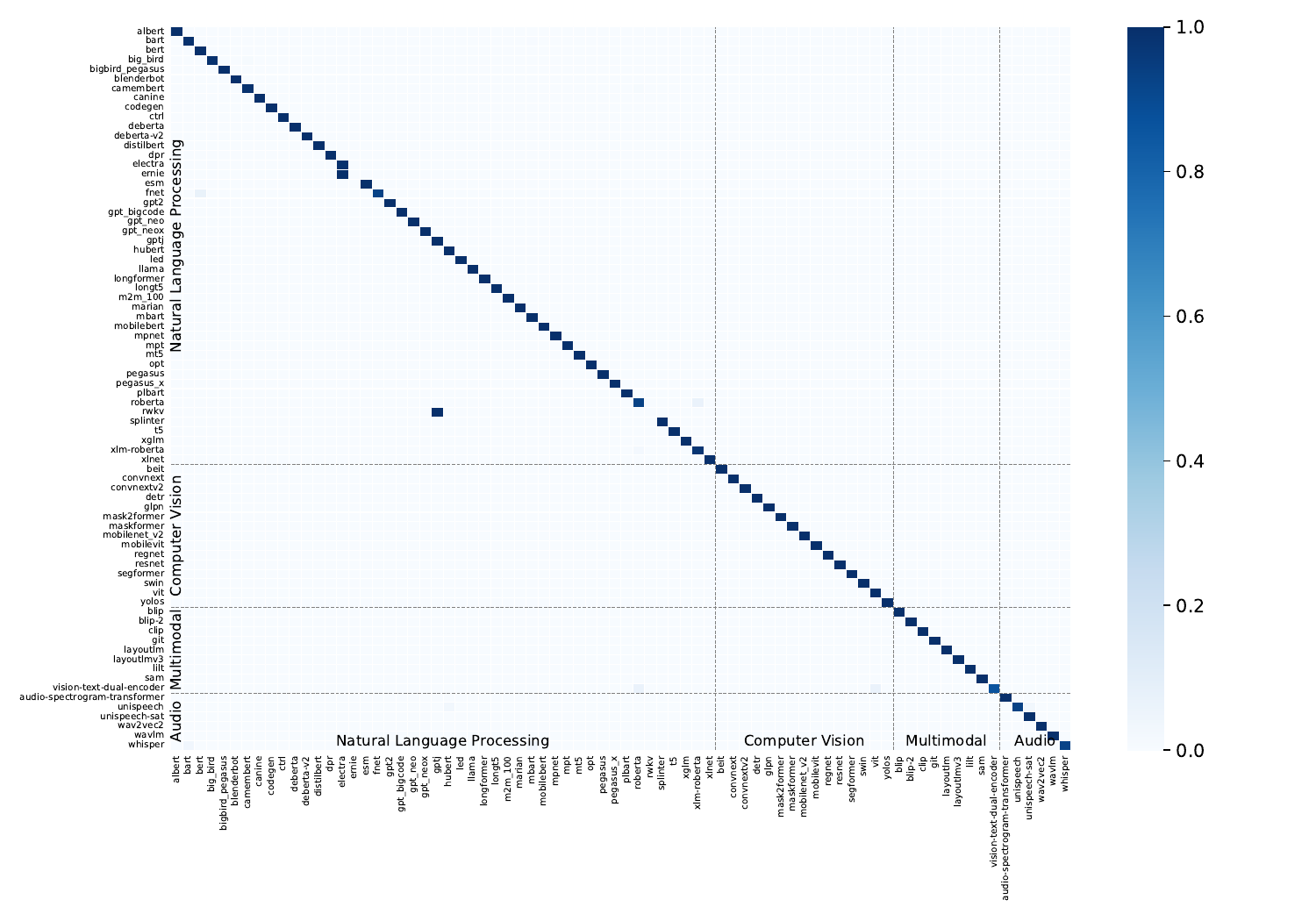}
    \caption{
    Confusion matrix categorized by \textbf{model\_type}.
    As seen along the diagonal of the confusion matrix, DARA accurately predicts most model types but encounters difficulties with audio models like \textit{unispeech} and \textit{wav2vec2}, which share the same architectural implementation and differ primarily in tokenization. A higher resolution version of this figure is available in the supplementary materials.
    }
    \label{fig:DARAEval-model_type}
\end{figure}

\begin{figure}[h]
    \centering
    \includegraphics[width=1.0\textwidth]{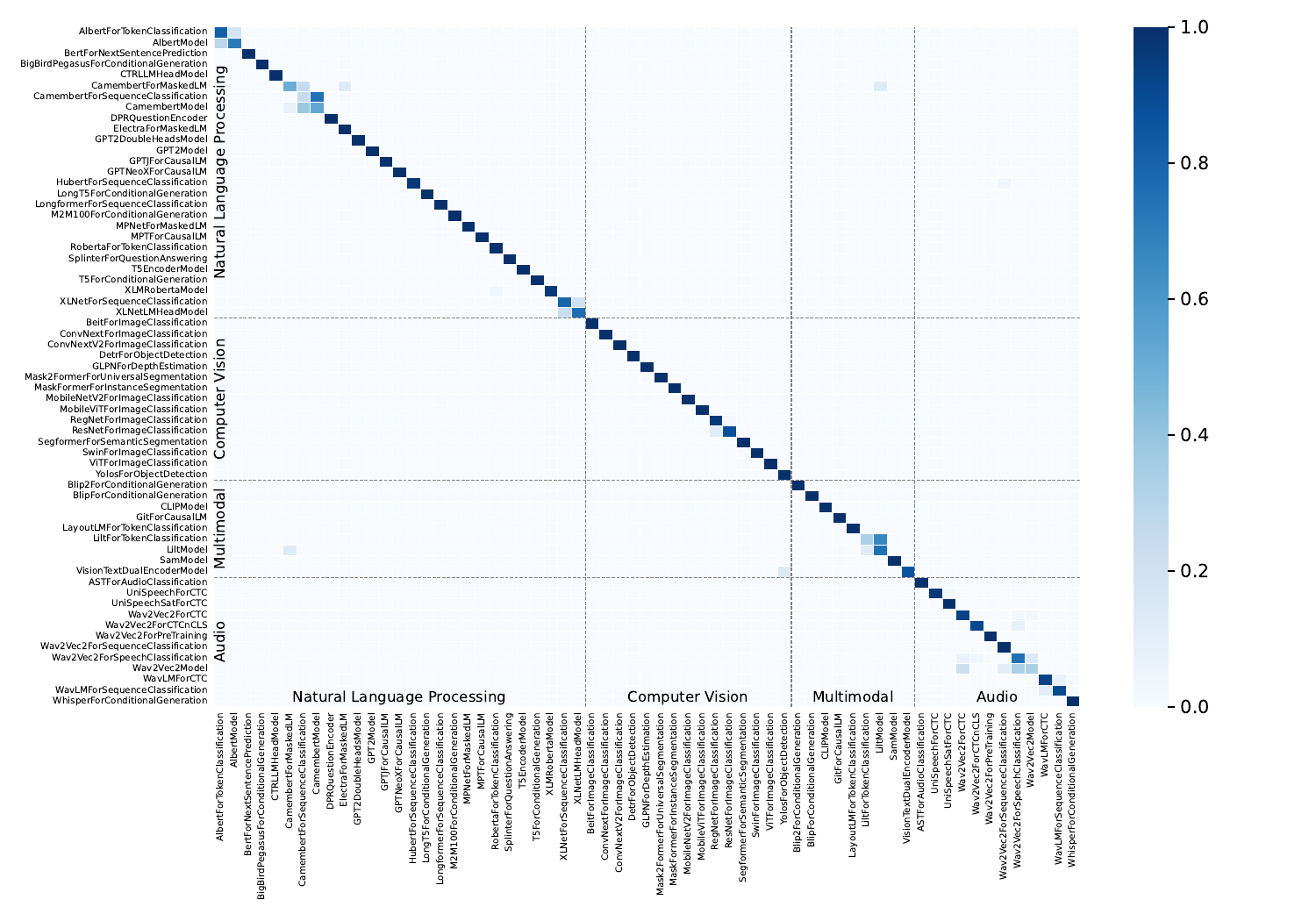}
    \caption{
    Confusion matrix categorized by \textbf{architecture}.
    As seen along the diagonal of the confusion matrix,
   DARA accurately predicts most architectures, particularly for computer vision models, but faces challenges with similar tasks such as \textit{QuestionAnswering} and \textit{SequenceClassification}, where the primary difference lies in the model heads.
   A higher resolution version of this figure is available in the supplementary materials.
    }
    \label{fig:DARAEval-architecture}
\end{figure}

\begin{figure}[h]
    \centering
    \includegraphics[width=1.0\textwidth]{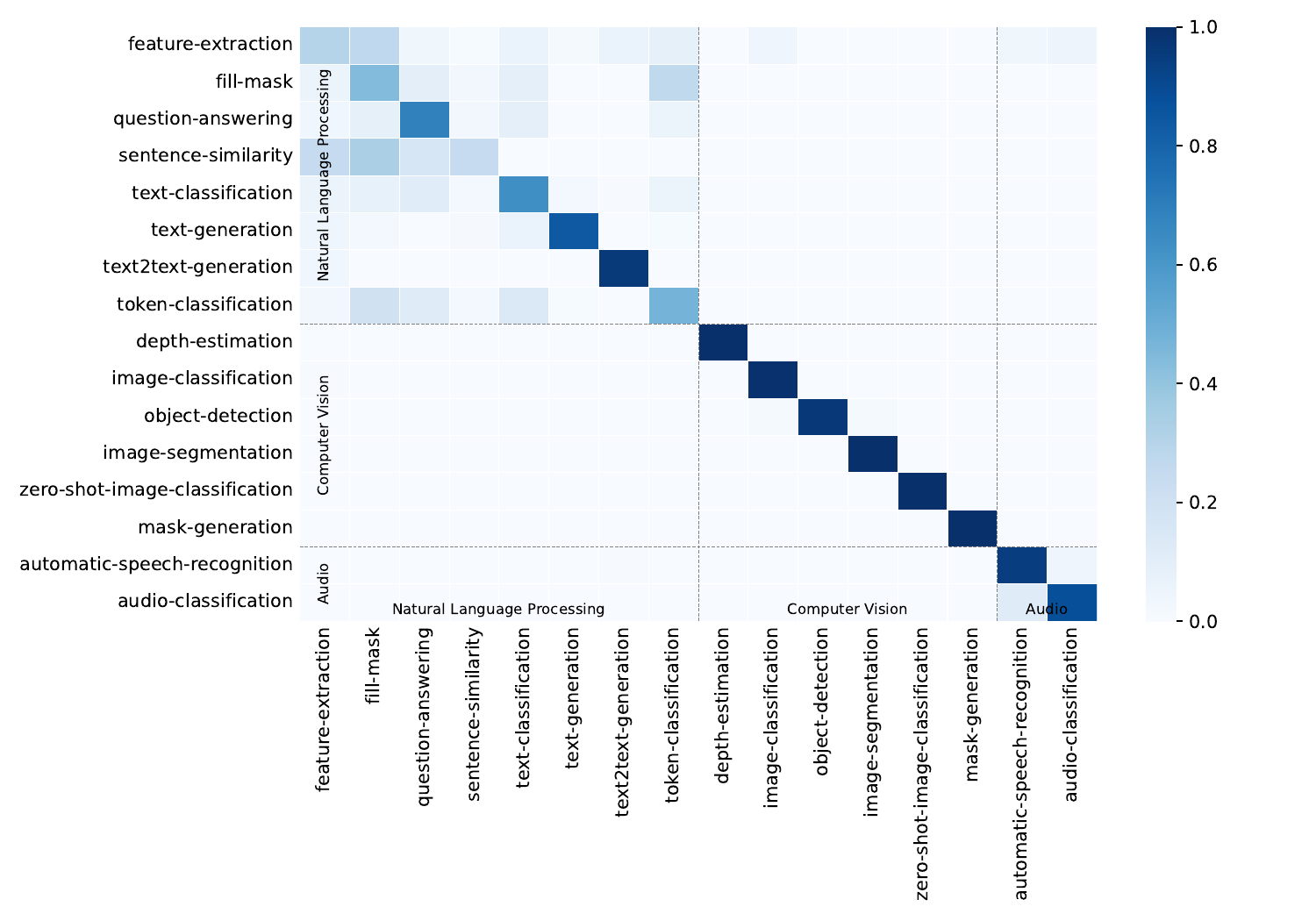}
    \caption{
    Confusion matrix categorized by \textbf{task}. As seen along the diagonal of the confusion matrix, DARA accurately predicts vision tasks (\eg depth estimation and image classification), but performs less effectively for audio and text models. \revision{For clarity and visualization purposes, this confusion matrix only includes models with a single designated task (\ie excluding multi-task models).}
    }
    \label{fig:DARAEval-task}
\end{figure}

\subsubsection{Analysis}

\myparagraph{\revision{Effectiveness of DARA}}
\cref{fig:DARAEval-model_type} shows the cases where DARA cannot predict correctly in \code{model\_type}.
In fact, these models share almost the same architectural designs and the only differences are either different training regimes, different datasets, or using different representations.
\revision{This indicates that while architectural features are sufficient for model\_type detection, they have limitations in distinguishing between architectures with similar structural patterns. To accurately differentiate between these architecturally similar but functionally distinct models, more sophisticated feature extraction approaches that capture training dynamics, data characteristics, and representational differences would be necessary.}

\cref{fig:DARAEval-architecture} shows the cases where DARA cannot predict correctly in \code{architectures.}
The same incorrect case arises for the four Audio models.
Furthermore, we can also find that DARA cannot correctly distinguish between different tasks as part of the \code{architecture}, such as \code{TokenClassification}, \code{QuestionAnswering}, and \code{SequenceClassification} in language models. 
Previous work indicates that the final layers of PTMs are highly task-specific~\citep{rogers2021primerinBERTology}. Therefore, due to the high similarity of PTMs' features across these tasks, with differences confined to the final layers, DARA struggled to effectively differentiate between these categories. 
However, the Longformer, with its ability to handle longer input sequences and capture more global architectural features, offers improved performance in predicting the correct model architectures.

\cref{fig:DARAEval-task} shows that DARA works well to predict the task label for computer vision tasks  (\eg \code{image-classification}, \code{image-segmentation}, and \code{object-detection}), while struggles to accurately predict the tasks of PTM packages, particularly in distinguishing between \code{audio-classification} and \code{automatic-speech-recognition}. This difficulty likely stems from the similarity in their architectural configurations, where the main difference lies in the head or output layer of each model. However, these tasks might be more distinguishable by incorporating the weight information of the PTM packages, which is beyond the scope of this work.

\review{Reviewer 3-Concern 15}{The authors suggested that 75\% accuracy might be sufficient to accept the engineering hypothesis. However, DARA achieves 50\% accuracy for task and architecture, and the authors already accepted the hypothesis. Thus, it looks like self-contradiction IMO.}
\revision{The evaluation results of DARA demonstrate varying effectiveness across different metadata categories. For model\_type classification, DARA achieves excellent performance with transformer-based approaches reaching up to 99.0\% accuracy, strongly supporting our hypothesis that architectural information alone can reliably detect naming inconsistencies in this category. However, for architecture classification, the best performance reaches only 72.0\% accuracy, and task classification performs even lower at approximately 76.3\% accuracy for top@1 predictions. While we initially suggested 75\% accuracy as sufficient for accepting the engineering hypothesis, our results indicate this threshold is met only for model\_type detection. For architecture and task classifications, the lower accuracy suggests architectural information alone provides useful but not definitive signals for naming inconsistency detection. This reveals a gradient of detectability across metadata types, with model structure being most reliably identified from architectural patterns, while more specific attributes like task require additional contextual information beyond what DARA currently extracts.
}

\begin{figure}
    \centering
    \includegraphics[width=0.99\linewidth]{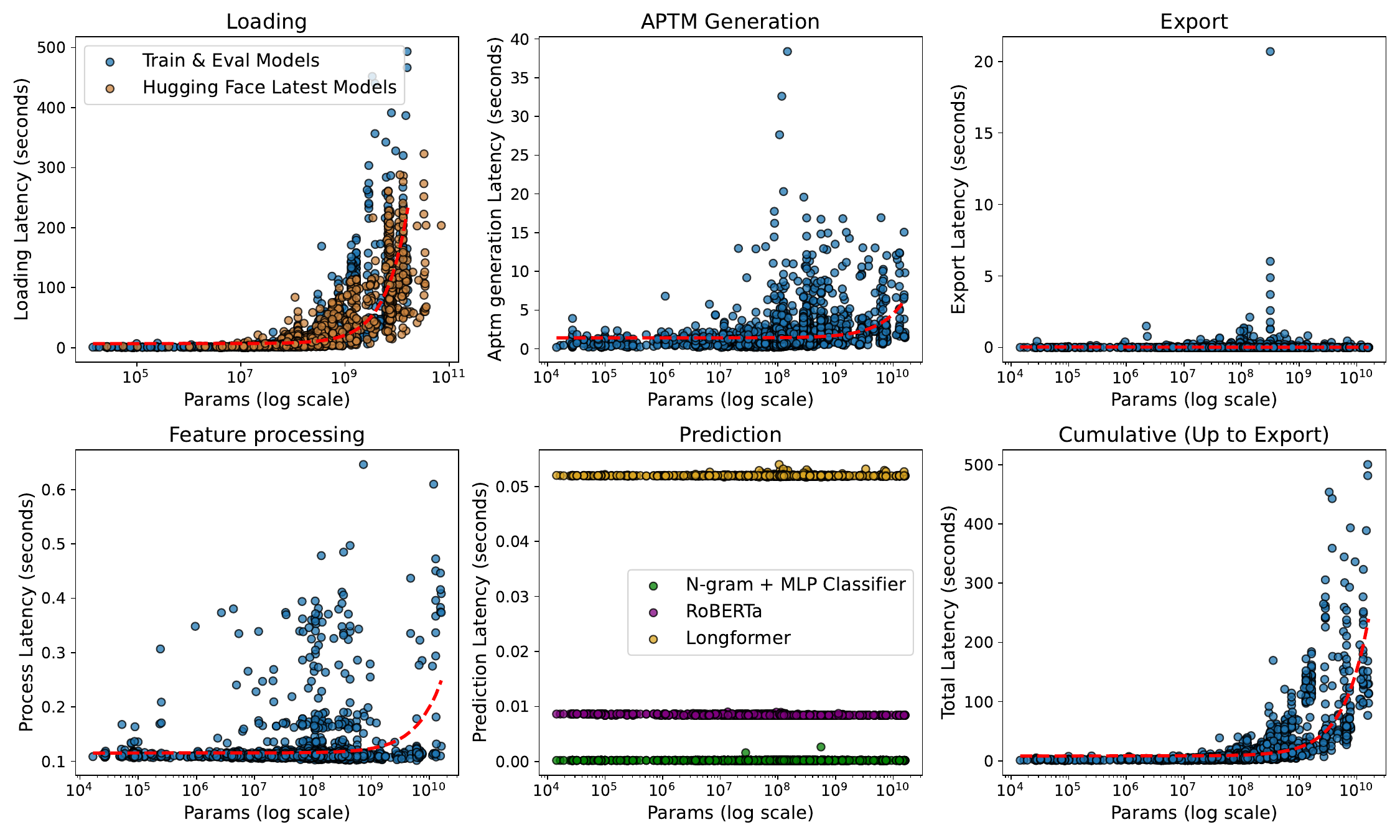}
    \caption{
   \revision{Latency measurements for each stage of \DNNDiff, including APTM extraction, feature export, feature processing, DARA prediction, model loading, and total execution time.
    \review{Reviewer 2-Concern 16}{How did the authors ensure that their findings are generalizable?}\review{Reviewer 3-Concern 6}{PeaTMOSS (Aug 2023) - One might argue that many recent PTMs are not considered for this work. Given the growth of PTM/LLM, one calendar year could be quite significant.}
    Red curves show fitted complexity trends based on our algorithmic analysis: quadratic for APTM extraction and feature processing; linear for feature export and prediction; and exponential for model loading.
    Orange markers indicate loading latencies of recent PTM/LLM models. 
    The close alignment of these points with our fitted distributions demonstrates that our results generalize to newer models.}
    }
    \label{fig:latency}
\end{figure}

\myparagraph{\revision{Complexity Analysis}}
\review{Reviewer 2-Concern 5}{What is the time DARA takes to detect naming inconsistencies?}
\review{Reviewer 3-Concern12}{Given the workflow of the proposed method, can we deliver the prediction during the upload of a PTM to HF? How long does it take to predict the misalignment? The current runtime behaviour of the proposed model was not reported to demonstrate its feasibility.}
\revision{We divide our time complexity analysis of DARA across five stages and provide both theoretical and empirical analysis. 
In our theoretical analysis, we use the following notation:
\begin{itemize}
    \item $n$: Number of learnable parameters in a given PTM.
    \item $V$: Number of layers in the PTM graph.
    \item $E$: Number of edges, representing connections between layers in the PTM graph.
\end{itemize}
We empirically analyze each stage by fitting observed latencies against theoretical complexity curves, reporting the best-fitting model with the residual sum of squares (RSS). The results are presented in \cref{fig:latency}.
All empirical measurements are conducted in terms of $n$, which correlates with and subsumes variations in $V$ and $E$.
\begin{enumerate}
    \item \textit{Model Loading:} 
    This stage is challenging to analyze, as it is influenced by both model architecture and Hugging Face's implementation of loading APIs.
    Empirically, the model‑loading phase grows faster, fitting an \(\mathrm{O}(n^{k})\) curve most closely (RSS = \(5.0091\)).
    \item \textit{APTM generation:} The APTM generation stage's complexity is based on the topological sort (Algorithm \ref{alg:sorting}), which is $\mathrm{O}(V+E)$ with $V$ the number of layers in a PTM graph and $E$ the edges between layers (\ie layer connections). However, the linear regression analysis yielded an $R^2$ near zero, indicating negligible correlation between latency and model properties such as parameter count, layer count, or edge count. We conclude that in practice, latency is dominated by constant-time overheads, rather than any traversal or size-dependent computation.
    Empirically, the best match for this stage is \(\mathrm{O}(\sqrt{n})\) (RSS = \(10.022\)), slightly outperforming a linear fit of $\mathrm{O}(n)$ (RSS = \(10.304\)), due to repeated dummy input calls and error handling.
    \item \textit{JSON export:} 
        JSON export mostly depends on the total number of items of features. The theoretical time complexity for layer connections is $\mathrm{O}(E)$, layer parameters is $\mathrm{O}(V)$. 
        Empirically, complexity follows $\mathrm{O}(n\log^2 n)$ (RSS = 1.2614), likely from bounded operation vocabularies and fixed vector structures.
    \item \textit{Feature Processing:} 
        Converting features to IDs using a lookup dictionary has a time complexity of $\mathrm{O}(V)$.
        Empirically, a linear fit $\mathrm{O}(n)$ (RSS = 13.1625) matches observed latency.
    \item \textit{Prediction:} Once features are extracted, prediction is a constant-time lookup and classification step, $\mathrm{O}(1)$. Empirically, this holds across all model variants, yielding minimal latency suitable for real-time use during PTM upload (\cref{tab:DARA-Latency}).
\end{enumerate}
\JD{I think ``entire pipeline'' is an odd term here since we are omitting two steps. Maybe pick another phrase. Figure 12 may need another phrase too since ``Total'' implies all stages?}
When considering all stages from model loading through export—which comprise the majority of the entire pipeline—the cumulative end-to-end complexity again aligns with \(\mathrm{O}(n^{k})\) (RSS = \(4.9973\)).
}

\myparagraph{\revision{Efficiency of DARA}}
\review{Reviewer 2-Concern 5}{The time DARA takes to detect naming inconsistencies}
\review{Reviewer 3-Concern 12}{Given the workflow of the proposed method, can we deliver the prediction during the upload of a PTM to HF? How long does it take to predict the misalignment? The current runtime behaviour of the proposed model was not reported to demonstrate its feasibility.}
\revision{\cref{fig:latency} and \cref{tab:DARA-Latency} present the latency and throughput measurements for different DARA variants. The N-gram + MLP classifiers exhibit the lowest latency at 0.19 ms per PTM and the highest throughput, exceeding 28,000 PTMs per second, making them highly suitable for real-time deployment. The CNN model also performs efficiently, processing over 17,000 PTMs per second with a latency of 0.56 ms.
Transformer-based models offer improved accuracy, particularly for \code{model\_type} and \code{architecture} classification, but incur significantly higher computational costs. RoBERTa variants show moderate latency (8.4 ms) and throughput (740–750 PTMs/s), while Longformer variants are substantially slower, with latency exceeding 51 ms and throughput dropping to as low as 61 PTMs/s.
These results confirm the computational feasibility of integrating DARA into the Hugging Face model upload workflow. Users would experience negligible delay when using N-gram or CNN-based variants, and acceptable latency with RoBERTa. The clear trade-off between accuracy and efficiency supports a tiered deployment strategy: fast models for broad coverage and real-time use, with more accurate transformer-based models reserved for high-stakes or ambiguous cases.
}

\begin{table}
    \centering
    \caption{\revision{Comparison of DARA types by latency and throughput. Latency was measured by averaging inference time across individual models. Throughput was measured independently using batched inputs to reflect real-world usage.
    The final column shows the overall F1 score (\textit{cf.} \cref{tab:DARAEval-TypeArch}, \cref{tab:DARAEval-Task}) 
    to allow comparison of the efficiency/accuracy tradeoff.
    N-gram approaches yield the lowest latency, while transformer-based methods incur higher latency due to greater model complexity but offer better F1 performance.
    }}
    \resizebox{0.99\linewidth}{!}{
\begin{tabular}{l|c|cc|ccc}
    \toprule
    \textbf{DARA Type} & \textbf{Variant / Training Setup} & \textbf{Latency} & \textbf{Throughput} & \multicolumn{3}{c}{\textbf{Macro F1 Score}} \\
    & & (ms/PTM) & (PTM/s) & \textbf{model\_type} & \textbf{task} & \textbf{architecture} \\
    \midrule
    \multirow{2}{*}{N-gram + MLP Classifier} 
    & layer connections & 0.19 & 29849.12 & 96.1$\pm$1.0 & \textbf{56.5$\pm$0.7} & 55.4$\pm$1.4 \\
    & layer connections + layer parameters & 0.19 & 28426.36 & 97.5$\pm$1.2 & 55.9$\pm$0.8 & 58.2$\pm$1.3 \\
    \midrule
    CNN & --- & 0.56 & 17127.99 & 88.0$\pm$1.8 & 53.8$\pm$2.6 & 62.1$\pm$2.4 \\
    \midrule
    \multirow{3}{*}{RoBERTa}
    & Vanilla Finetuning & 8.43 & 750.91 & 97.3$\pm$1.9 & 51.9$\pm$0.9 & 67.3$\pm$1.4 \\
    & Continued Pretraining & 8.34 & 740.11 & 97.3$\pm$2.0 & 52.2$\pm$3.5 & 66.3$\pm$1.1 \\
    & Finetuning w/ CL & 8.43 & 748.18 & 97.4$\pm$2.0 & 50.3$\pm$3.3 & 67.1$\pm$1.5 \\
    \midrule
    \multirow{3}{*}{Longformer}
    & Vanilla Finetuning & 51.60 & 64.96 & 97.7$\pm$2.1 & 55.1$\pm$2.8 & 67.4$\pm$2.6 \\
    & Continued Pretraining & 51.78 & 61.58 & 97.8$\pm$2.2 & 53.8$\pm$2.3 & 67.4$\pm$1.5 \\
    & Finetuning w/ CL & 51.90 & 64.58 & \textbf{97.9$\pm$1.7} & 45.4$\pm$6.2 & \textbf{67.5$\pm$1.5} \\
    \bottomrule
    \end{tabular}
    }
    \label{tab:DARA-Latency}
\end{table}

\myparagraph{\revision{Validating Engineering Hypothesis}}
\review{Reviewer 3-Concern 15}{The authors suggested that 75\% accuracy might be sufficient to accept the engineering hypothesis. However, DARA achieves 50\% accuracy for task and architecture, and the authors already accepted the hypothesis. Thus, it looks like self-contradiction IMO.}
\AUTHOR{We thank Reviewer 3 for highlighting this issue. We note that our updated results differ from the initially reported ones, and we now provide a detailed discussion clarifying to what extent and under what conditions our hypothesis about architectural features for metadata validation can be accepted.}
\revision{The evaluation results reveal important limitations in our initial hypothesis about architectural features being sufficient for metadata validation. While we initially proposed a 75\% accuracy threshold as acceptable for our engineering hypothesis, our results show a clear gradient of detectability across different metadata types. For model\_type detection, with accuracy reaching 99.0\%, the hypothesis is strongly supported. However, for architecture (72.0\%) and task classification (76.3\% top@1), the performance falls short of consistently meeting our target threshold. This indicates that architectural features alone provide valuable but incomplete signals for these more specific metadata categories. Rather than a binary acceptance or rejection of our hypothesis, these results suggest a more nuanced understanding: architectural patterns are highly reliable for broad model categorization but require supplementary information for more fine-grained classification tasks. This limitation highlights opportunities for future work in multimodal feature extraction that incorporates training dynamics and representational differences.}

\myparagraph{\revision{Generalizability of DARA}}
\review{Reviewer 2-Concern 16}{How did the authors ensure that their findings are generalizable?}\review{Reviewer 3-Concern 6}{PeaTMOSS (Aug 2023) - One might argue that many recent PTMs are not considered for this work. Given the growth of PTM/LLM, one calendar year could be quite significant.}
\revision{To analyze the generalizability of DARA to recent PTMs, we conducted a stratified random sampling across the Hugging Face repository. We identified all publicly available models with at least 100 downloads and categorized them into five domains: NLP, CV, audio, multimodal, and other. Using proportional allocation based on domain distribution, we extracted a stratified sample of 1,901 models from the full repository (N=1,609,131), achieving a 99\% confidence level with 3\% margin of error.
While our original dataset (\cref{sec:DARA-Eval}) contained models ranging from 12K to 16B parameters, this latest sampling expanded to models from 26K to 70B parameters. For each sampled model, we simulated loading for APTM generation. As shown in \cref{fig:latency}, the red curves represent fitted distributions for each plot. The close alignment between these newer models and our fitted distributions demonstrates that our performance results generalize effectively to more recent models.}

\section{Threats to Validity}
\label{sec: threats}


This section discusses three types of threats to validity~\citep{wohlin2012experimentation}.
Following the guidance of~\citet{verdecchia2023threats}, we emphasize substantive threats that could affect our findings.

\textbf{Construct threats} refer to potential limitations in how we operationalized the key concepts.
\textit{In the survey},
we operationalize the concept of PTM names in terms of the identifier and architecture. Alternative operationalizations might incorporate package type or package contents. However, per the registry naming taxonomy (\cref{fig:Naming}), we believe our operationalization covers the elements that involve judgment on the part of the namer. 

\textit{In the other parts}, the feature selection algorithm in DARA is designed based on an engineering hypothesis (\cref{sec:DARA-FeatureSelection}). We validated this hypothesis by achieving substantial performance, indicating that the underlying assumptions were correct.

\textbf{Internal threats} are factors that may compromise the validity of cause-and-effect relationships.
\textit{In the survey}, a key threat is potential bias in the qualitative analysis of the two open-ended text questions. To mitigate this, two researchers independently analyzed the data, and resolved discrepancies through discussion, ensuring consistency in interpretation (\cref{sec:Method-Survey-Analysis}). Another potential threat is the validity of our survey instrument. To address this, we conducted a pilot study with three actual Hugging Face users (\cref{sec:Method-Survey-Design}).

\textit{In the other parts}, we asked two researchers to manually label 200 naming conventions as ground truth in our measurement of naming conventions (\cref{sec:method-MeasurementEval} to evaluate our LLM pipeline. This process may introduce potential bias, but we measured inter-rater agreement, which indicated substantial consistency between the two researchers, mitigating this threat to some extent. Although DARA effectively detects metadata inconsistencies (\cref{sec:DARA-Results}), the algorithm's reliance on N-grams as features is a limitation, as it discards significant information about the PTM architecture and weights, reducing the depth of the analysis.

\textbf{External threats} are factors that may limit the generalizability of the findings.
%
\textit{In the survey}, our sample size of \SurveyNum achieves a 95\% confidence level with a margin of error of 10\%, based on the estimated number of Hugging Face users  (\cref{sec:Method-Survey-Recruitment}). However, we recognize that using Hugging Face does not require a user account, meaning there may be significantly more users. 
\review{Reviewer 2-Concern 3}{The survey was conducted using only 1.7\% of the Hugging Face dataset, with a low 5.4\% response rate predominantly from experienced users. This approach potentially introduces significant methodological biases, which could impact the validity of the findings. The authors should explain how they addressed these biases.
}
\review{Reviewer 2-Concern 16}{How did the authors ensure that their findings are generalizable?}
\revision{Furthermore, since most participants were experienced PTM contributors or users, our findings may reflect a degree of selection bias. Additionally, our mining study focused on commonly used PTM packages with over 50 monthly downloads, and the exclusion of less popular packages may also introduce bias.
These threats} could potentially affect the generalizability of our study's findings.
\revision{However, the particular bias that results --- towards experienced engineers and towards packages that people actually use --- strike us as desirable for a first study of PTM names. Future studies might investigate whether naming convention preferences and practices vary between experts and novices.}

\textit{In the other parts}, the dataset used for DARA evaluation is imbalanced, with a higher proportion of language and vision models and fewer multimodal and audio models. We recognize that this limitation may affect the generalizability of our evaluation. 
Additionally, we focused on attributes collected from PeaTMOSS, focusing primarily on data from the largest PTM ecosystem, Hugging Face (\cref{sec:method-MeasurementEval}, \cref{sec:DARA}). This introduces a slight risk that our findings may not generalize to other PTM ecosystems like ONNX Model Zoo or PyTorch Hub, \review{Reviewer 2-Concern 16}{How did the authors ensure that their findings are generalizable?}\review{Reviewer 3-Concern 6}{PeaTMOSS (Aug 2023) - One might argue that many recent PTMs are not considered for this work. Given the growth of PTM/LLM, one calendar year could be quite significant.}
\revision{and the August 2023 collection date might not capture the most recent PTMs. However, we expect that PTM practices remain relatively stable over a two-year period.}


\section{Discussion and Future Work}
\label{sec:discussion}

In this section, we describe the differences between PTM naming to traditional software naming (\cref{sec:Discussion-Compare}), the practical use cases of DARA (\cref{sec:DARA-UseCases}), and implication for PTM contributors (\cref{sec:Implication-PTMContributors}).
We also outline three directions for future research: developing automated tools for package naming standardization, improving model selection and reuse, and strengthening 
PTM supply chain security (\cref{sec:FutureWork}).

\subsection{\revision{Comparing to Traditional Software}}
\label{sec:Discussion-Compare}

Prior research has demonstrated that inconsistent naming of components and software packages has hampered software reuse for decades~\citep{krueger1992SoftwareReuse, griss1993softwareReusefromLibrarytoFactory}. Establishing more consistent and standardized naming conventions can improve software reuse~\citep{hofmeister2017shorterIdentifierNamesTakeLongertoComprehend, alpern2024NamingExperiments, Alsuhaibani2021NamingMethodsSurvey}.
Our study offers a deeper understanding of naming practices in PTM packages and introduces an automated validation tool to detect naming inconsistencies. However, much work remains to be done.

\revision{What's in a name? We suggest that examining naming conventions has given us insight into the hybrid nature of PTMs as both reusable software artifacts and cataloged components with embedded metadata. Unlike traditional software packages that prioritize simplicity or branding, PTM names often encode critical architectural and functional details --- more like a hardware spec sheet than a generic software identifier. This hybrid character underscores that PTMs are neither purely software nor hardware, but a unique class of reusable components whose names directly shape searchability, reliability, and reusability. As such, PTM naming is not just identifiers or metadata --- it is interface.}



\subsection{Use Cases of DARA and Similar Automated Naming Tools}
\label{sec:DARA-UseCases}

We describe DARA primarily as an \textbf{assessment tool} designed to detect naming inconsistencies, rather than a mechanism to enforce name changes. 
\ul{When DARA reports an anomaly, it signals a potential inconsistency between the actual model architecture and the metadata.} This discrepancy warrants caution from users or platform operators. DARA generates labels for \code{model\_type}, \code{task},
 and \code{architecture} metadata, with mismatches indicating potential issues in PTM packages. 

\review{Reviewer 3-Concern 12}{Given the workflow of the proposed method, can we deliver the prediction during the upload of a PTM to HF? How long does it take to predict the misalignment? The current runtime behaviour of the proposed model was not reported to demonstrate its feasibility.}
\review{Reviewer 2-Concern 7}{The authors should provide The effort required to use DARA effectively; The learning curve for users of the tool.
}
\review{Reviewer 2-Concern 10}{How could the DARA tool and study findings be integrated into existing workflows for PTM registries?}
\revision{Some versions of DARA are lightweight enough to integrate into the Hugging Face model upload workflow. During model upload, an N-gram or CNN-based DARA  could perform real-time consistency checks between the architectural features and the metadata provided by developers; they achieve throughput of ~29,000 PTMs/sec and ~17,000 PTMs/sec respectively (\cref{fig:latency}). This integration could follow a tiered approach:
\begin{itemize}
    \item \textbf{Pre-upload verification}: As developers complete the metadata form for their model, DARA can analyze the model architecture in the background and suggest appropriate values for model\_type, architecture, and task fields, potentially reducing manual errors.
    \item \textbf{Warning system}: When inconsistencies are detected, the platform can display non-blocking warnings to uploaders, allowing them to either confirm their unusual configuration or correct potential errors before publication.
    \item \textbf{Post-upload quality flagging}: For models already in the repository, periodic batch processing with more accurate (though computationally intensive) transformer-based DARA variants could identify and flag potential inconsistencies for platform moderators to review.
\end{itemize}
\noindent
This deployment strategy balances efficiency and accuracy while keeping false positives to a minimum, as evidenced by our F1 scores for model\_type (97.9\%) and architecture (67.5\%). The integration requires minimal effort from both platform operators (a single API endpoint) and model uploaders (who benefit from automated suggestions rather than facing additional requirements). The learning curve for using DARA is negligible, as it operates largely behind the scenes in the model upload workflow, with optional user-facing suggestions that follow familiar HF interface patterns.}

DARA can be employed in various ways to enhance the efficiency and security of model registries and address problems that were discussed in \cref{sec:background-roleofPTMNamingPractices}. Below are three specific use cases that illustrate its potential applications:

\begin{itemize}
\item \textbf{Model Validation}: Platforms like Hugging Face could utilize DARA to validate models, ensuring that the names and metadata align with the actual architectures. This would enhance the transparency and trustworthiness of model registries by detecting discrepancies~\citep{Montes2022DiscrepanciesAmongPTNN} and potential architectural backdoor attacks~\citep{langford2024architecturalNeuralBackdoors}. Supporting a more secure AI model supply chain is crucial, as highlighted in recent technical reports~\citep{Hepworth2024, HiddenLayer2024}.
\item \textbf{Label Generation and Validation}: 
\review{Reviewer 3-Concern 16}{Given the current workflow and classification/CNN architecture, it is not clear how DARA will generate labels for a PTM during its registration. Looks like an unsubstantiated claim.}
\revision{DARA directly addresses the insufficient model card information and metadata issues~\citep{Jiang2022PTMReuse, taraghi2024DLModelReuseinHF} by analyzing architectural features during model registration. When a developer uploads a new model, DARA's classification pipeline extracts architectural patterns from model definition files and compares them against its trained classifiers to generate probable metadata values (model\_type, architecture, task). These generated labels either validate developer-provided metadata or suggest appropriate alternatives when missing, improving model discoverability and reusability~\citep{di2024automatedcategorizationofPTM, Jiang2022PTMReuse}. The process is fully automated, requiring no manual intervention while significantly enhancing metadata quality across the registry.}
\item \textbf{Plagiarism Detection}: When a new model is published, DARA can be used to detect similarities between the new model architecture and existing architectures. This functionality helps in identifying model architecture plagiarism, a topic of recent discussions~\citep{OpenBMB2023}.
\end{itemize}

Applying DARA in these scenarios can enhance the transparency, security, and efficiency of managing and validating AI models in model registries. However, DARA is not presented as a complete solution but rather as a proof of concept that highlights opportunities in this emerging problem domain. Our work tested a key engineering hypothesis, demonstrating strong results in specific areas while acknowledging certain limitations. DARA’s implementation provides a foundation for further research and development, as discussed in \cref{sec:discussion}.

\revision{\subsection{Implications for PTM Contributors}}
\label{sec:Implication-PTMContributors}
\review{Reviewer 2-Concern 15}{Are there any lessons learned, suggestions, implications, or takeaways from this study that the authors would recommend to PTM contributors as part of the study's conclusions?
}
\revision{Our study offers actionable guidance for PTM contributors, by identifying the major ways in which naming practices impact model reuse, selection, and ecosystem trust.}

\revision{First, as detailed in \cref{sec:RQ1-Results}, contributors should enrich PTM names with semantic details -- such as architectural characteristics, parameter counts, and explicit reuse-relevant information -- to better guide users. Given that 88\% of survey participants noted that PTM naming conveys richer semantic content compared to traditional packages, if model authors (``name producers'') embrace this more descriptive naming style then it will simplify model selection.}

\revision{Second, the analysis in \cref{sec:RQ2-Results} shows a clear misalignment between user naming preferences and current naming practices. Users prefer names that reflect both the implementation (``how it works'') and the application domain (``what it does''), yet many PTM names emphasize only technical specifications. Contributors can improve this disconnect by standardizing names to incorporate both dimensions. Using automated validation tools, as demonstrated by our DARA pipeline in \cref{sec:DARA-Results} and discussed in \cref{sec:DARA-UseCases}, can help ensure that submitted names are consistent with the underlying model characteristics. In this way, automated linting integrated into PTM registries would not only facilitate a more uniform naming convention but also reduce downstream validation effort.
}

\revision{Third, our findings from \cref{sec:RQ3-Results} indicate that users currently rely on manual verification (\eg checking metadata and visual inspection) to detect naming inconsistencies --- a process that is both error-prone and time-consuming. By embracing standardized naming conventions informed by our empirical analysis and by adopting automated checks during the model upload process, PTM contributors can minimize these inconsistencies. Improved consistency not only aids in model selection and reuse but also contributes to stronger supply chain security. In practice, clear and standardized naming could serve as an early warning mechanism against potential typosquatting or package confusion attacks and other malicious practices in the model ecosystem~\citep{jiang2025Confuguard, Maurice2020OSSSupplyChainAttacks}.}

\subsection{Future Work}
\label{sec:FutureWork}

\subsubsection{Improving Naming Standardization through Automation}

Poor naming practices lead to poor comprehension~\citep{gresta2023namingPracticesinOOP, alpern2024NamingExperiments, Alsuhaibani2021NamingMethodsSurvey}. Our empirical analysis reveals that PTM naming differs from traditional package naming due to the substantial amount of semantic information embedded in PTM names (\cref{sec:RQ1-Results}). However, the quality and consistency of PTM naming is insufficient, primarily due to a lack of standardized practices. 

Automated \ul{validation} tools to detect inconsistencies in uploaded packages would streamline the PTM reuse process.
These tools would enhance the reliability of open-source packages while reducing the manual effort required for package validation by downstream users.
Our DARA system is the first such tool.
Although DARA effectively identifies real naming inconsistencies, its current feature extraction methods are limited (\cref{tab:DARAEval-TypeArch}, \cref{tab:DARAEval-Task}). For instance, DARA struggles to capture differences in training regimes and datasets between PTMs.
One approach to address this might be stronger feature extraction techniques that convert model architectures into embeddings. 
Contrastive learning~\citep{ni2021RelationshipbetweenCLandMetaL} offers a potential way to address some of these limitations.
Although it was not particularly successful in this context, we acknowledge that our application was exploratory.
Contrastive learning might be more effective when combined with more advanced feature extraction (\cref{sec:DARA-OtherFeatureExtraction}). We advocate for a more systematic investigation into the use of contrastive learning and advanced techniques, such as encoding directed graphs with transformers~\citep{geisler2023transformersMeetDirectedGraphs}, to enhance DARA’s inconsistency detection capabilities by generating more robust feature representations.

In parallel, automated \ul{linting} tools for package names could be introduced for PTM packages. While our work provides insights into PTM package naming, it does not yet address the best practices for standardization, akin to coding standards like Google’s style guide~\citep{google_style_guide}, PEP8~\citep{pep8}, or C++ coding standards~\citep{sutter2004c++CodingStandards}. As shown by~\citet{boogerd2008assessingCodingStandards}, the application of coding standards can enhance software reliability and reduce development effort, suggesting a similar benefit for naming conventions in package ecosystems.
Considerable work has been done on developing linters to support software development, often integrated into continuous integration pipelines to automate tasks~\citep{vassallo2020configuration, tomasdottir2018JAVALinterAdoption, habchi2018adoptingLintersforAndriodApps}. In a similar fashion, linters for package names could be integrated into package contribution workflows~\citep{sohacheski2021software, Jiang2022PTMSupplyChain}, improving standardization and enhancing both searchability and reliability. 
This approach seems particularly feasible for PTM registries, as our survey data suggests that contributors already perceive PTM names as being more standardized compared to traditional software packages (\cref{sec:RQ1-Results}, \cref{sec:RQ2-Results}). This existing trend toward standardization in PTM naming indicates that introducing automated tools like linters would be well-aligned with current practices, further promoting consistency and aiding discoverability.

\revision{Additionally, automated \ul{naming} tools can support the development of PTM packages, much like existing techniques for generating class and method names in traditional software~\citep{gao_neural_2019, allamanis_suggesting_2015}. Since PTM names often encode richer semantic information than conventional software names (\cref{sec:RQ1-Results}), automatically generating names based on static analysis of modeling and configuration code could help developers produce meaningful, standardized names. This would not only reduce engineering overhead but also improve the consistency and clarity of PTM package naming.}

\subsubsection{Improving Package Naming for Better Model Selection and Reuse}

The misalignment between engineers' expectations and PTM package naming conventions (\cref{sec:RQ2-Results}), coupled with inconsistent package metadata, increases the engineering effort and costs associated with model selection and reuse (\cref{sec:RQ3-Results}).

One approach to addressing the challenges of PTM selection and reuse is through automated machine learning (AutoML). AutoML techniques focus on automating aspects of model discovery and optimization, which can streamline the process and reduce manual effort~\citep{He2021AutoMLSurvey, wang_automl_2023}. 
We propose leveraging tools like DARA, which has already demonstrated its ability to utilize model architecture meta-features for detecting metadata inconsistencies. 
By extending DARA to extract additional meta-features, such as intermediate outputs, we can enable large-scale AutoML processes that enhance model selection efficiency and simplify PTM adoption. 
For instance, adapting the model selection approach developed by \citet{zhang2024modelSpider} for specific downstream tasks allows us to incorporate these meta-features, improving the selection process by considering engineering requirements like hardware constraints and domain adaptability~\citep{patil2025recommendingPTM4IoT}. 
Nguyen \etal further demonstrated the value of meta-features from PTM implementations on GitHub in boosting AutoML performance~\citep{Nguyen2022Manas}. Other studies have also explored innovative model selection methods to support PTM reuse and improve AutoML's ability to identify optimal models~\citep{You2021RankingandTuningPTMs, You2021LogME, Lu2019AutoDNNSelection4EdgeInference}.

\subsubsection{Securing Software Supply Chain by Leveraging Naming Information}
Poor comprehension opens the door to mistakes.
One class of attack related to naming is typosquatting attacks, which are known to occur in traditional software package registries~\citep{gu2023investigatingPackageRelatedSecurityThreatsinSoftwareRegistries, neupane2023beyondTyposquat}. 
For instance, attackers have exploited typographical errors in NPM package names to distribute malicious software, with several documented cases of typosquatting attacks targeting popular packages~\citep{reversinglabs2023typosquatting, infosecmag2023maliciousNPM, checkmarx2023typosquatting}.
Although we cannot point to a typosquatting incident for PTMs, \eg in \HF, we suggest that we can learn from issues in traditional software to predict --- and perhaps prevent --- similar recurrence in this new domain.
Solutions to hypothetical problems are not always necessary,
but solutions to \textit{predictable} problems strike us as a reasonable subject for research. 

Recent studies have underscored other vulnerabilities in PTM packages, including backdoors and malware injection~\citep{ langford2024architecturalNeuralBackdoors, Gu2019BadNets, Wang2022EvilModel2}.
These attacks threaten the AI model supply chain~\citep{HiddenLayer2024}. 
 For example, backdoors can be introduced into PTMs by exploiting architectural variations, making them challenging to detect~\citep{langford2024architecturalNeuralBackdoors}. Microsoft has emphasized that consistent naming practices can help expedite the detection and removal of such malware~\citep{AzureCounterfitMalwarePOC}. For instance, standardized naming conventions could enable PTM registries to develop automated tools that verify whether the package name accurately reflects the model’s architecture, helping to identify potential backdoors. These tools could build upon DARA, a system designed to detect mismatches between model architecture and metadata, providing a solid foundation for enhanced security measures in the future.

\revision{Recent work has also shown that typosquatting attack, or package confusion attack, which is the most common traditional software supply chain attack, can also affect the AI model supply chain~\citep{protectai_huggingface, jiang2025Confuguard}.
Our results suggest that existing techniques for typosquat detection may not apply directly to PTMs.
Most existing typoquatting detection techniques rely on syntactic variations, but ignore attacks involving name semantics\citep{taylor2020defending, neupane2023beyondTyposquat}.
We conjecture that such approaches are useful for traditional packages, whose names are quite distinct and not semantically meaningful (cf.~\cref{tab:top_packages}).
In traditional software packages, small tweaks to a package name are a signal of typosquatting.
For PTMs, our data show that small tweaks to a name are part of the naming convention for PTM names (\cref{sec:RQ1-Results}, \cref{sec:RQ2-Results}), 
PTM names contain much more semantic information and are much more structured than are the names of traditional software packages.
This distinction indicates that conventional typosquatting detection approaches will detect some classes of typosquatting attacks for PTMs, but will miss attacks that leverage this property of PTM names.
We suggest that techniques of \textbf{semantically‑enriched} string similarity -- for example, combining character‑level edit distance with token‑level embedding distances -- and \textbf{transformer‑based} anomaly classification might be applied to PTMs, because their semantically aware classification system can group and distinguish models based on rich architectural and task metadata.
However, those techniques may need to be adapted for the following aspects of PTM names: multi‑part tokenization (\eg separating architecture, task, dataset and hyperparameters), domain‑specific abbreviations and acronyms, versioning conventions embedded within names, and numeric parameter encodings (\cref{sec:RQ4-Results}).}

\section{Conclusion}
\label{sec:conclusion}

Pre-Trained Models (PTMs) are a new unit of reuse in software engineering, where naming practices
significantly impact their reusability.
Similarly to prior reuse units, engineers struggle to name PTMs well.
We conducted the first study of PTM naming practices by a mixed-method approach, including a survey and a repository mining study.
Our finding shows that PTM naming is different from traditional naming  in terms of evolution practices and reuse decisions. We also show that engineers prefer PTM naming with architecture/function, and design purpose is important in PTM naming.
PTM users look for naming inconsistencies using metadata and manual inspection.  Our tool applies a novel feature extraction method and indicated that we can effectively detect naming inconsistencies with only architectural information.
This study opens up several avenues for future work: establishing best practices for PTM naming, enhancing model search capabilities, and improving automated detection of naming inconsistencies.
We have identified a context where new types of names are emerging, reflecting the evolving landscape of software engineering.

\section{Declarations}

\subsection{Funding}
This research was supported by gifts from Google and Cisco, and by NSF awards \#2107230, \#2229703, \#2107020, and \#2104319.

\subsection{Ethical Approval}
Our research involving human participants, specifically the survey study, was conducted under a protocol approved by Purdue University's Institutional Review Board (IRB), with protocol number \#\emph{IRB-2022-606}.

\subsection{Informed Consent}
All participants provided written informed consent prior to taking part in our survey study. All anonymized responses are included in the published artifact.

\subsection{Author Contributions}
The contributions of each author are as follows:
\begin{itemize}
  \item \textbf{Wenxin Jiang:} Lead author; conducted the literature review and defined the problem scope; designed and administered the survey study and its analysis; designed DARA and all evaluation experiments.
  \item \textbf{Mingyu Kim:} Led the implementation of naming convention extraction and manual evaluation in the repository mining study; implemented advanced feature‐extraction and anomaly‐detection methods within DARA; implemented all evaluation experiments and visualizations relevant to DARA.
  \item \textbf{Chingwo Cheung:} Developed the repository‐mining prototype for naming‐element extraction; designed and implemented DARA’s abstract architecture extraction.
  \item \textbf{Heesoo Kim:} Co-designed and evaluated the repository mining study.
  \item \textbf{George K. Thiruvathukal:} Provided feedback on study design, evaluation methodology, and problem statement.
  \item \textbf{James C. Davis:} Supervised all aspects of study conduct; performed qualitative analysis of survey data (Tables 6-7); guided DARA’s evaluation process.
\end{itemize}

\subsection{Data Availability} \label{sec:Artifact}
The artifact of this work is available at \url{https://github.com/PurdueDualityLab/PTM-Naming}. The artifact includes:
\begin{itemize}
    \item \textbf{Survey study (\cref{sec:theme1-method-survey})}: We provide the complete survey instrument, along with the anonymized raw data and data analysis sheets used in the analysis.
    \item \textbf{Repository Mining (\cref{sec:Repo-Mining})}: We include the prompts, evaluation and data
    \item \textbf{DARA (\cref{sec:RQ4-Results})}: We include the end-to-end pipeline, fine-tuning code, feature extraction, visualization, and scripts to generate all figures and tables.

\end{itemize}

\subsection{Conflict of Interest}
The authors declare that they have no conflict of interest.

\subsection{Clinical Trial Number}

Not applicable.







%
%

\bibliographystyle{spbasic}      
\bibliography{main}

\begin{thebibliography}{165}
\providecommand{\natexlab}[1]{#1}
\providecommand{\url}[1]{{#1}}
\providecommand{\urlprefix}{URL }
\expandafter\ifx\csname urlstyle\endcsname\relax
  \providecommand{\doi}[1]{DOI~\discretionary{}{}{}#1}\else
  \providecommand{\doi}{DOI~\discretionary{}{}{}\begingroup
  \urlstyle{rm}\Url}\fi
\providecommand{\eprint}[2][]{\url{#2}}

\bibitem[{Sur(2021)}]{SurveySystem2021}
 (2021) Sample size calculator. \url{https://www.surveysystem.com/sscalc.htm}

\bibitem[{Abdalkareem et~al.(2017)Abdalkareem, Nourry, Wehaibi, Mujahid, and
  Shihab}]{Abdalkareem2017TrivialPackages}
Abdalkareem R, Nourry O, Wehaibi S, Mujahid S, Shihab E (2017) Why do
  developers use trivial packages? an empirical case study on npm. In: European
  {Software} {Engineering} {Conference}/{Foundations} of {Software}
  {Engineering} ({ESEC}/{FSE})

\bibitem[{Abdellatif et~al.(2020)Abdellatif, Zeng, Elshafei, Shihab, and
  Shang}]{abdellatif2020simplifyingtheSearchofNPMPackages}
Abdellatif A, Zeng Y, Elshafei M, Shihab E, Shang W (2020) Simplifying the
  search of npm packages. Information and Software Technology 126:106365

\bibitem[{Abiodun et~al.(2018)Abiodun, Jantan, Omolara, Dada, Mohamed, and
  Arshad}]{abiodun2018SOTADNNAppSurvey}
Abiodun OI, Jantan A, Omolara AE, Dada KV, Mohamed NA, Arshad H (2018)
  State-of-the-art in artificial neural network applications: A survey. Heliyon
  4(11)

\bibitem[{Ali et~al.(2020)Ali, Shiaeles, Bendiab, and Ghita}]{ali2020malgra}
Ali M, Shiaeles S, Bendiab G, Ghita B (2020) Malgra: Machine learning and
  n-gram malware feature extraction and detection system. Electronics
  9(11):1777

\bibitem[{Allamanis et~al.(2015)Allamanis, Barr, Bird, and
  Sutton}]{allamanis_suggesting_2015}
Allamanis M, Barr ET, Bird C, Sutton C (2015) Suggesting accurate method and
  class names. In: Proceedings of the 2015 10th {Joint} {Meeting} on
  {Foundations} of {Software} {Engineering}, ACM, Bergamo Italy, pp 38--49,
  \doi{10.1145/2786805.2786849},
  \urlprefix\url{https://dl.acm.org/doi/10.1145/2786805.2786849}

\bibitem[{Alpern et~al.(2024)Alpern, Lazer, Tzachor, Hakim, Weissbuch, and
  Feitelson}]{alpern2024NamingExperiments}
Alpern R, Lazer I, Tzachor I, Hakim H, Weissbuch S, Feitelson DG (2024)
  Reproducing, extending, and analyzing naming experiments. arXiv

\bibitem[{Alsuhaibani et~al.(2021)Alsuhaibani, Newman, Decker, Collard, and
  Maletic}]{Alsuhaibani2021NamingMethodsSurvey}
Alsuhaibani R, Newman C, Decker M, Collard M, Maletic J (2021) On the {Naming}
  of {Methods}: {A} {Survey} of {Professional} {Developers}. In: 2021
  {IEEE}/{ACM} 43rd {International} {Conference} on {Software} {Engineering}
  ({ICSE}), pp 587--599, \doi{10.1109/ICSE43902.2021.00061}

\bibitem[{Ayodele(2010)}]{ayodele2010typesofMLAlgorithms}
Ayodele TO (2010) Types of machine learning algorithms. New advances in machine
  learning 3(19-48):5--1

\bibitem[{{Azure and contributors}(2023)}]{AzureCounterfitMalwarePOC}
{Azure and contributors} (2023) Abusing ml model file formats to create malware
  on ai systems: A proof of concept.
  \url{https://github.com/Azure/counterfit/wiki/Abusing-ML-model-file-formats-to-create-malware-on-AI-systems:-A-proof-of-concept#detection}

\bibitem[{Beltagy et~al.(2020)Beltagy, Peters, and
  Cohan}]{beltagy2020longformer}
Beltagy I, Peters ME, Cohan A (2020) Longformer: The long-document transformer.
  arXiv preprint arXiv:200405150

\bibitem[{Bengio et~al.(2013)Bengio, Courville, and
  Vincent}]{bengio2013representationLearning}
Bengio Y, Courville A, Vincent P (2013) Representation learning: A review and
  new perspectives. IEEE transactions on pattern analysis and machine
  intelligence 35(8):1798--1828

\bibitem[{Bhatia et~al.(2023)Bhatia, Eghan, Grichi, Cavanagh, Jiang, and
  Adams}]{bhatia2023towardsMLpipelines}
Bhatia A, Eghan EE, Grichi M, Cavanagh WG, Jiang ZM, Adams B (2023) Towards a
  change taxonomy for machine learning pipelines: Empirical study of ml
  pipelines and forks related to academic publications. Empirical Software
  Engineering 28(3):60

\bibitem[{Bober-Irizar et~al.(2023)Bober-Irizar, Shumailov, Zhao, Mullins, and
  Papernot}]{bober-irizar_architectural_2023}
Bober-Irizar M, Shumailov I, Zhao Y, Mullins R, Papernot N (2023) Architectural
  {Backdoors} in {Neural} {Networks}. In: 2023 {IEEE}/{CVF} {Conference} on
  {Computer} {Vision} and {Pattern} {Recognition} ({CVPR}), IEEE, Vancouver,
  BC, Canada, pp 24595--24604, \doi{10.1109/CVPR52729.2023.02356},
  \urlprefix\url{https://ieeexplore.ieee.org/document/10204575/}

\bibitem[{Boogerd and Moonen(2008)}]{boogerd2008assessingCodingStandards}
Boogerd C, Moonen L (2008) Assessing the value of coding standards: An
  empirical study. In: 2008 IEEE International conference on software
  maintenance, IEEE, pp 277--286

\bibitem[{Brown et~al.(2020{\natexlab{a}})Brown, Mann, Ryder, Subbiah, Kaplan,
  Dhariwal, Neelakantan, Shyam, Sastry, Askell et~al.}]{Brown2020gpt}
Brown T, Mann B, Ryder N, Subbiah M, Kaplan JD, Dhariwal P, Neelakantan A,
  Shyam P, Sastry G, Askell A, et~al. (2020{\natexlab{a}}) Language models are
  few-shot learners. Advances in neural information processing systems
  33:1877--1901

\bibitem[{Brown et~al.(2020{\natexlab{b}})Brown, Mann, Ryder, Subbiah, Kaplan,
  Dhariwal, Neelakantan, Shyam, Sastry, Askell
  et~al.}]{brown2020languageModelsAreFewShotLearners}
Brown T, Mann B, Ryder N, Subbiah M, Kaplan JD, Dhariwal P, Neelakantan A,
  Shyam P, Sastry G, Askell A, et~al. (2020{\natexlab{b}}) Language models are
  few-shot learners. Advances in neural information processing systems
  33:1877--1901

\bibitem[{Bubeck et~al.(2023)Bubeck, Chandrasekaran, Eldan, Gehrke, Horvitz,
  Kamar, Lee, Lee, Li, Lundberg, Nori, Palangi, Ribeiro, and
  Zhang}]{Bubeck2023AGISystem}
Bubeck S, Chandrasekaran V, Eldan R, Gehrke J, Horvitz E, Kamar E, Lee P, Lee
  YT, Li Y, Lundberg S, Nori H, Palangi H, Ribeiro MT, Zhang Y (2023) Sparks of
  {Artificial} {General} {Intelligence}: {Early} experiments with {GPT}-4.
  \urlprefix\url{http://arxiv.org/abs/2303.12712}

\bibitem[{Butler et~al.(2010)Butler, Wermelinger, Yu, and
  Sharp}]{Butler2012MethodNameConsistency}
Butler S, Wermelinger M, Yu Y, Sharp H (2010) Exploring the {Influence} of
  {Identifier} {Names} on {Code} {Quality}: {An} {Empirical} {Study}. In: 2010
  14th {European} {Conference} on {Software} {Maintenance} and {Reengineering},
  pp 156--165, \doi{10.1109/CSMR.2010.27}

\bibitem[{Butler et~al.(2011)Butler, Wermelinger, Yu, and
  Sharp}]{butler_mining_2011}
Butler S, Wermelinger M, Yu Y, Sharp H (2011) Mining java class naming
  conventions. In: 2011 27th {IEEE} {International} {Conference} on {Software}
  {Maintenance} ({ICSM}), pp 93--102, \doi{10.1109/ICSM.2011.6080776}, iSSN:
  1063-6773

\bibitem[{Capra et~al.(2020)Capra, Bussolino, Marchisio, Masera, Martina, and
  Shafique}]{capra2020hardwareandSWOptimization}
Capra M, Bussolino B, Marchisio A, Masera G, Martina M, Shafique M (2020)
  Hardware and software optimizations for accelerating deep neural networks:
  Survey of current trends, challenges, and the road ahead. IEEE Access
  8:225134--225180

\bibitem[{Casta{\~n}o et~al.(2023)Casta{\~n}o, Mart{\'\i}nez-Fern{\'a}ndez,
  Franch, and Bogner}]{castano2023analyzingEvolandMaintenanceofMLModelsonHF}
Casta{\~n}o J, Mart{\'\i}nez-Fern{\'a}ndez S, Franch X, Bogner J (2023)
  Analyzing the evolution and maintenance of ml models on hugging face. arXiv
  preprint arXiv:231113380

\bibitem[{{Checkmarx Research Team}(2023)}]{checkmarx2023typosquatting}
{Checkmarx Research Team} (2023) A new stealthier type of typosquatting attack
  spotted targeting npm.
  \urlprefix\url{https://checkmarx.com/blog/a-new-stealthier-type-of-typosquatting-attack-spotted-targeting-npm/}

\bibitem[{Chmielinski et~al.(2022)Chmielinski, Newman, Taylor, Joseph, Thomas,
  Yurkofsky, and Qiu}]{chmielinski2022datasetNutritionLabel}
Chmielinski KS, Newman S, Taylor M, Joseph J, Thomas K, Yurkofsky J, Qiu YC
  (2022) The dataset nutrition label (2nd gen): Leveraging context to mitigate
  harms in artificial intelligence. arXiv preprint arXiv:220103954

\bibitem[{Chung et~al.(2014)Chung, Gulcehre, Cho, and Bengio}]{chung2014GRU}
Chung J, Gulcehre C, Cho K, Bengio Y (2014) Empirical evaluation of gated
  recurrent neural networks on sequence modeling. arXiv preprint arXiv:14123555

\bibitem[{Davis et~al.(2023)Davis, Jajal, Jiang, Schorlemmer, Synovic, and
  Thiruvathukal}]{davis2023JVA}
Davis JC, Jajal P, Jiang W, Schorlemmer TR, Synovic N, Thiruvathukal GK (2023)
  Reusing deep learning models: Challenges and directions in software
  engineering. In: Proceedings of the IEEE John Vincent Atanasoff Symposium on
  Modern Computing (JVA’23)

\bibitem[{Decan et~al.(2019)Decan, Mens, and
  Grosjean}]{decan2019empiricalComparisonofDependencyNetworkEvolutioninSevenSWPkgEcosystems}
Decan A, Mens T, Grosjean P (2019) An empirical comparison of dependency
  network evolution in seven software packaging ecosystems. Empirical Software
  Engineering 24(1):381--416

\bibitem[{{DeepSeek AI}(2024)}]{deepseek2024hf}
{DeepSeek AI} (2024) Deepseek ai on hugging face.
  \urlprefix\url{https://huggingface.co/deepseek-ai}, accessed: 2025-04-14

\bibitem[{Deissenboeck and
  Pizka(2006)}]{deissenboeck2006conciseandConsistentNaming}
Deissenboeck F, Pizka M (2006) Concise and consistent naming. Software Quality
  Journal 14:261--282

\bibitem[{Devlin et~al.(2019)Devlin, Chang, Lee, and
  Toutanova}]{devlin2019bert}
Devlin J, Chang MW, Lee K, Toutanova K (2019) Bert: Pre-training of deep
  bidirectional transformers for language understanding. In: Proceedings of the
  2019 conference of the North American chapter of the association for
  computational linguistics: human language technologies, volume 1 (long and
  short papers), pp 4171--4186

\bibitem[{Di~Sipio et~al.(2024)Di~Sipio, Rubei, Di~Rocco, Di~Ruscio, and
  Nguyen}]{di2024automatedcategorizationofPTM}
Di~Sipio C, Rubei R, Di~Rocco J, Di~Ruscio D, Nguyen PT (2024) Automated
  categorization of pre-trained models for software engineering: A case study
  with a hugging face dataset. arXiv preprint arXiv:240513185

\bibitem[{Feitelson et~al.(2020)Feitelson, Mizrahi, Noy, Shabat, Eliyahu, and
  Sheffer}]{feitelson2020HowDeveloperChooseNames}
Feitelson DG, Mizrahi A, Noy N, Shabat AB, Eliyahu O, Sheffer R (2020) How
  developers choose names. IEEE Transactions on Software Engineering
  48(1):37--52

\bibitem[{Frakes and Kang(2005)}]{frakes2005SoftwareReuseResearch}
Frakes WB, Kang K (2005) Software reuse research: Status and future. IEEE
  transactions on Software Engineering 31(7):529--536

\bibitem[{Gao et~al.(2019)Gao, Chen, Xing, Ma, Song, and Lin}]{gao_neural_2019}
Gao S, Chen C, Xing Z, Ma Y, Song W, Lin SW (2019) A {Neural} {Model} for
  {Method} {Name} {Generation} from {Functional} {Description}. In: 2019 {IEEE}
  26th {International} {Conference} on {Software} {Analysis}, {Evolution} and
  {Reengineering} ({SANER}), pp 414--421, \doi{10.1109/SANER.2019.8667994},
  iSSN: 1534-5351

\bibitem[{Garg(2023)}]{garg2023powerofHFAI}
Garg G (2023) The power of hugging face ai.
  \url{https://medium.com/@gargg/the-power-of-hugging-face-ai-4f6558ee0874}

\bibitem[{Geisler et~al.(2023)Geisler, Li, Mankowitz, Cemgil, G{\"u}nnemann,
  and Paduraru}]{geisler2023transformersMeetDirectedGraphs}
Geisler S, Li Y, Mankowitz DJ, Cemgil AT, G{\"u}nnemann S, Paduraru C (2023)
  Transformers meet directed graphs. In: International Conference on Machine
  Learning (ICML'23), PMLR, pp 11144--11172

\bibitem[{Gonz{\'a}lez and Van Der~Meer(2004)}]{gonzalez2004standard}
Gonz{\'a}lez R, Van Der~Meer K (2004) Standard metadata applied to software
  retrieval. Journal of Information Science 30(4):300--309

\bibitem[{{Google}(2024{\natexlab{a}})}]{google2024hf}
{Google} (2024{\natexlab{a}}) Google on hugging face.
  \urlprefix\url{https://huggingface.co/google}, accessed: 2025-04-14

\bibitem[{{Google}(2024{\natexlab{b}})}]{google_style_guide}
{Google} (2024{\natexlab{b}}) {Google Style Guide}.
  \url{https://google.github.io/styleguide/}

\bibitem[{Gresta et~al.(2023)Gresta, Durelli, and
  Cirilo}]{gresta2023namingPracticesinOOP}
Gresta R, Durelli V, Cirilo E (2023) Naming practices in object-oriented
  programming: An empirical study. Journal of Software Engineering Research and
  Development pp 5--1

\bibitem[{Griss(1993)}]{griss1993softwareReusefromLibrarytoFactory}
Griss ML (1993) Software reuse: From library to factory. IBM systems journal
  32(4):548--566

\bibitem[{Gu et~al.(2019)Gu, Liu, Dolan-Gavitt, and Garg}]{Gu2019BadNets}
Gu T, Liu K, Dolan-Gavitt B, Garg S (2019) Badnets: Evaluating backdooring
  attacks on deep neural networks. IEEE Access 7:47230--47244

\bibitem[{Gu et~al.(2023)Gu, Ying, Pu, Hu, Chai, Wang, Gao, and
  Duan}]{gu2023investigatingPackageRelatedSecurityThreatsinSoftwareRegistries}
Gu Y, Ying L, Pu Y, Hu X, Chai H, Wang R, Gao X, Duan H (2023) Investigating
  package related security threats in software registries. In: 2023 IEEE
  Symposium on Security and Privacy (SP), IEEE, pp 1578--1595

\bibitem[{Guest et~al.(2011)Guest, MacQueen, and Namey}]{guest2011applied}
Guest G, MacQueen KM, Namey EE (2011) Applied thematic analysis. sage
  publications

\bibitem[{{Guido van Rossum et al.}(2001)}]{pep8}
{Guido van Rossum et al} (2001) {PEP 8 – Style Guide for Python Code}.
  \url{https://peps.python.org/pep-0008/}

\bibitem[{Gunel et~al.(2020)Gunel, Du, Conneau, and
  Stoyanov}]{gunel2020supervised}
Gunel B, Du J, Conneau A, Stoyanov V (2020) Supervised contrastive learning for
  pre-trained language model fine-tuning. arXiv preprint arXiv:201101403

\bibitem[{Gururangan et~al.(2020)Gururangan, Marasovi{\'c}, Swayamdipta, Lo,
  Beltagy, Downey, and Smith}]{gururangan2020dontstoppretraining}
Gururangan S, Marasovi{\'c} A, Swayamdipta S, Lo K, Beltagy I, Downey D, Smith
  NA (2020) Don't stop pretraining: Adapt language models to domains and tasks.
  In: Proceedings of the 58th Annual Meeting of the Association for
  Computational Linguistics (ACL)

\bibitem[{Géron(2019)}]{geron2019hands}
Géron A (2019) Hands-On Machine Learning with Scikit-Learn, Keras, and
  TensorFlow: Concepts, Tools, and Techniques to Build Intelligent Systems, 2nd
  edn. O'Reilly Media

\bibitem[{Habchi et~al.(2018)Habchi, Blanc, and
  Rouvoy}]{habchi2018adoptingLintersforAndriodApps}
Habchi S, Blanc X, Rouvoy R (2018) On adopting linters to deal with performance
  concerns in android apps. In: Proceedings of the 33rd ACM/IEEE International
  Conference on Automated Software Engineering, pp 6--16

\bibitem[{Han et~al.(2021)Han, Zhang, Ding, Gu, Liu, Huo, Qiu, Yao, Zhang,
  Zhang, Han, Huang, Jin, Lan, Liu, Liu, Lu, Qiu, Song, Tang, Wen, Yuan, Zhao,
  and Zhu}]{Han2021PTM}
Han X, Zhang Z, Ding N, Gu Y, Liu X, Huo Y, Qiu J, Yao Y, Zhang A, Zhang L, Han
  W, Huang M, Jin Q, Lan Y, Liu Y, Liu Z, Lu Z, Qiu X, Song R, Tang J, Wen JR,
  Yuan J, Zhao WX, Zhu J (2021) Pre-trained models: {Past}, present and future.
  AI Open 2:225--250

\bibitem[{He et~al.(2021{\natexlab{a}})He, Lee, Raychev, and
  Vechev}]{he2021learningtoFindNamingIssueswithBigCodeandSmallSupervision}
He J, Lee CC, Raychev V, Vechev M (2021{\natexlab{a}}) Learning to find naming
  issues with big code and small supervision. In: Proceedings of the 42nd ACM
  SIGPLAN International Conference on Programming Language Design and
  Implementation, pp 296--311

\bibitem[{He et~al.(2021{\natexlab{b}})He, Zhao, and Chu}]{He2021AutoMLSurvey}
He X, Zhao K, Chu X (2021{\natexlab{b}}) {AutoML}: {A} survey of the
  state-of-the-art. Knowledge-Based Systems

\bibitem[{Heineman and Councill(2001)}]{heineman2001componentBasedSE}
Heineman GT, Councill WT (2001) Component-based software engineering. Putting
  the pieces together, addison-westley 5:1

\bibitem[{Henninger(1994)}]{henninger1994usingIterativeRefinementtoFindReusableSW}
Henninger S (1994) Using iterative refinement to find reusable software. IEEE
  software 11(5):48--59

\bibitem[{Hepworth et~al.(2024)Hepworth, Olive, Dasgupta, Le, Lodato, Maruseac,
  Meiklejohn, Chaudhuri, and Minkus}]{Hepworth2024}
Hepworth I, Olive K, Dasgupta K, Le M, Lodato M, Maruseac M, Meiklejohn S,
  Chaudhuri S, Minkus T (2024) Securing the ai software supply chain. Technical
  report, Google

\bibitem[{HiddenLayer(2024)}]{HiddenLayer2024}
HiddenLayer (2024) Ai threat landscape report.
  \url{https://21998286.fs1.hubspotusercontent-na1.net/hubfs/21998286/HiddenLayer%20AI%20Threat%20Landscape%20Report%202024.pdf},
  retrieved from HiddenLayer

\bibitem[{Hofmeister et~al.(2017{\natexlab{a}})Hofmeister, Siegmund, and
  Holt}]{hofmeister2017shorterIdentifierNamesTakeLongertoComprehend}
Hofmeister J, Siegmund J, Holt DV (2017{\natexlab{a}}) Shorter identifier names
  take longer to comprehend. In: 2017 IEEE 24th International conference on
  software analysis, evolution and reengineering (SANER), IEEE, pp 217--227

\bibitem[{Hofmeister et~al.(2017{\natexlab{b}})Hofmeister, Siegmund, and
  Holt}]{hofmeister2017shorterIIDNamesTakeLongertoComprehend}
Hofmeister J, Siegmund J, Holt DV (2017{\natexlab{b}}) Shorter identifier names
  take longer to comprehend. In: 2017 IEEE 24th International conference on
  software analysis, evolution and reengineering (SANER), IEEE, pp 217--227

\bibitem[{H{\o}st and {\O}stvold(2009)}]{host2009debuggingMethodNames}
H{\o}st EW, {\O}stvold BM (2009) Debugging method names. In: European
  Conference on Object-Oriented Programming, Springer, pp 294--317

\bibitem[{Huang and Lin(2023)}]{huang2023cigar}
Huang J, Lin B (2023) Cigar: Contrastive learning for github action
  recommendation. In: 2023 IEEE 23rd International Working Conference on Source
  Code Analysis and Manipulation (SCAM), IEEE, pp 61--71

\bibitem[{{Hugging Face}(2023)}]{HFDoc}
{Hugging Face} (2023) Hugging face documentations.
  \urlprefix\url{https://huggingface.co/docs}

\bibitem[{{Hugging Face, Inc.}(2024)}]{huggingface2024termsofService}
{Hugging Face, Inc} (2024) Hugging face terms of service.
  \urlprefix\url{https://huggingface.co/terms-of-service}, accessed: 2024-08-18

\bibitem[{{Infosecurity Magazine}(2023)}]{infosecmag2023maliciousNPM}
{Infosecurity Magazine} (2023) Malicious npm package uses new stealthier
  typosquatting attack.
  \urlprefix\url{https://www.infosecurity-magazine.com/news/malicious-npm-package-uses/}

\bibitem[{Islam et~al.(2023)Islam, Elmekki, Elsebai, Bentahar, Drawel, Rjoub,
  and Pedrycz}]{islam2023comprehensiveSurveyonTransformerApplications}
Islam S, Elmekki H, Elsebai A, Bentahar J, Drawel N, Rjoub G, Pedrycz W (2023)
  A comprehensive survey on applications of transformers for deep learning
  tasks. Expert Systems with Applications p 122666

\bibitem[{Jain and Verma(2021)}]{jain2021contrastive}
Jain A, Verma Y (2021) Contrastive learning of visual-semantic embeddings.
  arXiv preprint arXiv:211008872

\bibitem[{Jajal et~al.(2024)Jajal, Jiang, Tewari, Kocinare, Woo, Sarraf, Lu,
  Thiruvathukal, and Davis}]{jajal2023ONNXFailureStudy}
Jajal P, Jiang W, Tewari A, Kocinare E, Woo J, Sarraf A, Lu YH, Thiruvathukal
  GK, Davis JC (2024) Interoperability in deep learning: A user survey and
  failure analysis of onnx model converters. the 33rd ACM SIGSOFT International
  Symposium on Software Testing and Analysis (ISSTA'24)

\bibitem[{Jiang et~al.(2022{\natexlab{a}})Jiang, Synovic, Sethi, Indarapu,
  Hyatt, Schorlemmer, Thiruvathukal, and Davis}]{Jiang2022PTMSupplyChain}
Jiang W, Synovic N, Sethi R, Indarapu A, Hyatt M, Schorlemmer TR, Thiruvathukal
  GK, Davis JC (2022{\natexlab{a}}) An empirical study of artifacts and
  security risks in the pre-trained model supply chain. In: Proceedings of the
  2022 ACM Workshop on Software Supply Chain Offensive Research and Ecosystem
  Defenses, pp 105--114

\bibitem[{Jiang et~al.(2022{\natexlab{b}})Jiang, Synovic, Sethi, Indarapu,
  Hyatt, Schorlemmer, Thiruvathukal, and Davis}]{2022JiangEmpirical}
Jiang W, Synovic N, Sethi R, Indarapu A, Hyatt M, Schorlemmer TR, Thiruvathukal
  GK, Davis JC (2022{\natexlab{b}}) An empirical study of artifacts and
  security risks in the pre-trained model supply chain. In: ACM Workshop on
  Software Supply Chain Offensive Research and Ecosystem Defenses (SCORED'22),
  p 105–114, \doi{10.1145/3560835.3564547},
  \urlprefix\url{https://doi.org/10.1145/3560835.3564547}

\bibitem[{Jiang et~al.(2023{\natexlab{a}})Jiang, Banna, Vivek, Goel, Synovic,
  Thiruvathukal, and Davis}]{Jiang2023CVReengineering}
Jiang W, Banna V, Vivek N, Goel A, Synovic N, Thiruvathukal GK, Davis JC
  (2023{\natexlab{a}}) Challenges and practices of deep learning model
  reengineering: A case study on computer vision. arXiv preprint
  arXiv:230307476

\bibitem[{Jiang et~al.(2023{\natexlab{b}})Jiang, Synovic, Hyatt, Schorlemmer,
  Sethi, Lu, Thiruvathukal, and Davis}]{Jiang2022PTMReuse}
Jiang W, Synovic N, Hyatt M, Schorlemmer TR, Sethi R, Lu YH, Thiruvathukal GK,
  Davis JC (2023{\natexlab{b}}) An empirical study of pre-trained model reuse
  in the hugging face deep learning model registry. In: {IEEE}/{ACM} 45th
  {International} {Conference} on {Software} {Engineering} (ICSE'23)

\bibitem[{Jiang et~al.(2023{\natexlab{c}})Jiang, Synovic, Jajal, Schorlemmer,
  Tewari, Pareek, Thiruvathukal, and Davis}]{jiang2023ptmtorrent}
Jiang W, Synovic N, Jajal P, Schorlemmer TR, Tewari A, Pareek B, Thiruvathukal
  GK, Davis JC (2023{\natexlab{c}}) Ptmtorrent: A dataset for mining
  open-source pre-trained model packages. Proceedings of the 20th International
  Conference on Mining Software Repositories (MSR'23)

\bibitem[{Jiang et~al.(2024)Jiang, Yasmin, Jones, Synovic, Kuo, Tian,
  Thiruvathukal, , and Davis}]{Jiang2024peatmoss}
Jiang W, Yasmin J, Jones J, Synovic N, Kuo J, Tian Y, Thiruvathukal GK, , Davis
  JC (2024) Peatmoss: A dataset and initial analysis of pre-trained models in
  open-source software. In: Proceedings of the 21th Annual Conference on Mining
  Software Repositories (MSR'24)

\bibitem[{Jiang et~al.(2025)Jiang, {\c{C}}akar, Lysenko, and
  Davis}]{jiang2025Confuguard}
Jiang W, {\c{C}}akar B, Lysenko M, Davis JC (2025) Detecting active and
  stealthy typosquatting threats in package registries. arXiv preprint
  arXiv:250220528

\bibitem[{Johnson and Onwuegbuzie(2004)}]{MixedMethodsResearch}
Johnson RB, Onwuegbuzie AJ (2004) Mixed {Methods} {Research}: {A} {Research}
  {Paradigm} {Whose} {Time} {Has} {Come}. Educational Researcher

\bibitem[{Jones et~al.(2024)Jones, Jiang, Synovic, Thiruvathukal, and
  Davis}]{jones2024HFQualitativeValidation}
Jones J, Jiang W, Synovic N, Thiruvathukal GK, Davis JC (2024) What do we know
  about hugging face? a systematic literature review and quantitative
  validation of qualitative claims. In: Proceedings of the 18th International
  Conference on Empirical Software Engineering and Measurement (ESEM)

\bibitem[{Kathikar et~al.(2023)Kathikar, Nair, Lazarine, Sachdeva, and
  Samtani}]{kathikar2023assessingHFVulnerabilities}
Kathikar A, Nair A, Lazarine B, Sachdeva A, Samtani S (2023) Assessing the
  Vulnerabilities of the Open-Source Artificial Intelligence (AI) Landscape: A
  Large-Scale Analysis of the Hugging Face Platform

\bibitem[{Khalid and Zim{\'a}nyi(2024)}]{khalid2024repairingRawMetadata}
Khalid H, Zim{\'a}nyi E (2024) Repairing raw metadata for metadata management.
  Information Systems 122:102344

\bibitem[{Khosla et~al.(2021)Khosla, Teterwak, Wang, Sarna, Tian, Isola,
  Maschinot, Liu, and Krishnan}]{khosla2021supervisedcontrastivelearning}
Khosla P, Teterwak P, Wang C, Sarna A, Tian Y, Isola P, Maschinot A, Liu C,
  Krishnan D (2021) Supervised contrastive learning.
  \urlprefix\url{https://arxiv.org/abs/2004.11362}, \eprint{2004.11362}

\bibitem[{Kim(2014)}]{kim2014cnn}
Kim Y (2014) Convolutional neural networks for sentence classification. In:
  Proceedings of the 2014 Conference on Empirical Methods in Natural Language
  Processing (EMNLP), pp 1746--1751

\bibitem[{Kitchenham and Pfleeger(2008)}]{kitchenham2008personalSurvey}
Kitchenham BA, Pfleeger SL (2008) Personal opinion surveys. In: Guide to
  advanced empirical software engineering, Springer, pp 63--92

\bibitem[{Kotonya and Lee(2014)}]{kotonya2014teachingReuseDrivenSE}
Kotonya G, Lee J (2014) Teaching reuse-driven software engineering through
  innovative role playing. In: Companion Proceedings of the 36th International
  Conference on Software Engineering, pp 276--282

\bibitem[{Krueger(1992)}]{krueger1992SoftwareReuse}
Krueger CW (1992) Software reuse. ACM Computing Surveys (CSUR) 24(2):131--183

\bibitem[{Langford et~al.(2024)Langford, Shumailov, Zhao, Mullins, and
  Papernot}]{langford2024architecturalNeuralBackdoors}
Langford H, Shumailov I, Zhao Y, Mullins R, Papernot N (2024) Architectural
  neural backdoors from first principles. arXiv preprint arXiv:240206957

\bibitem[{Lau(2006)}]{lau2006softwareComponentModels}
Lau KK (2006) Software component models. In: Proceedings of the 28th
  international conference on Software engineering, pp 1081--1082

\bibitem[{Lawrie et~al.(2006)Lawrie, Morrell, Feild, and
  Binkley}]{lawrie2006whatsinaName}
Lawrie D, Morrell C, Feild H, Binkley D (2006) What's in a name? a study of
  identifiers. In: 14th IEEE international conference on program comprehension
  (ICPC'06), IEEE, pp 3--12

\bibitem[{Lawrie et~al.(2007)Lawrie, Morrell, Feild, and
  Binkley}]{Lawrie2007IdentifierNamingStudy}
Lawrie D, Morrell C, Feild H, Binkley D (2007) Effective identifier names for
  comprehension and memory. Innovations in Systems and Software Engineering
  3(4):303--318, \doi{10.1007/s11334-007-0031-2},
  \urlprefix\url{http://link.springer.com/10.1007/s11334-007-0031-2}

\bibitem[{Le and Zhang(2022)}]{le2022logBasedAnomalyDetection}
Le VH, Zhang H (2022) Log-based anomaly detection with deep learning: How far
  are we? In: Proceedings of the 44th international conference on software
  engineering, pp 1356--1367

\bibitem[{Le~Scao et~al.(2022)Le~Scao, Fan, Akiki, Pavlick, Ili{\'c}, Hesslow,
  Castagn{\'e}, Luccioni, Yvon, Gall{\'e} et~al.}]{le2022bloom}
Le~Scao T, Fan A, Akiki C, Pavlick E, Ili{\'c} S, Hesslow D, Castagn{\'e} R,
  Luccioni AS, Yvon F, Gall{\'e} M, et~al. (2022) Bloom: A 176b-parameter
  open-access multilingual language model

\bibitem[{LeCun et~al.(2015)LeCun, Bengio, and Hinton}]{lecun2015deepLearning}
LeCun Y, Bengio Y, Hinton G (2015) Deep learning. Nature 521(7553):436--444,
  \doi{10.1038/nature14539}

\bibitem[{Lester et~al.(2021)Lester, Al-Rfou, and
  Constant}]{Lester2021PromptTuning}
Lester B, Al-Rfou R, Constant N (2021) The {Power} of {Scale} for
  {Parameter}-{Efficient} {Prompt} {Tuning}.
  \urlprefix\url{http://arxiv.org/abs/2104.08691}

\bibitem[{Li et~al.(2024)Li, Tang, Zhao, Nie, and
  Wen}]{li2024PTLM4TextGenerationSurvey}
Li J, Tang T, Zhao WX, Nie JY, Wen JR (2024) Pre-trained language models for
  text generation: A survey. ACM Computing Surveys 56(9):1--39

\bibitem[{Li et~al.(2022)Li, Hai, Bozzon, and Katsifodimos}]{li_metadata_2022}
Li Z, Hai R, Bozzon A, Katsifodimos A (2022) Metadata {Representations} for
  {Queryable} {ML} {Model} {Zoos}.
  \urlprefix\url{http://arxiv.org/abs/2207.09315}, arXiv:2207.09315 [cs]

\bibitem[{Liu et~al.(2016)Liu, Liu, Staicu, Pradel, and
  Luo}]{liu2016ExploringandExploitingSimBetweenArgandParameterNames}
Liu H, Liu Q, Staicu CA, Pradel M, Luo Y (2016) Nomen est omen: Exploring and
  exploiting similarities between argument and parameter names. In: Proceedings
  of the 38th International Conference on Software Engineering, pp 1063--1073

\bibitem[{Liu et~al.(2019{\natexlab{a}})Liu, Demirel, and Liang}]{liu2019ngram}
Liu S, Demirel MF, Liang Y (2019{\natexlab{a}}) N-gram graph: Simple
  unsupervised representation for graphs, with applications to molecules.
  Advances in neural information processing systems 32

\bibitem[{Liu et~al.(2019{\natexlab{b}})Liu, Ott, Goyal, Du, Joshi, Chen, Levy,
  Lewis, Zettlemoyer, and Stoyanov}]{liu2019roberta}
Liu Y, Ott M, Goyal N, Du J, Joshi M, Chen D, Levy O, Lewis M, Zettlemoyer L,
  Stoyanov V (2019{\natexlab{b}}) Roberta: A robustly optimized bert
  pretraining approach. arXiv preprint arXiv:190711692

\bibitem[{Lu et~al.(2019)Lu, Yang, Chen, and
  Ren}]{Lu2019AutoDNNSelection4EdgeInference}
Lu B, Yang J, Chen LY, Ren S (2019) Automating {Deep} {Neural} {Network}
  {Model} {Selection} for {Edge} {Inference}. In: 2019 {IEEE} {First}
  {International} {Conference} on {Cognitive} {Machine} {Intelligence}
  ({CogMI}), pp 184--193, \doi{10.1109/CogMI48466.2019.00035}

\bibitem[{Ma et~al.(2022)Ma, Rong, and Huang}]{ma2022GNN}
Ma H, Rong Y, Huang J (2022) Graph neural networks: Scalability. Graph Neural
  Networks: Foundations, Frontiers, and Applications pp 99--119

\bibitem[{Melegati et~al.(2019)Melegati, Wang, and
  Abrahamsson}]{melegati2019hypotheses}
Melegati J, Wang X, Abrahamsson P (2019) Hypotheses engineering: first
  essential steps of experiment-driven software development. In: 2019 IEEE/ACM
  Joint 4th International Workshop on Rapid Continuous Software Engineering and
  1st International Workshop on Data-Driven Decisions, Experimentation and
  Evolution (RCoSE/DDrEE), IEEE, pp 16--19

\bibitem[{{Meta AI}(2024)}]{facebook2024hf}
{Meta AI} (2024) Meta ai on hugging face.
  \urlprefix\url{https://huggingface.co/facebook}, accessed: 2025-04-14

\bibitem[{Montes et~al.(2022)Montes, Peerapatanapokin, Schultz, Guo, Jiang, and
  Davis}]{Montes2022DiscrepanciesAmongPTNN}
Montes D, Peerapatanapokin P, Schultz J, Guo C, Jiang W, Davis JC (2022)
  Discrepancies among pre-trained deep neural networks: a new threat to model
  zoo reliability. In: European Software Engineering Conference and Symposium
  on the Foundations of Software Engineering (ESEC/FSE-IVR track).

\bibitem[{Neupane et~al.(2023)Neupane, Holmes, Wyss, Davidson, and
  De~Carli}]{neupane2023beyondTyposquat}
Neupane S, Holmes G, Wyss E, Davidson D, De~Carli L (2023) Beyond
  typosquatting: an in-depth look at package confusion. In: 32nd USENIX
  Security Symposium (USENIX Security 23), pp 3439--3456

\bibitem[{Nguyen et~al.(2022)Nguyen, Islam, Pan, and Rajan}]{Nguyen2022Manas}
Nguyen G, Islam MJ, Pan R, Rajan H (2022) Manas: mining software repositories
  to assist {AutoML}. In: International {Conference} on {Software}
  {Engineering} ({ICSE}), ACM, Pittsburgh Pennsylvania, pp 1368--1380,
  \doi{10.1145/3510003.3510052},
  \urlprefix\url{https://dl.acm.org/doi/10.1145/3510003.3510052}

\bibitem[{Ni et~al.(2021)Ni, Shu, Souri, Goldblum, and
  Goldstein}]{ni2021RelationshipbetweenCLandMetaL}
Ni R, Shu M, Souri H, Goldblum M, Goldstein T (2021) The close relationship
  between contrastive learning and meta-learning. In: International Conference
  on Learning Representations (ICLR'21)

\bibitem[{{NPM}(2023)}]{NPM_package_guidelines}
{NPM} (2023) Package name guidelines.
  \urlprefix\url{https://docs.npmjs.com/package-name-guidelines}

\bibitem[{{NVIDIA}(2024)}]{nemo_issue10006}
{NVIDIA} (2024) Inconsistent model config and architecture - issue \#10006.
  \url{https://github.com/NVIDIA/NeMo/issues/10006}, accessed: 2025-04-14

\bibitem[{Ohm et~al.(2020)Ohm, Plate, Sykosch, and
  Meier}]{Maurice2020OSSSupplyChainAttacks}
Ohm M, Plate H, Sykosch A, Meier M (2020) Backstabber’s {Knife} {Collection}:
  {A} {Review} of {Open} {Source} {Software} {Supply} {Chain} {Attacks}. In:
  Maurice C, Bilge L, Stringhini G, Neves N (eds) Detection of {Intrusions} and
  {Malware}, and {Vulnerability} {Assessment}, vol 12223, Springer
  International Publishing, Cham, pp 23--43,
  \urlprefix\url{http://link.springer.com/10.1007/978-3-030-52683-2_2}

\bibitem[{Olsson and Bosch(2014)}]{olsson2014opinions}
Olsson HH, Bosch J (2014) From opinions to data-driven software r\&d: A
  multi-case study on how to close the'open loop'problem. In: 2014 40th
  EUROMICRO Conference on Software Engineering and Advanced Applications, IEEE,
  pp 9--16

\bibitem[{ONNX(2023)}]{ONNXModelZoo}
ONNX (2023) Onnx model zoo. \url{https://github.com/onnx/models}

\bibitem[{{OpenBMB}(2023)}]{OpenBMB2023}
{OpenBMB} (2023) Issue \#196 - project author team stay tuned: I found out that
  the llama3-v project is stealing a lot of academic work from minicpm-llama3-v
  2.5. \url{https://github.com/OpenBMB/MiniCPM-V/issues/196}, miniCPM-V. GitHub

\bibitem[{Patil et~al.(2025)Patil, Jiang, Peng, Lugo, Kalu, LeBlanc, Smith,
  Heo, Aou, and Davis}]{patil2025recommendingPTM4IoT}
Patil PV, Jiang W, Peng H, Lugo D, Kalu KG, LeBlanc J, Smith L, Heo H, Aou N,
  Davis JC (2025) Recommending pre-trained models for iot devices. In:
  Proceedings of the International Workshop on Software Engineering Research in
  Practice (SERIP), to appear

\bibitem[{Peng(2021)}]{longformer_issue185}
Peng B (2021) About the model name.
  \url{https://github.com/allenai/longformer/issues/185}

\bibitem[{Pradel and
  Gross(2011)}]{pradel2011detectingAnomaliesEquallyTypedMethodArgs}
Pradel M, Gross TR (2011) Detecting anomalies in the order of equally-typed
  method arguments. In: Proceedings of the 2011 International Symposium on
  Software Testing and Analysis, pp 232--242

\bibitem[{Pradel and Sen(2018)}]{pradel2018deepbugs}
Pradel M, Sen K (2018) Deepbugs: A learning approach to name-based bug
  detection. Proceedings of the ACM on Programming Languages 2(OOPSLA):1--25

\bibitem[{{Preferred Networks}(2021)}]{PyTorchGraphs2021}
{Preferred Networks} (2021) How computational graphs are constructed in
  pytorch.
  \urlprefix\url{https://pytorch.org/blog/computational-graphs-constructed-in-pytorch/}

\bibitem[{{ProtectAI}(2024)}]{protectai_huggingface}
{ProtectAI} (2024) Unveiling ai supply chain attacks on hugging face.
  \urlprefix\url{https://protectai.com/threat-research/unveiling-ai-supply-chain-attacks-on-hugging-face}

\bibitem[{Pytorch(2021)}]{PytorchHub}
Pytorch (2021) Pytorch hub. \urlprefix\url{https://pytorch.org/hub/}

\bibitem[{Qi et~al.(2023)Qi, Sun, Gao, Zhang, Li, and Liu}]{qi2023reusing}
Qi B, Sun H, Gao X, Zhang H, Li Z, Liu X (2023) Reusing deep neural network
  models through model re-engineering. In: 2023 IEEE/ACM 45th International
  Conference on Software Engineering (ICSE), IEEE, pp 983--994

\bibitem[{Radford et~al.(2019)Radford, Wu, Child, Luan, Amodei, and
  Sutskever}]{Radford2019LMareUnsupervisedMultitaskLearners}
Radford A, Wu J, Child R, Luan D, Amodei D, Sutskever I (2019) Language
  {Models} are {Unsupervised} {Multitask} {Learners}. OpenAI blog 1(8):9

\bibitem[{Ren et~al.(2019)Ren, Xu, Wang, Yi, Huang, Kou, Xing, Yang, Tong, and
  Zhang}]{ren2019timeSeriesAnomalyDetection}
Ren H, Xu B, Wang Y, Yi C, Huang C, Kou X, Xing T, Yang M, Tong J, Zhang Q
  (2019) Time-series anomaly detection service at microsoft. In: Proceedings of
  the 25th ACM SIGKDD international conference on knowledge discovery \& data
  mining, pp 3009--3017

\bibitem[{{ReversingLabs Threat Research
  Team}(2023)}]{reversinglabs2023typosquatting}
{ReversingLabs Threat Research Team} (2023) r77 rootkit typosquatting: Npm
  threat research.
  \urlprefix\url{https://www.reversinglabs.com/blog/r77-rootkit-typosquatting-npm-threat-research}

\bibitem[{Rogers et~al.(2021)Rogers, Kovaleva, and
  Rumshisky}]{rogers2021primerinBERTology}
Rogers A, Kovaleva O, Rumshisky A (2021) A primer in bertology: What we know
  about how bert works. Transactions of the Association for Computational
  Linguistics 8:842--866

\bibitem[{Sabor et~al.(2017)Sabor, Hamou-Lhadj, and Larsson}]{sabor2017durfex}
Sabor KK, Hamou-Lhadj A, Larsson A (2017) Durfex: a feature extraction
  technique for efficient detection of duplicate bug reports. In: 2017 IEEE
  international conference on software quality, reliability and security (QRS),
  IEEE, pp 240--250

\bibitem[{Saieva et~al.(2023)Saieva, Chakraborty, and
  Kaiser}]{saieva2023contrastive}
Saieva A, Chakraborty S, Kaiser G (2023) On contrastive learning of semantic
  similarity forcode to code search. arXiv preprint arXiv:230503843

\bibitem[{Salda{\~n}a(2021)}]{saldana2021coding}
Salda{\~n}a J (2021) The coding manual for qualitative researchers. SAGE
  publications Ltd

\bibitem[{Schelter et~al.(2018)Schelter, Biessmann, Januschowski, Salinas,
  Seufert, and Szarvas}]{Amazon2018MLModelManagementChallenge}
Schelter S, Biessmann F, Januschowski T, Salinas D, Seufert S, Szarvas G (2018)
  On {Challenges} in {Machine} {Learning} {Model} {Management}. Bulletin of the
  IEEE Computer Society Technical Committee on Data Engineering

\bibitem[{Seacord et~al.(1998)Seacord, Hissam, and
  Wallnau}]{seacord1998agoraASearchEngine4SWComponents}
Seacord RC, Hissam SA, Wallnau KC (1998) Agora: A search engine for software
  components. IEEE Internet computing 2(6):62

\bibitem[{Sejfia and Sch{\"a}fer(2022)}]{sejfia2022practical}
Sejfia A, Sch{\"a}fer M (2022) Practical automated detection of malicious npm
  packages. In: Proceedings of the 44th International Conference on Software
  Engineering, pp 1681--1692

\bibitem[{Sens et~al.(2024)Sens, Knopp, Peldszus, and
  Berger}]{sens2024largeScaleStudyofModelIntegrationinMLSystem}
Sens Y, Knopp H, Peldszus S, Berger T (2024) A large-scale study of model
  integration in ml-enabled software systems. arXiv preprint arXiv:240806226

\bibitem[{shersoni610 and Li(2024)}]{llava_issue1136}
shersoni610, Li C (2024) {Issue \#1136: Naming llava models}.
  \url{https://github.com/haotian-liu/LLaVA/issues/1136}

\bibitem[{Sohacheski et~al.(2021)Sohacheski, Lurie, and
  Mark}]{sohacheski2021software}
Sohacheski DB, Lurie Y, Mark S (2021) Software identifier naming conventions \&
  dictionary. WSEAS Transactions on Computer Research 9:21--32

\bibitem[{Sommerville(2015)}]{sommerville2015software}
Sommerville I (2015) Software Engineering, 10th edn. Pearson Education Limited

\bibitem[{Soto-Valero et~al.(2019)Soto-Valero, Benelallam, Harrand, Barais, and
  Baudry}]{soto2019emergenceSWDiversityinMaven}
Soto-Valero C, Benelallam A, Harrand N, Barais O, Baudry B (2019) The emergence
  of software diversity in maven central. In: 2019 IEEE/ACM 16th International
  Conference on Mining Software Repositories (MSR), IEEE, pp 333--343

\bibitem[{Storey et~al.(2025)Storey, Hoda, Milani, and
  Baldassarre}]{storey2025guidingPrinciples4MixedMethodsResearchinSE}
Storey MA, Hoda R, Milani AMP, Baldassarre MT (2025) Guiding principles for
  using mixed methods research in software engineering. Empirical Software
  Engineering (EMSE)

\bibitem[{Sutter and Alexandrescu(2004)}]{sutter2004c++CodingStandards}
Sutter H, Alexandrescu A (2004) C++ coding standards: 101 rules, guidelines,
  and best practices. Pearson Education

\bibitem[{Szyperski et~al.(2002)Szyperski, Gruntz, and
  Murer}]{szyperski2002componentSoftware}
Szyperski C, Gruntz D, Murer S (2002) Component Software: Beyond
  Object-Oriented Programming, 2nd edn. Addison-Wesley

\bibitem[{Tan et~al.(2024)Tan, Li, Chen, Liu, and
  Zhang}]{tan2024challengesofUsingPTMs}
Tan X, Li T, Chen R, Liu F, Zhang L (2024) Challenges of using pre-trained
  models: the practitioners' perspective. arXiv preprint arXiv:240414710

\bibitem[{Taraghi et~al.(2024)Taraghi, Dorcelus, Foundjem, Tambon, and
  Khomh}]{taraghi2024DLModelReuseinHF}
Taraghi M, Dorcelus G, Foundjem A, Tambon F, Khomh F (2024) Deep learning model
  reuse in the huggingface community: Challenges, benefit and trends. arXiv
  preprint arXiv:240113177

\bibitem[{Taye(2023)}]{taye2023theoretical}
Taye MM (2023) Theoretical understanding of convolutional neural network:
  Concepts, architectures, applications, future directions. Computation
  11(3):52

\bibitem[{Taylor et~al.(2020{\natexlab{a}})Taylor, Vaidya, Davidson, De~Carli,
  and Rastogi}]{taylor_defending_2020}
Taylor M, Vaidya R, Davidson D, De~Carli L, Rastogi V (2020{\natexlab{a}})
  Defending {Against} {Package} {Typosquatting}. In: Kutyłowski M, Zhang J,
  Chen C (eds) Network and {System} {Security}, Springer International
  Publishing, Cham, Lecture {Notes} in {Computer} {Science}, pp 112--131,
  \doi{10.1007/978-3-030-65745-1_7}

\bibitem[{Taylor et~al.(2020{\natexlab{b}})Taylor, Vaidya, Davidson, De~Carli,
  and Rastogi}]{taylor2020defending}
Taylor M, Vaidya R, Davidson D, De~Carli L, Rastogi V (2020{\natexlab{b}})
  Defending against package typosquatting. In: Network and System Security:
  14th International Conference, NSS 2020, Melbourne, VIC, Australia, November
  25--27, 2020, Proceedings 14, Springer, pp 112--131

\bibitem[{TensorFlow(2023)}]{TFHub}
TensorFlow (2023) Tensorflow hub.
  \urlprefix\url{https://www.tensorflow.org/hub}

\bibitem[{Terdchanakul et~al.(2017)Terdchanakul, Hata, Phannachitta, and
  Matsumoto}]{terdchanakul2017bugReportClsusingNgram}
Terdchanakul P, Hata H, Phannachitta P, Matsumoto K (2017) Bug or not? bug
  report classification using n-gram idf. In: 2017 IEEE international
  conference on software maintenance and evolution (ICSME), IEEE, pp 534--538

\bibitem[{T{\'o}masd{\'o}ttir et~al.(2018)T{\'o}masd{\'o}ttir, Aniche, and
  Van~Deursen}]{tomasdottir2018JAVALinterAdoption}
T{\'o}masd{\'o}ttir KF, Aniche M, Van~Deursen A (2018) The adoption of
  javascript linters in practice: A case study on eslint. IEEE Transactions on
  Software Engineering 46(8):863--891

\bibitem[{Tsay et~al.(2020)Tsay, Braz, Hirzel, Shinnar, and
  Mummert}]{IBM2020AIMMX}
Tsay J, Braz A, Hirzel M, Shinnar A, Mummert T (2020) Aimmx: Artificial
  intelligence model metadata extractor. In: Proceedings of the 17th
  international conference on mining software repositories, pp 81--92

\bibitem[{Vassallo et~al.(2020)Vassallo, Proksch, Jancso, Gall, and
  Di~Penta}]{vassallo2020configuration}
Vassallo C, Proksch S, Jancso A, Gall HC, Di~Penta M (2020) Configuration
  smells in continuous delivery pipelines: a linter and a six-month study on
  gitlab. In: Proceedings of the 28th ACM Joint Meeting on European Software
  Engineering Conference and Symposium on the Foundations of Software
  Engineering, pp 327--337

\bibitem[{Vaswani(2017)}]{vaswani2017attention}
Vaswani A (2017) Attention is all you need. Advances in Neural Information
  Processing Systems

\bibitem[{Verdecchia et~al.(2023)Verdecchia, Engstr{\"o}m, Lago, Runeson, and
  Song}]{verdecchia2023threats}
Verdecchia R, Engstr{\"o}m E, Lago P, Runeson P, Song Q (2023) Threats to
  validity in software engineering research: A critical reflection. Information
  and Software Technology 164:107329

\bibitem[{Violos et~al.(2018)Violos, Tserpes, Varlamis, and
  Varvarigou}]{violos2018textclassificationUsingNgram}
Violos J, Tserpes K, Varlamis I, Varvarigou T (2018) Text classification using
  the n-gram graph representation model over high frequency data streams.
  Frontiers in Applied Mathematics and Statistics 4:41

\bibitem[{Wang et~al.(2023{\natexlab{a}})Wang, Chen, and
  Zhou}]{wang_automl_2023}
Wang C, Chen Z, Zhou M (2023{\natexlab{a}}) {AutoML} from {Software}
  {Engineering} {Perspective}: {Landscapes} and {Challenges}. In: 2023
  {IEEE}/{ACM} 20th {International} {Conference} on {Mining} {Software}
  {Repositories} ({MSR}), IEEE, Melbourne, Australia, pp 39--51,
  \doi{10.1109/MSR59073.2023.00019},
  \urlprefix\url{https://ieeexplore.ieee.org/document/10173951/}

\bibitem[{Wang and Manning(2012)}]{wang2012baselines}
Wang S, Manning CD (2012) Baselines and bigrams: Simple, good sentiment and
  topic classification. In: Proceedings of the 50th Annual Meeting of the
  Association for Computational Linguistics (ACL), pp 90--94

\bibitem[{Wang et~al.(2023{\natexlab{b}})Wang, Wen, Lin, Liu, Bissyandé, and
  Mao}]{Wang2023PreImplementationMethodNamePrediction4OOP}
Wang S, Wen M, Lin B, Liu Y, Bissyandé TF, Mao X (2023{\natexlab{b}})
  Pre-{Implementation} {Method} {Name} {Prediction} for {Object}-{Oriented}
  {Programming}. ACM Transactions on Software Engineering and Methodology
  (TOSEM) \doi{10.1145/3597203},
  \urlprefix\url{https://dl.acm.org/doi/10.1145/3597203}

\bibitem[{Wang et~al.(2022)Wang, Liu, Cui, Yin, and Wang}]{Wang2022EvilModel2}
Wang Z, Liu C, Cui X, Yin J, Wang X (2022) {EvilModel} 2.0: {Bringing} {Neural}
  {Network} {Models} into {Malware} {Attacks}. Computers \& Security
  \doi{10.1016/j.cose.2022.102807}

\bibitem[{Winkler et~al.(2021)Winkler, Agarwal, Tung, Ugalde, Jung, and
  Davis}]{winkler2021replication}
Winkler J, Agarwal A, Tung C, Ugalde DR, Jung YJ, Davis JC (2021) A replication
  of ‘deepbugs: a learning approach to name-based bug detection’. In:
  Proceedings of the 29th ACM Joint Meeting on European Software Engineering
  Conference and Symposium on the Foundations of Software Engineering, pp
  1604--1604

\bibitem[{Wittern et~al.(2016)Wittern, Suter, and
  Rajagopalan}]{Wittern2016JSPackageEcosystem}
Wittern E, Suter P, Rajagopalan S (2016) A look at the dynamics of the
  {JavaScript} package ecosystem. In: International {Conference} on {Mining}
  {Software} {Repositories} ({MSR})

\bibitem[{Wohlin et~al.(2012)Wohlin, Runeson, H{\"o}st, Ohlsson, Regnell,
  Wessl{\'e}n et~al.}]{wohlin2012experimentation}
Wohlin C, Runeson P, H{\"o}st M, Ohlsson MC, Regnell B, Wessl{\'e}n A, et~al.
  (2012) Experimentation in software engineering, vol 236. Springer

\bibitem[{Wu et~al.(2017)Wu, Cao, Chen, Wei, Zhong, and Huang}]{wu2017appcheck}
Wu G, Cao Y, Chen W, Wei J, Zhong H, Huang T (2017) Appcheck: a crowdsourced
  testing service for android applications. In: 2017 IEEE International
  Conference on Web Services (ICWS), IEEE, pp 253--260

\bibitem[{Wu et~al.(2020)Wu, Pan, Chen, Long, Zhang, and
  Philip}]{wu2020comprehensiveSurveyonGNN}
Wu Z, Pan S, Chen F, Long G, Zhang C, Philip SY (2020) A comprehensive survey
  on graph neural networks. IEEE transactions on neural networks and learning
  systems 32(1):4--24

\bibitem[{You et~al.(2021{\natexlab{a}})You, Liu, Wang, Jordan, and
  Long}]{You2021RankingandTuningPTMs}
You K, Liu Y, Wang J, Jordan MI, Long M (2021{\natexlab{a}}) Ranking and
  {Tuning} {Pre}-trained {Models}: {A} {New} {Paradigm} of {Exploiting} {Model}
  {Hubs}. The Journal of Machine Learning Research (JMLR) 23(1):9400--9446,
  \urlprefix\url{http://arxiv.org/abs/2110.10545}

\bibitem[{You et~al.(2021{\natexlab{b}})You, Liu, Wang, and
  Long}]{You2021LogME}
You K, Liu Y, Wang J, Long M (2021{\natexlab{b}}) {LogME}: {Practical}
  {Assessment} of {Pre}-trained {Models} for {Transfer} {Learning}. In:
  International {Conference} on {Machine} {Learning} ({ICML}), PMLR, pp
  12133--12143, \urlprefix\url{https://proceedings.mlr.press/v139/you21b.html}

\bibitem[{Zaigrajew and Zieba(2022)}]{zaigrajew2022contrastive}
Zaigrajew V, Zieba M (2022) Contrastive learning for multi-label
  classification. In: Proceedings of Conference on Neural Information
  Processing Systems, New Orleans, pp 1--8

\bibitem[{Zamfirescu-Pereira et~al.(2023)Zamfirescu-Pereira, Wong, Hartmann,
  and Yang}]{zamfirescu2023johnny}
Zamfirescu-Pereira J, Wong RY, Hartmann B, Yang Q (2023) Why johnny can’t
  prompt: how non-ai experts try (and fail) to design llm prompts. In:
  Proceedings of the 2023 CHI Conference on Human Factors in Computing Systems,
  pp 1--21

\bibitem[{Zhang et~al.(2015)Zhang, Zhao, and LeCun}]{zhang2015character}
Zhang X, Zhao J, LeCun Y (2015) Character-level convolutional networks for text
  classification. In: Advances in Neural Information Processing Systems
  (NeurIPS)

\bibitem[{Zhang et~al.(2024)Zhang, Huang, Ding, Zhan, and
  Ye}]{zhang2024modelSpider}
Zhang YK, Huang TJ, Ding YX, Zhan DC, Ye HJ (2024) Model spider: Learning to
  rank pre-trained models efficiently. Advances in Neural Information
  Processing Systems 36

\bibitem[{Zhao et~al.(2024)Zhao, Wang, Zhao, Hou, Wang, Gao, Zhang, Wei, and
  Wang}]{zhao2024modelsarecode}
Zhao J, Wang S, Zhao Y, Hou X, Wang K, Gao P, Zhang Y, Wei C, Wang H (2024)
  Models are codes: Towards measuring malicious code poisoning attacks on
  pre-trained model hubs. In: Proceedings of the 39th IEEE/ACM International
  Conference on Automated Software Engineering, pp 2087--2098

\bibitem[{Zhu et~al.(2015)Zhu, Zuo, Zhang, Hu, and Shiu}]{zhu2015unsupervised}
Zhu P, Zuo W, Zhang L, Hu Q, Shiu SC (2015) Unsupervised feature selection by
  regularized self-representation. Pattern Recognition 48(2):438--446

\end{thebibliography}

\end{document}